\documentclass[12pt,english]{article}
\usepackage{amsmath}
\usepackage{mathptmx}
\usepackage{newtxmath}

\usepackage[T1]{fontenc}
\usepackage[latin9]{inputenc}
\usepackage[letterpaper]{geometry}
\usepackage{color}
\usepackage{babel}
\usepackage{float}
\usepackage{graphicx}
\usepackage{setspace}
\usepackage[authoryear]{natbib}
\usepackage[unicode=true,pdfusetitle,
 bookmarks=true,bookmarksnumbered=false,bookmarksopen=false,
 breaklinks=true,pdfborder={0 0 0},pdfborderstyle={},backref=false,colorlinks=true]
 {hyperref}
\hypersetup{
 citecolor=DarkBlue,linkcolor=DarkBlue,urlcolor=DarkBlue}

\makeatletter

\providecommand{\tabularnewline}{\\}

\usepackage[svgnames]{xcolor}
\usepackage{bbm}

\@ifundefined{showcaptionsetup}{}{%
 \PassOptionsToPackage{caption=false}{subfig}}
\usepackage{subfig}
\makeatother

\begin{document}
\title{ICT Capital\textendash Skill Complementarity and Wage Inequality:\\
{\Large{}Evidence from OECD Countries}\thanks{We are grateful to Richard Blundell, Yongsung Chang, Richard Rogerson,
Naoki Takayama, anonymous referees, and seminar participants at Hitotsubashi
University, the EALE SOLE AASLE World Conference, and the Society
for Economic Measurement conference for their helpful comments. Yamada
gratefully acknowledges support from JSPS KAKENHI grant numbers 17H04782
and 21H00724.}\textit{\Large{}\bigskip{}
}}
\author{Hiroya Taniguchi\thanks{Kyoto University. \texttt{taniguchi.hiroya.24z@st.kyoto-u.ac.jp}}
\and Ken Yamada\thanks{Kyoto University. \texttt{yamada@econ.kyoto-u.ac.jp}}\textit{\Large{}\bigskip{}
}}
\date{January 2022}
\maketitle
\begin{abstract}
\begin{onehalfspace}
Although wage inequality has evolved in advanced countries over recent
decades, it remains unknown the extent to which changes in wage inequality
and their differences across countries are attributable to specific
capital and labor quantities. We examine this issue by estimating
a sector-level production function extended to allow for capital\textendash skill
complementarity and factor-biased technological change using cross-country
and cross-industry panel data. Our results indicate that most of the
changes in the skill premium are attributable to the relative quantities
of ICT equipment, skilled labor, and unskilled labor in the goods
and service sectors of the majority of advanced countries.\bigskip{}
\end{onehalfspace}

\global\long\def\E{\mathbb{E}}%
\end{abstract}
\captionsetup[subfigure]{labelformat=empty}

\newpage{}

\captionsetup[subfigure]{font=normalsize}

\captionsetup[subtable]{font=normalsize}

\section{Introduction}

One of the greatest changes in production activities over recent decades
is the introduction and expansion of information and communication
technology (ICT). The prevalence of new technology could result in
a change in the relative demand for different types of labor, and
thus, a change in their relative wages. According to the concept of
capital\textendash skill complementarity \citep{Griliches_RESTAT69},\footnote{See \citet{Acemoglu_JEL02}, \citet{Bond_VanReenen_HoE07}, and \citet{Goldin_Katz_bk10}
for surveys on capital\textendash skill (and technology\textendash skill)
complementarity.} if ICT equipment is more complementary to skilled labor than unskilled
labor, the increased use of ICT would raise the relative demand for
skilled to unskilled labor, and thus, the relative wages of skilled
to unskilled labor. Nevertheless, it is unknown the extent to which
the increased use of ICT can account for changes in the relative wages
of skilled to unskilled labor in many advanced countries.

We examine this issue by estimating a sector-level production function
extended to allow for complementarity between ICT and skills. Our
approach builds upon the aggregate production function developed by
\citet{Fallon_Layard_JPE75} and \citet*{Krusell_Ohanian_RiosRull_Violante_EM00}.
We focus on the relative wages of skilled to unskilled labor, which
is referred to as the skill premium or sometimes as the college premium.
The skill premium is a key metric used to evaluate the level of wage
inequality and its differences across countries. Changes in the skill
premium can be interpreted as the outcome of the race between supply
shifts due to advances in education and demand shifts due to developments
in technology \citep*{Tinbergen_K74,Goldin_Katz_bk10}, and thus,
can be decomposed into a part due to shifts in the relative supply
of skills and a part due to shifts in the relative demand for skills.
We refer to the supply shifts as the relative labor quantity effect.
The advantage of our approach is that it enables us to decompose the
demand shifts into a part attributable to observed factors such as
specific capital and labor quantities and a part attributable to unobserved
factors such as factor-augmenting technology. We refer to the former
as the capital\textendash skill complementarity effect and the latter
as the relative factor-augmenting technology effect. Consequently,
we can evaluate the extent to which changes in the skill premium are
attributable to the expansion of ICT equipment. The impact of policies
that increase skilled labor depends on the mechanisms through which
the skill premium has changed over recent decades. This study aims
to expand our understanding of the sources and mechanisms of changes
in the skill premium by measuring the quantitative contribution of
the three effects to changes in the skill premium in the goods and
service sectors of advanced countries.

For this purpose, we make use of cross-country and cross-industry
panel data from 14 OECD countries for the years 1970 to 2005. This
brings two major advantages to the analysis. First, we can exploit
large variation in relative factor quantities across countries over
time, which helps identify the elasticities of substitution between
different types of capital and labor in production. In doing so, we
exploit demand shocks in the service (goods) sector as a source of
exogenous variation in factor supply in the goods (service) sector.
Second, we can examine the sources and mechanisms of the differences
in changes in the skill premium across countries. It is unclear from
the literature the extent to which international differences in changes
in the skill premium can be attributed to specific capital and labor
quantities. Using the estimated production function, we decompose
the sources of differences in changes in the skill premium across
countries, as well as those of changes in the skill premium for each
country and sector.

The main findings of this study can be summarized as follows. First,
the estimated elasticity of substitution between ICT equipment and
skilled labor is much smaller than that between ICT equipment and
unskilled labor. This result supports the capital\textendash skill
complementarity hypothesis that ICT equipment is more complementary
to skilled labor than unskilled labor. Second, the capital\textendash skill
complementarity effect associated with a rise in ICT equipment is
large enough to account for a rise in the skill premium in the goods
and service sectors of most OECD countries. The capital\textendash skill
complementarity effect is greater for the goods sector than the service
sector in most OECD countries. Third, when ICT equipment is distinguished
from non-ICT capital, changes in the skill premium are attributable
mainly to observed capital and labor quantities in the goods and service
sectors of the majority of OECD countries. Fourth, as its consequences,
a large part of the differences in changes in the skill premium among
those countries can be explained by the relative quantities of ICT
equipment, skilled labor, and unskilled labor. Fifth, a rise in the
relative demand for skilled to unskilled labor is attributable to
a fall in the rental price of ICT equipment in the goods and service
sectors of most OECD countries. Finally, a rise in skilled labor has
a greater effect on reducing the skill premium in the presence than
in the absence of capital\textendash skill complementarity. This result
points to the possibility that the contribution of higher education
expansion to reducing the skill premium would be understated if capital\textendash skill
complementarity is not taken into account.

The remainder of this paper is organized as follows. The next section
briefly reviews the related literature. Section \ref{sec: model}
presents the sector-level production function used to account for
changes in the skill premium. Section \ref{sec: data} describes the
data and variables used in the analysis. Section \ref{sec: estimation}
outlines the econometric specifications and techniques. Section \ref{sec: results}
discusses the estimation results and their implications. The final
section provides a summary and conclusions.

\section{Related Literature}

A cornerstone for the analysis of the skill premium has been the aggregate
production function. \citet{Bound_Johnson_AER92} and \citet{Katz_Murphy_QJE92}
estimate an aggregate production function with two types of labor
(skilled and unskilled labor) to understand the sources and mechanisms
of changes in the skill premium in the United States from the 1960s
or 1970s to the 1980s. These studies demonstrate that changes in the
skill premium are partially attributable to the relative quantity
of skilled labor but mostly to skill-biased technological change.
For a given elasticity of substitution between skilled and unskilled
labor, \citet{Acemoglu_EJ03} and \citet{Caselli_Coleman_AER06} measure
relative labor-augmenting technology in many countries using an aggregate
production function with two types of labor. These studies consider
cross-country differences in changes in the skill premium as a consequence
of the differences in the direction of technological change.

\citet{Krusell_Ohanian_RiosRull_Violante_EM00} develop and estimate
a four-factor production function in which not only skilled labor
is distinguished from unskilled labor, but also capital equipment
is distinguished from capital structure. They demonstrate that capital
equipment is more complementary to skilled labor than unskilled labor,
and attribute the rise in the skill premium in the United States from
the years 1963 to 1992 mainly to a rise in capital equipment. \citet{Lindquist_SJE05}
finds similar results to \citet{Krusell_Ohanian_RiosRull_Violante_EM00}
in Sweden. \citet{Caselli_Coleman_AERPP02} measure labor-augmenting
technology in the United States during the same period using the same
production function as in \citet{Krusell_Ohanian_RiosRull_Violante_EM00}.
They find that technological change was biased towards skilled labor.

A cornerstone for the analysis of the impact of ICT on demand for
skills has been the factor-share equations derived from the translog
cost or production function. \citet*{Autor_Katz_Krueger_QJE98} find
that the wage-bill share of skilled labor increased with a rise in
the proportion of workers using computers in the United States between
the years 1984 and 1993. \citet*{Michaels_Natraj_VanReenen_RESTAT14}
find that the wage-bill share of high-skilled labor increased with
a rise in ICT equipment in 11 OECD countries between the years 1980
and 2004, while the wage-bill share of medium-skilled labor decreased.
\citet{RuizArranz_wp03} confirms the presence of capital\textendash skill
complementarity when information technology equipment is distinguished
from other types of capital. However, her results suggest that more
than half of the rise in the skill premium in the United States between
the years 1965 and 1999 is attributable to unobserved factors. \citet{PerezLaborda_PerezSebastian_JoM20}
find similar results to \citet{RuizArranz_wp03} in many sectors of
the United States between the years 1970 and 2005 when ICT equipment
is distinguished from non-ICT capital.

A growing body of research focuses on the allocation of skills to
tasks to examine the impact of new technology on the labor market.
\citet*{Autor_Levy_Murnane_QJE03} find that the increased share of
non-routine tasks and the decreased share of routine tasks are associated
with the widespread use of computers. Their results suggest that more
than half of the increased demand for skilled relative to unskilled
labor in the United States between the years 1970 and 1998 is attributable
to changes in the task composition. \citet*{Acemoglu_Autor_HLE11}
emphasize the importance of distinguishing between skills and tasks
to shed light on some phenomena such as the polarization of employment
and wages, while recognizing the value of the canonical model of the
race between education and technology for the analysis of wage inequality.
\citet{Eden_Gaggl_RED18} estimate an aggregate production function
in which capital is divided into ICT and non-ICT capital and labor
is divided into routine and non-routine tasks. They demonstrate that
about half of the decline in the labor income share in the United
States between the years 1950 and 2013 is attributable to a rise in
the income share of ICT equipment.

\citet{Acemoglu_JEL02} notes that skill-biased technological change
is a threat to the identification of capital\textendash skill complementarity
when using time-series data from a single country.\footnote{See also \citet*{Diamond_McFadden_Rodriguez_CEA78} for identification
issues.} \citet*{Duffy_Papageorgiou_PerezSebastian_RESTAT04} estimate a three-factor
production function with one type of capital and two types of labor
using cross-country panel data from the Penn World Tables 5.6, and
partially confirm the capital\textendash skill complementarity hypothesis.
However, the data used in their study do not contain information on
wages. \citet{Karabarbounis_Neiman_QJE14} estimate the elasticity
of substitution between capital and labor using cross-country panel
data from multiple sources on the labor share of income and the relative
price of investment to consumption goods, and demonstrate that about
half of the global decline in the labor share of income between the
years 1975 and 2012 is attributable to a fall in the relative price
of investment.

Compared with previous studies, this study aims to examine the extent
to which the expansion of ICT equipment can account for changes in
the skill premium for each country and sector and the differences
in such changes across countries. For this purpose, we estimate a
sector-level production function with two types of capital (ICT and
non-ICT capital) and two types of labor (skilled and unskilled labor)
using data from a panel of OECD countries.

\section{The Model\label{sec: model}}

We start our analysis by considering a four-factor production function
with capital\textendash skill complementarity. Perhaps surprisingly,
the production function presented here has not been estimated using
cross-country panel data.

\subsection{Production function}

The economy consists of two sectors: goods and services. We assume
that the output ($y$) is produced by a constant-returns-to-scale
technology using ICT capital ($k_{i}$), non-ICT capital ($k_{o}$),
skilled labor ($\ell_{h}$), and unskilled labor ($\ell_{u}$) for
each sector. Following \citet*{Fallon_Layard_JPE75} and \citet{Krusell_Ohanian_RiosRull_Violante_EM00},
the four-factor production function is specified as
\begin{equation}
y=Ak_{o}^{\alpha}\left\{ \lambda\left[\mu k_{i}^{\rho}+\left(1-\mu\right)\ell_{h}^{\rho}\right]^{\frac{\sigma}{\rho}}+\left(1-\lambda\right)\ell_{u}^{\sigma}\right\} ^{\frac{1-\alpha}{\sigma}}\qquad\text{for }\sigma,\rho<1,\label{eq: F(Ki,Ko,Lh,Lu;A)}
\end{equation}
where $A$ is factor-neutral technology. The parameter $\sigma$ governs
the degree of substitution between the $k_{i}$-$\ell_{h}$ composite
and $\ell_{u}$, while the parameter $\rho$ governs the degree of
substitution between $k_{i}$ and $\ell_{h}$. Production technology
exhibits capital\textendash skill complementarity if ICT capital is
more complementary to, or less substitutable with, skilled labor than
unskilled labor ($\sigma>\rho$). The specification of the production
function \eqref{eq: F(Ki,Ko,Lh,Lu;A)} is consistent with the data
in the sense that the parameter estimates satisfy $\sigma<1$ and
$\rho<1$.\footnote{The alternative specification, in which $\ell_{h}$ and $\ell_{u}$
are replaced with each other, is not, as also confirmed by \citet*{Fallon_Layard_JPE75},
\citet{Krusell_Ohanian_RiosRull_Violante_EM00}, and \citet{Duffy_Papageorgiou_PerezSebastian_RESTAT04}.} The advantage of this approach over the translog approach is that
it enables us to incorporate many factors with a small number of parameters
and directly relate the skill premium to relative factor quantities.\footnote{Although it is possible to indirectly measure the contribution of
factor quantities to changes in the skill premium using the estimated
factor-share equations, the translog approach basically estimates
the effects of factor prices on factor income shares.}

The four-factor production function can also be represented as
\begin{equation}
f\left(k_{i},k_{o},\ell_{h},\ell_{u}\right)=k_{o}^{\alpha}\left\{ \left[\left(A_{i}k_{i}\right)^{\rho}+\left(A_{h}\ell_{h}\right)^{\rho}\right]^{\frac{\sigma}{\rho}}+\left(A_{u}\ell_{u}\right)^{\sigma}\right\} ^{\frac{1-\alpha}{\sigma}},\label{eq: F(Ki,Ko,Lh,Lu;Ai,Ah,Au)}
\end{equation}
where factor-augmenting technology has the following form: $A_{i}=A^{\frac{1}{1-\alpha}}\lambda^{\frac{1}{\sigma}}\mu^{\frac{1}{\rho}}$,
$A_{h}=A^{\frac{1}{1-\alpha}}\lambda^{\frac{1}{\sigma}}(1-\mu)^{\frac{1}{\rho}}$,
and $A_{u}=A^{\frac{1}{1-\alpha}}(1-\lambda)^{\frac{1}{\sigma}}$.
Let $w_{h}$ and $w_{u}$ denote the wages of skilled and unskilled
labor, respectively, and $r_{i}$ and $r_{o}$ denote the rental prices
of ICT and non-ICT capital, respectively. Profit maximization entails
equating the value of marginal product with the marginal cost.
\begin{eqnarray}
w_{h} & = & \omega_{h}\frac{\partial f}{\partial\ell_{h}},\label{eq: FOC_Lh_4}\\
w_{u} & = & \omega_{u}\frac{\partial f}{\partial\ell_{u}},\label{eq: FOC_Lu_4}\\
r_{i} & = & \omega_{i}\frac{\partial f}{\partial k_{i}},\label{eq: FOC_Ki_4}\\
r_{o} & = & \omega_{o}\frac{\partial f}{\partial k_{o}},\label{eq: FOC_Ko_4}
\end{eqnarray}
where $\omega$ is the wedge that represents the deviation from the
profit-maximizing conditions in competitive markets. We allow the
size of the wedge to differ across factor markets. Appendix \ref{subsec: FOC}
provides a detailed derivation of the first-order conditions.

The first-order conditions \eqref{eq: FOC_Lh_4} and \eqref{eq: FOC_Lu_4}
imply that the relative wages of skilled to unskilled labor are proportional
to the marginal rate of technical substitution of unskilled for skilled
labor. After simple algebra, the relative wages of skilled to unskilled
labor are given by
\begin{equation}
\ln\left(\frac{w_{h}}{w_{u}}\right)=\sigma\ln\left(\frac{A_{h}}{A_{u}}\right)+\frac{\sigma-\rho}{\rho}\ln\left[\left(\frac{A_{i}k_{i}}{A_{h}\ell_{h}}\right)^{\rho}+1\right]-\left(1-\sigma\right)\ln\left(\frac{\ell_{h}}{\ell_{u}}\right)+\ln\left(\frac{\omega_{h}}{\omega_{u}}\right).\label{eq: Wh/Wu}
\end{equation}
Within a reasonable range of parameter values ($0<\sigma<1$ and $\sigma>\rho$),
the skill premium ($w_{h}/w_{u}$) increases with the ratio of skilled
to unskilled labor-augmenting technology ($A_{h}/A_{u}$), the ratio
of ICT capital- to skilled labor-augmenting technology ($A_{i}/A_{h}$),
and the ratio of ICT capital to skilled labor ($k_{i}/\ell_{h}$)
but decreases with the ratio of skilled to unskilled labor ($\ell_{h}/\ell_{u}$),
holding the relative wedge ($\omega_{h}/\omega_{u}$) constant. We
refer to the sum of the first and second effects associated with unobserved
technology ($A_{h}/A_{u}$ and $A_{i}/A_{h}$) as the relative factor-augmenting
technology effect. We refer to the third and fourth effects associated
with observed factors ($k_{i}/\ell_{h}$ and $\ell_{h}/\ell_{u}$)
as the capital\textendash skill complementarity effect and the relative
labor quantity effect, respectively. The capital\textendash skill
complementarity effect is proportional in magnitude to the difference
in substitution parameters ($\sigma-\rho$), as well as $k_{i}/\ell_{h}$.
The relative labor quantity effect is inversely proportional in magnitude
to the substitution parameter $\sigma$, as well as proportional in
magnitude to $\ell_{h}/\ell_{u}$. Whether the skill premium increases
or decreases depends on the outcome of the race between demand shifts
driven by developments in technology and supply shifts driven by advances
in education. Demand shifts arise from the capital\textendash skill
complementarity effect and the relative factor-augmenting technology
effect, while supply shifts arise from the relative labor quantity
effect. If the capital\textendash skill complementarity effect is
ignored by assuming that ICT equipment is equally substitutable with
skilled and unskilled labor (i.e., $\sigma=\rho$), the relative labor-augmenting
technology effect is likely to be overstated because the residual
unexplained by observed factors increases.

From a policy perspective, the expansion of tertiary education can
prevent a rise in the skill premium as a result of raising skilled
labor. Equation \eqref{eq: Wh/Wu} implies that the expansion of tertiary
education has a dual effect on the skill premium as it not only raises
the relative labor quantity effect but also reduces the capital\textendash skill
complementarity effect. If the capital\textendash skill complementarity
effect is not taken into account, the role of education policies is
likely to be understated because one of the two effects disappears.

The first-order conditions \eqref{eq: FOC_Lh_4} and \eqref{eq: FOC_Ki_4}
imply that the ratio of the wages of skilled labor to the rental price
of ICT capital is given by
\begin{equation}
\ln\left(\frac{w_{h}}{r_{i}}\right)=\rho\ln\left(\frac{A_{h}}{A_{i}}\right)-\left(1-\rho\right)\ln\left(\frac{\ell_{h}}{k_{i}}\right)+\ln\left(\frac{\omega_{h}}{\omega_{i}}\right).\label{eq: Wh/Ri}
\end{equation}
Equation \eqref{eq: Wh/Ri} is used to estimate the substitution parameter
$\rho$, while equation \eqref{eq: Wh/Wu} is used to estimate the
substitution parameter $\sigma$.

\subsection{Factor demand function}

The production function described above involves more than two factors.
We measure the degree of substitution between different types of capital
and labor using the \citet{Morishima_KH67} elasticity of substitution.\footnote{See \citet{Blackorby_Russell_AER89} for discussions on the desirable
properties of Morishima elasticity relative to other types of elasticities.} The Morishima elasticity of substitution between two factors $a$
and $b$ is defined as
\begin{equation}
\epsilon_{ab}=-\frac{\partial\ln\left(\left.x_{a}\left(\boldsymbol{p},y,\boldsymbol{A}\right)\right/x_{b}\left(\boldsymbol{p},y,\boldsymbol{A}\right)\right)}{\partial\ln\left(\left.p_{a}\right/p_{b}\right)},\label{eq: morishima}
\end{equation}
where $x_{a}$ denotes demand for factor $a$, and $p_{a}$ denotes
the price of input $a$. Factor demand depends on the vector of factor
prices ($\boldsymbol{p}$), output ($y$), and the vector of factor-augmenting
technologies ($\boldsymbol{A}$). Appendix \ref{subsec: factor} provides
the exact expression for the factor demand function and the Morishima
elasticities.

The relative demand for skilled to unskilled labor can be expressed
in a simpler form:
\begin{equation}
\frac{\ell_{h}}{\ell_{u}}=f\left(w_{h},w_{u},r_{i};\frac{A_{h}}{A_{u}},\frac{A_{i}}{A_{h}}\right).\label{eq: Lh/Lu}
\end{equation}
This equation is used to measure the contribution of factor prices
to changes in the relative demand for skilled labor.

\section{Data\label{sec: data}}

We start this section by describing the sample and variables used
in the analysis. We then discuss changes in factor prices and quantities
in the United States and other OECD countries.

\subsection{Sample and variables}

The data used in the analysis are from the EU KLEMS database. This
database is constructed from information collected by national statistical
offices and is grounded in national accounts statistics \citep{OMahony_Timmer_EJ09}.
The advantage of the EU KLEMS database is that it provides detailed
and internationally comparable information on the prices and quantities
of capital and labor in many OECD countries. We use the March 2008
version because it contains the longest time series from the years
1970 to 2005. There is a difference across countries in the number
of years for which data are available. We include as many countries
and years as possible in the sample. Our sample comprises 14 OECD
countries: Australia, Austria, the Czech Republic, Denmark, Finland,
Germany, Italy, Japan, the Netherlands, Portugal, Slovenia, Sweden,
the United Kingdom, and the United States.\footnote{The main results remain almost unchanged irrespective of whether the
Czech Republic and Slovenia are included or excluded.} Each economy is divided into two sectors. The goods sector consists
of 17 industries, while the service sector consists of 13 industries.\footnote{The goods sector includes five broad categories: agriculture, hunting,
forestry, and fishing; mining and quarrying; manufacturing; electricity,
gas, and water supply; and construction. The service sector includes
nine broad categories: wholesale and retail trade; hotels and restaurants;
transport and storage, and communication; financial intermediation;
real estate, renting, and business activities; public administration
and defense and compulsory social security; education; health and
social work; and other community and social and personal services.} In total, the sample comprises 682 country-sector-year observations.

Labor is divided into high-, medium-, and low-skilled labor. High-skilled
labor consists of workers who completed college, medium-skilled labor
consists of workers who entered college or completed high-school education,
and low-skilled labor consists of workers who dropped out of high
school or attended only compulsory education. As is standard in the
analysis of the skill premium, we classify high-skilled labor as skilled
labor and medium- and low-skilled labor as unskilled labor. Wages
are calculated at each skill level by dividing total labor compensation
by total hours worked. It would be worth mentioning that the relative
wages reported here need not be the same as those in other studies
because all workers (including part-time, self-employed, and family
workers without age restrictions) in all industries (except private
households with employed persons) and all jobs (including side jobs)
are included in the calculation. Nonetheless, the pattern of changes
in the skill premium in the United States aligns with the well-known
trend that the skill premium declined in the 1970s and rose from the
1980s to the 2000s. We adjust for changes in labor composition and
efficiency that can occur due to demographic changes in calculating
the wages and quantities of labor. Appendix \ref{subsec: adjustment}
provides a detailed description of the adjustment.

Capital is divided into capital equipment (e.g., machines) and capital
structure (e.g., buildings). Capital equipment is further divided
into ICT and non-ICT equipment. We classify ICT equipment as ICT capital
and non-ICT equipment and capital structure as non-ICT capital. The
share of ICT equipment was close to zero, ranging from 0.2 to 2.4
percent, in the 1970s in all countries but increased to a level ranging
from 5 to 21 percent in the 2000s. The rental price of capital, also
known as the user cost of capital, is calculated without the assumption
of competitive markets, as described in \citet{Jorgenson_AERPP63}
and \citet{Niebel_Saam_RIW16}. Appendix \ref{subsec: rental_price}
provides a detailed description of the calculation.

All variables measured in monetary values are converted into U.S.
dollars using the purchasing power parity index and deflated using
the gross value-added deflator as described in \citet*{Timmer_etal_bk07}.
The base year is 1995.

\begin{figure}[h]
\caption{Skill premium in OECD countries\label{fig: Wh/Wu_oecd}}

\begin{centering}
\subfloat[{\small{}(a) Increasing trend}]{
\centering{}\includegraphics[scale=0.5]{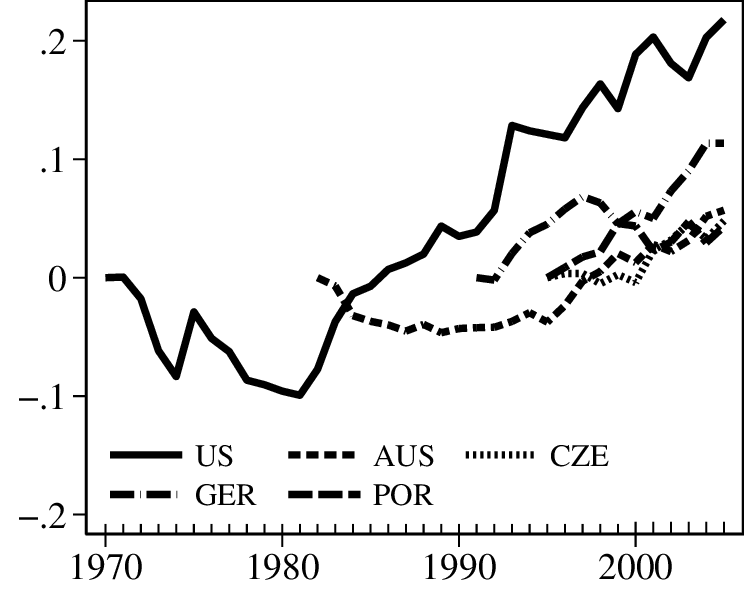}}\qquad{}\subfloat[{\small{}(b) No clear trend}]{
\centering{}\includegraphics[scale=0.5]{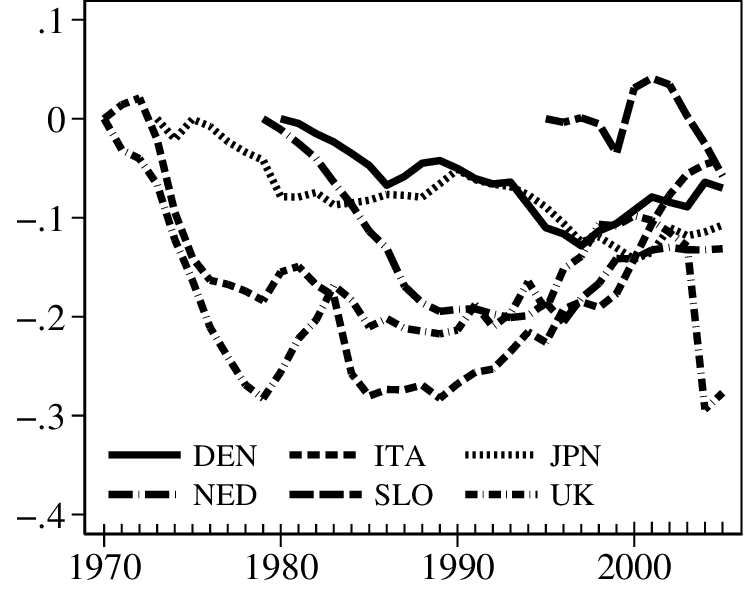}}
\par\end{centering}
\begin{centering}
\subfloat[{\small{}(c) Decreasing trend}]{
\centering{}\includegraphics[scale=0.5]{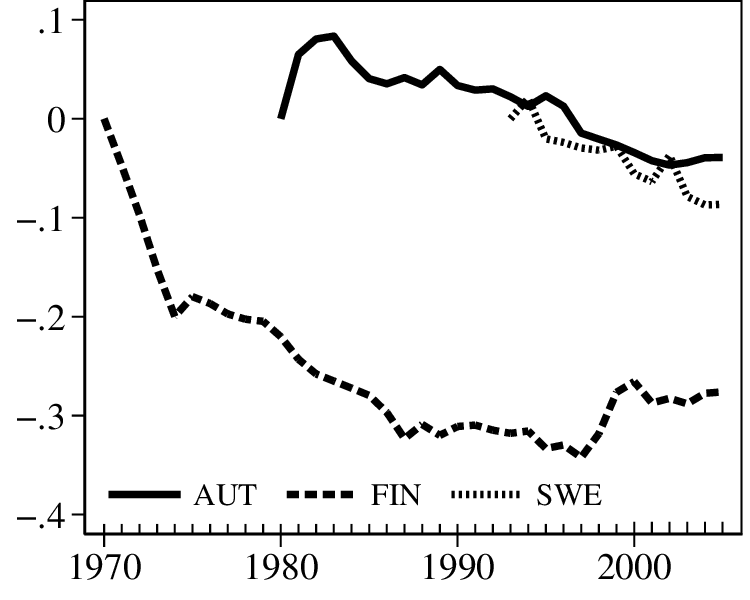}}
\par\end{centering}
\textit{\footnotesize{}Notes}{\footnotesize{}: AUS: Australia, AUT:
Austria, CZE: the Czech Republic, DEN: Denmark, FIN: Finland, GER:
Germany, ITA: Italy, JPN: Japan, NED: the Netherlands, POR: Portugal,
SLO: Slovenia, SWE: Sweden, UK: the United Kingdom, and US: the United
States. All series are logged and normalized to zero in the initial
year of observation.}{\footnotesize\par}
\end{figure}

\subsection{Trends in factor prices and quantities\label{subsec: trends_us}}

Trends in the skill premium vary across countries (Figure \ref{fig: Wh/Wu_oecd}).
The 14 countries are broadly divided into three groups for ease of
visibility. The skill premium exhibits an increasing trend in five
countries (Australia, the Czech Republic, Germany, Portugal, and the
United States), no clear trend in six countries (Denmark, Italy, Japan,
the Netherlands, Slovenia, and the United Kingdom), and a decreasing
trend in three countries (Austria, Finland, and Sweden). As shown
later, trends in the skill premium do not differ substantially across
sectors.

\begin{figure}[h]
\caption{Rental prices of ICT and non-ICT capital in the United States\label{fig: Ri/Ro_us}}

\begin{centering}
\subfloat[{\small{}(a) }Goods sector]{
\centering{}\includegraphics[scale=0.5]{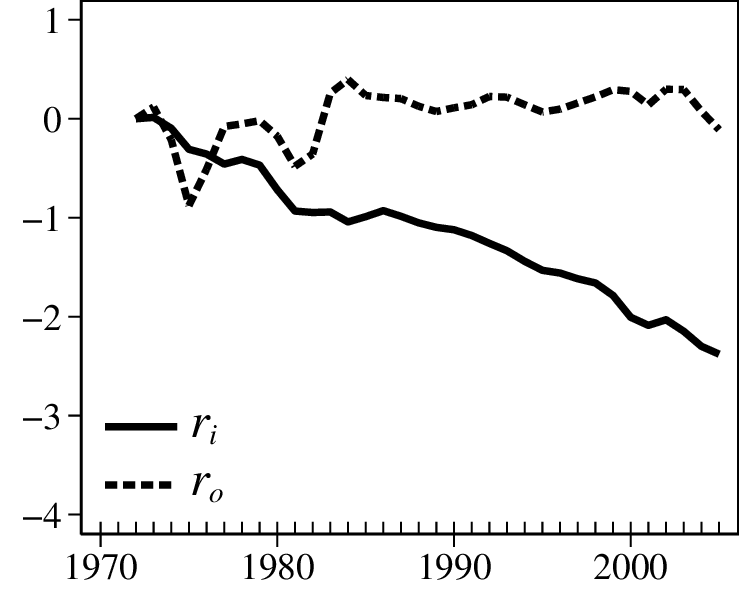}}\qquad{}\subfloat[{\small{}(b) }Service sector]{
\centering{}\includegraphics[scale=0.5]{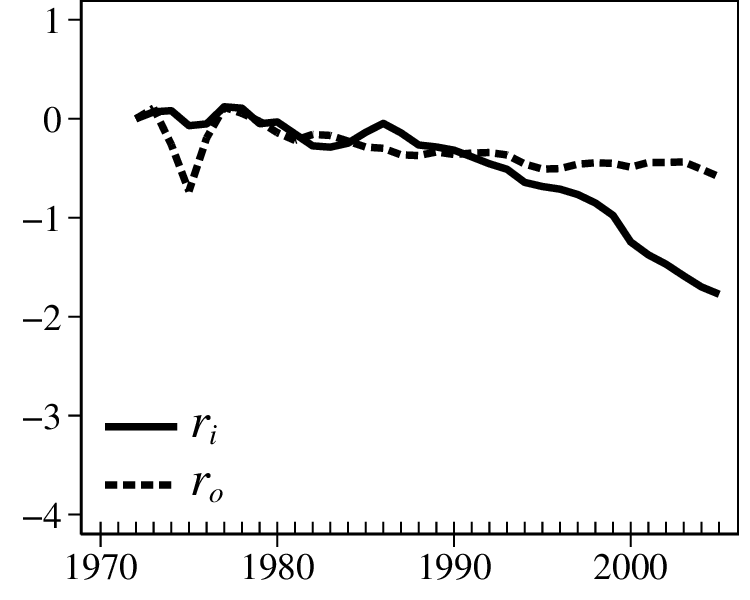}}
\par\end{centering}
\textit{\footnotesize{}Notes}{\footnotesize{}: The bold and dashed
lines indicate the rental price of ICT capital ($r_{i}$) and the
rental price of non-ICT capital ($r_{o}$), respectively. All series
are logged and normalized to zero in the initial year.}{\footnotesize\par}
\end{figure}

Trends in the rental price of capital vary across capital types in
the United States (Figure \ref{fig: Ri/Ro_us}). The rental price
of ICT capital fell steadily from the 1970s to the 2000s in the goods
sector and from the mid-1980s to the 2000s in the service sector.
Meanwhile, the rental price of non-ICT capital was roughly unchanged
in the goods sector and slightly declined in the service sector from
the 1970s to the 2000s. In all countries and sectors, the rental price
of ICT capital declined much more than that of non-ICT capital (see
Figures \ref{fig: Ri/Ro_oecd_goods} and \ref{fig: Ri/Ro_oecd_service}
in Appendix \ref{subsec: trends_oecd} for countries other than the
United States). The rate of decline in the rental price of ICT capital
was greater for the goods sector than the service sector in the United
States but not necessarily so in other countries.

\begin{figure}[h]
\caption{Relative factor quantities in the United States\label{fig: LhLuKiKe_us}}

\begin{centering}
\subfloat[{\small{}(a) }Goods sector]{
\centering{}\includegraphics[scale=0.5]{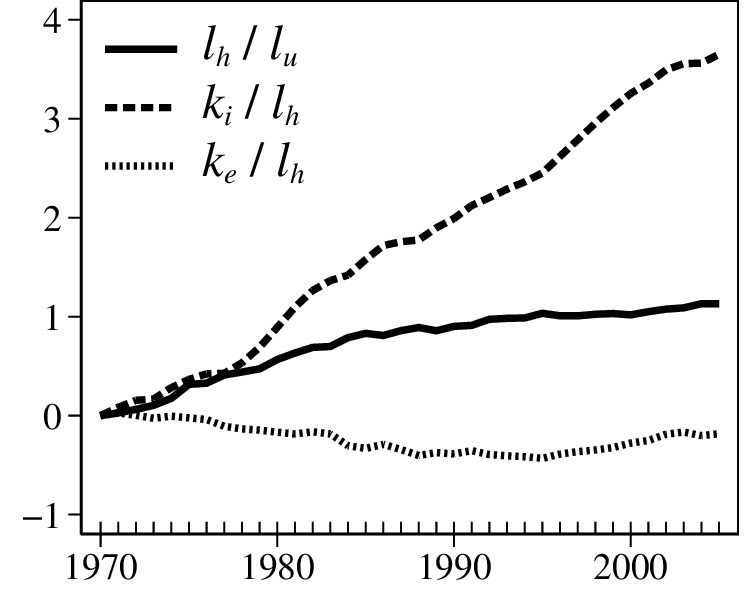}}\qquad{}\subfloat[{\small{}(b) }Service sector]{
\centering{}\includegraphics[scale=0.5]{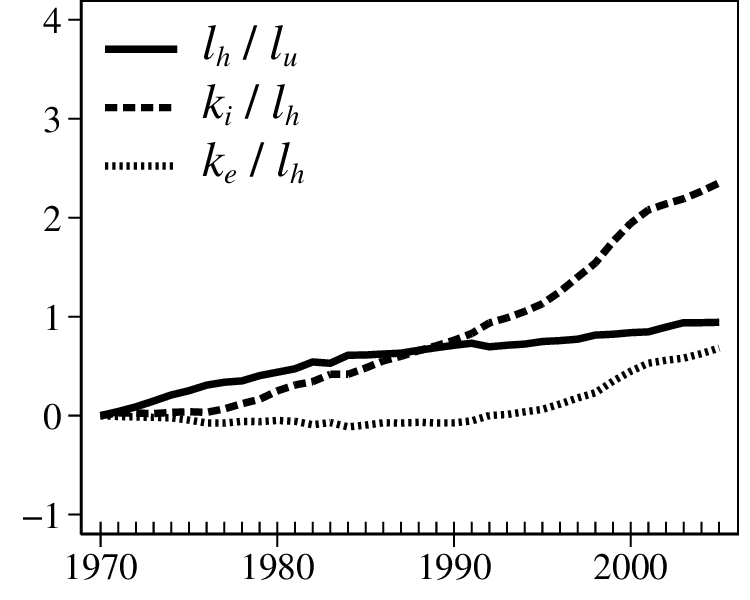}}
\par\end{centering}
\textit{\footnotesize{}Notes}{\footnotesize{}: The bold, dashed, and
dotted lines indicate the ratios of skilled to unskilled labor ($\ell_{h}/\ell_{u}$),
ICT capital to skilled labor ($k_{i}/\ell_{h}$), and capital equipment
to skilled labor, ($k_{e}/\ell_{h}$), respectively. All series are
logged and normalized to zero in the initial year.}{\footnotesize\par}
\end{figure}

Trends in capital and labor quantities vary across capital and labor
types in the United States (Figure \ref{fig: LhLuKiKe_us}). The relative
quantity of skilled to unskilled labor increased from the 1970s to
the 2000s. The relative quantity of ICT equipment to skilled labor
increased steadily and substantially after the mid-1970s, although
the relative quantity of total capital equipment to skilled labor
did not change substantially from the 1970s to the 2000s. The negative
co-movement between the rental price and quantity of ICT equipment
indicates technological change \citep*{Greenwood_Hercowitz_Krusell_AER97}.
The capital\textendash skill complementarity effect is proportional
in magnitude to the ratio of ICT capital to skilled labor, while the
relative labor quantity effect is proportional in magnitude to the
ratio of skilled to unskilled labor. The observed trends in relative
factor quantities suggest that the two effects would work in opposite
directions. The same applies to all countries and sectors (see Figures
\ref{fig: LhLuKiKe_oecd_goods} and \ref{fig: LhLuKiKe_oecd_service}
in Appendix \ref{subsec: trends_oecd} for countries other than the
United States). However, the magnitude and timing of changes in relative
factor quantities vary greatly across countries and sectors. The rate
of increase in the relative quantity of ICT equipment to skilled labor
was greater in the goods sector than in the service sector in the
United States but smaller in the goods sector than in the service
sector in many other countries.

\section{Estimation\label{sec: estimation}}

We start this section by specifying the estimating equations. We then
discuss how we identify and estimate the parameters in the production
function. We end this section by briefly describing how we decompose
the sources of changes in the skill premium.

\subsection{Econometric specifications}

\subsubsection{Factor-biased technological change}

We incorporate factor-biased technological change, as well as capital\textendash skill
complementarity, in the production function. Technological change
is said to be factor-biased (factor-neutral) if either or both of
(neither) relative capital- and (nor) labor-augmenting technology
changes over time. Technological change is said to be skill-biased
if the ratio of skilled to unskilled labor-augmenting technology ($A_{h}/A_{u}$)
increases over time. We can account for skill-biased technological
change by allowing the share parameter $\lambda$ to increase over
time. More flexibly, we can account for factor-biased technological
change by allowing the two share parameters $\lambda$ and $\mu$
to change over time. The share parameters are specified to ensure
that they lie between zero and one as follows:
\begin{equation}
\lambda_{cnt}=\frac{\exp\left(\sum_{\varsigma=0}^{S_{cn}^{\lambda}}\lambda_{\varsigma,cn}t^{\varsigma}\right)}{1+\exp\left(\sum_{\varsigma=0}^{S_{cn}^{\lambda}}\lambda_{\varsigma,cn}t^{\varsigma}\right)}\quad\text{and}\quad\mu_{cnt}=\frac{\exp\left(\sum_{\varsigma=0}^{S_{cn}^{\mu}}\mu_{\varsigma,cn}t^{\varsigma}\right)}{1+\exp\left(\sum_{\varsigma=0}^{S_{cn}^{\mu}}\mu_{\varsigma,cn}t^{\varsigma}\right)},\label{eq: lambda=000026mu}
\end{equation}
where $c$, $n$, and $t$ are the indices for countries, sectors,
and years, respectively. Changes in the direction and magnitude of
factor-biased technological change can be accounted for by adding
higher-order trend terms. Differences in the speed and timing of factor-biased
technological change can be accounted for by allowing the coefficients
of trend terms to vary across countries and sectors. We choose the
orders of polynomials ($S_{cn}^{\lambda}$ and $S_{cn}^{\mu}$) to
fit the trends in the respective relative factor prices ($w_{h}/w_{u}$
and $w_{h}/r_{i}$) for each country and sector. This parameterization
is sufficiently flexible to account for the trends in the relative
factor prices for each country and sector, as shown below.

\subsubsection{Wedge}

Moreover, we allow the degree of competitiveness to vary across countries
and sectors over time. The wedge can be decomposed as
\begin{equation}
\omega_{j,cnt}=\varpi_{j,cn}\varphi_{j,cnt}v_{j,cnt}^{\omega}\qquad\text{for}\enskip j\in\left\{ h,u,i,o\right\} ,\label{eq: wedge_4}
\end{equation}
where $\varpi$ is a time-invariant effect specific to each country
and sector, $\varphi$ is a time-varying effect of wage-setting institutions,
and $v^{\omega}$ is an idiosyncratic shock to the wedge. Any time-invariant
country-specific factors can be eliminated by taking logs and differences
over time, even though there are substantial and persistent differences
in non-competitive and institutional factors across countries. Conceivably,
however, the wedge may decrease as wage-setting institutions become
less centralized. If the decrease in the wedge due to wage-setting
institutions differs for skilled and unskilled labor, changes in the
skill premium can be influenced by wage-setting institutions. We allow
for this possibility in a robustness check by incorporating the collective
bargaining coverage (i.e., the percentage of employees with the right
to bargain) as a time-varying factor in the wedge. Information about
the collective bargaining coverage can be obtained from OECD.Stat
throughout the sample period.

\subsubsection{Estimating equations}

After substituting equations \eqref{eq: lambda=000026mu} and \eqref{eq: wedge_4}
into the marginal-rate-of-technical-substitution conditions \eqref{eq: Wh/Wu}
and \eqref{eq: Wh/Ri} and taking differences over time, we obtain
the following estimating equations:
\begin{multline}
\Delta\ln\left(\frac{w_{h,cnt}}{w_{u,cnt}}\right)=\sum_{\varsigma=1}^{S_{cn}^{\lambda}}\lambda_{\varsigma,cn}\Delta t^{\varsigma}-\frac{\sigma}{\rho}\Delta\ln\left[1+\exp\left(\sum_{\varsigma=0}^{S_{cn}^{\mu}}\mu_{\varsigma,cn}t^{\varsigma}\right)\right]\\
+\frac{\sigma-\rho}{\rho}\Delta\ln\left[\exp\left(\sum_{\varsigma=0}^{S_{cn}^{\mu}}\mu_{\varsigma,cn}t^{\varsigma}\right)\left(\frac{k_{i,cnt}}{\ell_{h,cnt}}\right)^{\rho}+1\right]-\left(1-\sigma\right)\Delta\ln\left(\frac{\ell_{h,cnt}}{\ell_{u,cnt}}\right)+\Delta\ln\left(\frac{\varphi_{h,cnt}}{\varphi_{u,cnt}}\right)+\Delta v_{1,cnt},\label{eq: D.Wh/Wu_4}
\end{multline}
\begin{equation}
\Delta\ln\left(\frac{w_{h,cnt}}{r_{i,cnt}}\right)=-\sum_{\varsigma=1}^{S_{cn}^{\mu}}\mu_{\varsigma,cn}\Delta t^{\varsigma}-\left(1-\rho\right)\Delta\ln\left(\frac{\ell_{h,cnt}}{k_{i,cnt}}\right)+\Delta\ln\left(\frac{\varphi_{h,cnt}}{\varphi_{i,cnt}}\right)+\Delta v_{2,cnt},\label{eq: D.Wh/Ri_4}
\end{equation}
where the error terms are $\Delta v_{1,cnt}=\Delta\ln(v_{h,cnt}^{\omega}/v_{u,cnt}^{\omega})$
and $\Delta v_{2,cnt}=\Delta\ln(v_{h,cnt}^{\omega}/v_{i,cnt}^{\omega})$.
One of the substitution parameters ($\rho$) appears in both equations.
The error terms are correlated across equations as the idiosyncratic
shock $v_{h}^{\omega}$ is a component of the error terms in both
equations. Thus, it is efficient to estimate the system of equations
\eqref{eq: D.Wh/Wu_4} and \eqref{eq: D.Wh/Ri_4} jointly.

Two things could be worth mentioning. First, the error terms arise
not from shocks to markup in the output market but from differences
in shocks to the wedge across input markets. The error terms can also
arise from changes in measurement errors in relative factor prices
and quantities. Second, estimating equations \eqref{eq: D.Wh/Wu_4}
and \eqref{eq: D.Wh/Ri_4} remain of the same form even if the wedge
has time trends. In that case, the interpretation of decomposition
results requires caution in that the relative factor-augmenting technology
effect could be attributed in part to changes in the factor-specific
wedge. However, following the convention in the literature, we use
the term \textquotedblleft technology\textquotedblright{} to refer
to unobserved factors.

\subsubsection{Instrumental variables}

The substitution parameters ($\sigma$ and $\rho$) can be identified
from variation in relative factor quantities across countries, sectors,
and time. We allow for correlations between changes in factor quantities
and idiosyncratic shocks, and estimate the substitution parameters
using the shift\textendash share instrument:
\begin{equation}
\Delta\ln z_{cnt}^{b}=\sum_{d\in\mathcal{D}_{n^{\prime}}}\frac{z_{dcn^{\prime}t_{0}}}{\sum_{d^{\prime}\in\mathcal{D}_{n^{\prime}}}z_{d^{\prime}cn^{\prime}t_{0}}}\Delta\ln\left(\sum_{c^{\prime}\neq c\in\mathcal{C}}z_{dc^{\prime}n^{\prime}t}\right)\qquad\text{for }z\in\left\{ k_{i},\ell_{h},\ell_{u}\right\} ,\label{eq: Bartik}
\end{equation}
where $d$ is an index for subsectors (or equivalently industries),
and $\mathcal{C}$ and $\mathcal{D}_{n}$ are the sets of countries
and the set of subsectors in sector $n$, respectively. The shift\textendash share
instrument is the interaction between the growth rate of factors in
each industry (\textit{shift}), which measures global shocks to industries,
and the lagged industry share of factors in each country (\textit{share}),
which measures local exposure to industry shocks.\footnote{\citet*{GoldsmithPinkham_Sorkin_Swift_AER20} show that the two-stage
least squares estimator with the shift\textendash share instrument
is numerically equivalent to the generalized method of moments estimator
using local industry shares as excluded instruments, whereas \citet*{Borusyak_Hull_Jaravel_RESf}
show that it is also numerically equivalent to the two-stage least
squares estimator using global industry growth rates as an excluded
instrument in the the exposure-weighted industry-level regression.
The former result implies that identification comes from the exogeneity
of exposure shares, whereas the latter result implies that identification
comes from the exogeneity of industry shocks. It follows that the
shift\textendash share instrument can be valid if either lagged local
industry shares or global industry growth rates are exogenous to idiosyncratic
shocks, conditional on a set of regressors.} We construct the shift\textendash share instrument using the data
from the service (goods) sector to deal with the endogeneity of factor
quantities in the goods (service) sector. The idea behind this is
that a demand shock in one sector serves as a supply shifter in another
sector to identify a demand curve \citep{Oberfield_Raval_EM21}. We
evaluate the time-invariant measure of local industry shares at the
initial year ($t=t_{0}$) to ensure the exogeneity of industry shares,
and use the leave-one-out measure of global industry growth rates
to ensure the exogeneity of industry shocks.

\subsection{Generalized method of moments}

We estimate the system of equations \eqref{eq: D.Wh/Wu_4} and \eqref{eq: D.Wh/Ri_4}
jointly using the generalized method of moments (GMM). This approach
is semi-parametric as it does not impose a distributional assumption
on the error terms. We treat all capital and labor quantities as endogenous
variables. Specifically, we use the change in the log ratio of the
shift\textendash share instruments, such as $\Delta\ln(\ell_{h}^{b}/\ell_{u}^{b})$
and $\Delta\ln(k_{i}^{b}/\ell_{h}^{b})$, as instrumental variables
for the change in the log ratio of factor quantities, such as $\Delta\ln(\ell_{h}/\ell_{u})$
and $\Delta\ln(k_{i}/\ell_{h})$. We take five-year differences in
the system of equations \eqref{eq: D.Wh/Wu_4} and \eqref{eq: D.Wh/Ri_4}.

Let $\boldsymbol{\theta}$ denote the set of parameters to be estimated.
The vector of parameters is $\boldsymbol{\theta}=(\sigma,\rho,\lambda_{1,cn},\ldots,\lambda_{S_{c}^{\lambda},cn},$
$\mu_{0,cn},\ldots,\mu_{S_{cn}^{\mu},cn})$. The GMM estimator $\widehat{\theta}$
is chosen to minimize the quadratic form:
\begin{equation}
\widehat{\boldsymbol{\theta}}=\arg\min_{\boldsymbol{\theta}}\enskip g_{N}\left(\boldsymbol{\theta}\right)^{\prime}\,\boldsymbol{W}_{N}\,g_{N}\left(\boldsymbol{\theta}\right),
\end{equation}
where $g_{N}\left(\boldsymbol{\theta}\right)$ is a vector of the
moment conditions, and $\boldsymbol{W}_{N}$ is a weighting matrix.
Let $\boldsymbol{z}$ denote a vector of exogenous variables that
include the instrumental variables and trend terms. The elements of
$g_{N}\left(\theta\right)$ are $N^{-1}\sum_{c=1}^{14}\sum_{n=1}^{2}\sum_{t=1}^{T_{c}}\boldsymbol{z}_{m,cnt}v_{m,cnt}$
for $m=1,2$, where $T_{c}$ is the number of years used for estimation
in country $c$, and $N$ is the sum of $T_{c}$ across countries
multiplied by the number of sectors. The GMM estimator is consistent
as the sample size ($N$) approaches infinity.

\subsection{Decomposition}

After estimating the production function, we decompose changes in
the skill premium into the capital\textendash skill complementarity
effect associated with $k_{i}/\ell_{h}$, the relative labor quantity
effect associated with $\ell_{h}/\ell_{u}$, and the relative factor-augmenting
technology effect associated with $A_{h}/A_{u}$ or $A_{i}/A_{h}$
using equation \eqref{eq: Wh/Wu}. The three effects can be further
decomposed into parts due to each factor and technology. The issue
here is that the decomposition results generally depend on the order
of decomposition as equation \eqref{eq: Wh/Wu} is not additively
linear in $k_{i}$, $\ell_{h}$, or $A_{i}/A_{h}$. The same issue
arises when we decompose the sources of changes in the relative demand
for skilled to unskilled labor using equation \eqref{eq: Lh/Lu}.
We implement the Shapley decomposition to address the issue of path
dependence \citep{Shorrocks_JoEI13}. Appendix \ref{subsec: shapley}
provides a detailed description of the decomposition.

\section{Results\label{sec: results}}

We first present the estimates of the elasticities of substitution
in production. We then discuss the sources and mechanisms of changes
in the skill premium for each country and sector and of the differences
in changes in the skill premium across countries.

\subsection{Production function estimates}

\subsubsection{Capital\textendash Skill complementarity}

The four-factor production function involves two substitution parameters
($\sigma$ and $\rho$) and two share parameters ($\lambda$ and $\mu$).
Table \ref{tab: F(Ki,Ko,Lh,Lu)} presents the estimates of the substitution
parameters. Although we focus on the results when we allow technological
change to be factor-biased, we also report the results when we assume
technological change to be factor-neutral for reference. The estimates
of the substitution parameters are greater in absolute value in the
former case than in the latter case. Subsequent figures present the
estimates of the share parameters, which consist of higher-order trend
terms and their country- and sector-specific coefficients, in the
form of the relative factor-augmenting technology effect.

\begin{table}[h]
\caption{Production function estimates\label{tab: F(Ki,Ko,Lh,Lu)}}

\begin{centering}
\smallskip{}
\par\end{centering}
\begin{centering}
\begin{tabular}{r@{\extracolsep{0pt}.}lr@{\extracolsep{0pt}.}l}
\hline 
\multicolumn{2}{c}{$\sigma$} & \multicolumn{2}{c}{$\rho$}\tabularnewline
\hline 
\multicolumn{4}{c}{Factor-biased}\tabularnewline
\multicolumn{4}{c}{technological change}\tabularnewline
0&888 & \textendash 0&162\tabularnewline
(0&029) & (0&051)\tabularnewline
{[}0&026{]} & {[}0&057{]}\tabularnewline
\hline 
\multicolumn{4}{c}{Factor-neutral}\tabularnewline
\multicolumn{4}{c}{technological change}\tabularnewline
0&642 & \textendash 0&053\tabularnewline
(0&248) & (0&047)\tabularnewline
{[}0&130{]} & {[}0&053{]}\tabularnewline
\hline 
\end{tabular}\smallskip{}
\par\end{centering}
\textit{\footnotesize{}Notes}{\footnotesize{}: Standard errors in
parentheses are clustered at the country-sector level, and those in
square brackets are Newey\textendash West adjusted with the optimal
lag length \citep{Newey_West_RES94}.}{\footnotesize\par}
\end{table}

Our results indicate that ICT equipment is much more complementary
to skilled labor than unskilled labor even in the presence of factor-biased
technological change, which strongly confirms the ICT capital\textendash skill
complementarity hypothesis (i.e., $\sigma>\rho$). The estimate of
$\rho$, which governs the degree of substitution between $k_{i}$
and $\ell_{h}$, is close to that in \citet{Krusell_Ohanian_RiosRull_Violante_EM00},
while the estimate of $\sigma$, which governs the degree of substitution
between the $k_{i}$-$\ell_{h}$ composite and $\ell_{u}$, is greater
than that in \citet{Krusell_Ohanian_RiosRull_Violante_EM00}. Consequently,
the estimated difference between the two substitution parameters ($\sigma-\rho$),
to which the capital\textendash skill complementarity effect is proportional,
is slightly greater here. \citet{Polgreen_Silos_RED08} show that
the estimate of $\sigma$ becomes greater when using data from the
National Income and Product Accounts (NIPA) than when using the data
in \citet{Krusell_Ohanian_RiosRull_Violante_EM00}. Our estimate of
$\sigma$ is close to what \citet{Polgreen_Silos_RED08} obtain from
the NIPA data.

\subsubsection{Robustness checks}

We examine whether the results presented above are robust to the influence
of international trade and wage-setting institutions. Table \ref{tab: robustness}
presents the extent to which the estimates of the substitution parameters
vary when splitting the sample by sector and when controlling for
the collective bargaining coverage.

The first, second, and third rows of Table \ref{tab: robustness}
report the estimated substitution parameters for all, goods, and service
sectors, respectively. The estimates of $\sigma$ and $\rho$ are
similar for the goods and service sectors. In addition, the estimated
difference between the two substitution parameters ($\sigma-\rho$),
to which the capital\textendash skill complementarity effect is proportional,
is almost the same for the goods and service sectors. The ICT capital\textendash skill
complementarity hypothesis appears to hold irrespective of the influence
of international trade.

The last row of Table \ref{tab: robustness} indicates that the estimated
substitution parameters remain almost unchanged even after allowing
the wedge to vary with the degree of centralized wage bargaining.
The coefficient of the collective bargaining coverage is insignificant,
indicating that the skill premium is not associated with the collective
bargaining coverage, conditional on the capital\textendash skill complementarity
effect, relative labor quantity effect, and relative factor-augmenting
technology effect.

\begin{table}[h]
\caption{Robustness checks\label{tab: robustness}}

\begin{centering}
\begin{tabular}{lcc}
\hline 
 & $\sigma$ & $\rho$\tabularnewline
\cline{2-3} \cline{3-3} 
All sectors & 0.888 & \textendash 0.162\tabularnewline
 & (0.029) & (0.051)\tabularnewline
Goods sector & 0.847 & \textendash 0.114\tabularnewline
 & (0.036) & (0.113)\tabularnewline
Service sector & 0.889 & \textendash 0.123\tabularnewline
 & (0.032) & (0.043)\tabularnewline
Collective bargaining coverage & 0.873 & \textendash 0.119\tabularnewline
 & (0.025) & (0.045)\tabularnewline
\hline 
\end{tabular}
\par\end{centering}
\textit{\footnotesize{}\hspace{3.5cm}Notes}{\footnotesize{}: Standard
errors in parentheses are clustered at the country-sector level.}{\footnotesize\par}
\end{table}

\subsubsection{Morishima elasticities}

Table \ref{tab: morishima_4} presents the estimates of the Morishima
elasticities of substitution among ICT equipment, skilled labor, and
unskilled labor. ICT equipment is much more substitutable with unskilled
labor than skilled labor, while skilled labor is more substitutable
with unskilled labor than ICT equipment. The former result confirms
the presence of capital\textendash skill complementarity. The latter
result implies that the degree of labor\textendash labor substitution
is greater than that of capital\textendash labor substitution. The
Morishima elasticities are asymmetric. The estimated elasticity of
$\ell_{u}$ with respect to $r_{i}$ is smaller than that of $k_{i}$
with respect to $w_{u}$.

\begin{table}[h]
\caption{Morishima elasticities of substitution\label{tab: morishima_4}}
\smallskip{}

\begin{centering}
\begin{tabular}{cr@{\extracolsep{0pt}.}lr@{\extracolsep{0pt}.}lr@{\extracolsep{0pt}.}l}
\hline 
 & \multicolumn{2}{c}{$k_{i}$} & \multicolumn{2}{c}{$\ell_{h}$} & \multicolumn{2}{c}{$\ell_{u}$}\tabularnewline
$k_{i}$ & \multicolumn{2}{c}{} & 0&860 & 8&943\tabularnewline
 & \multicolumn{2}{c}{} & (0&038) & (2&303)\tabularnewline
$\ell_{h}$ & 0&860 & \multicolumn{2}{c}{} & 8&943\tabularnewline
 & (0&038) & \multicolumn{2}{c}{} & (2&303)\tabularnewline
$\ell_{u}$ & 2&324 & 7&479 & \multicolumn{2}{c}{}\tabularnewline
 & (0&629) & (2&004) & \multicolumn{2}{c}{}\tabularnewline
\hline 
\end{tabular}\medskip{}
\par\end{centering}
\textit{\footnotesize{}Notes}{\footnotesize{}: Elasticities are evaluated
at the sample means. Standard errors in parentheses are clustered
at the country-sector level.}{\footnotesize\par}
\end{table}

\begin{figure}[H]
\caption{Actual and predicted skill premium ($w_{h}/w_{u}$) in the goods sector\label{fig: Wh/Wu_4_oecd_goods}}

\begin{centering}
\subfloat[{\small{}(a) }United States]{
\centering{}\includegraphics[scale=0.35]{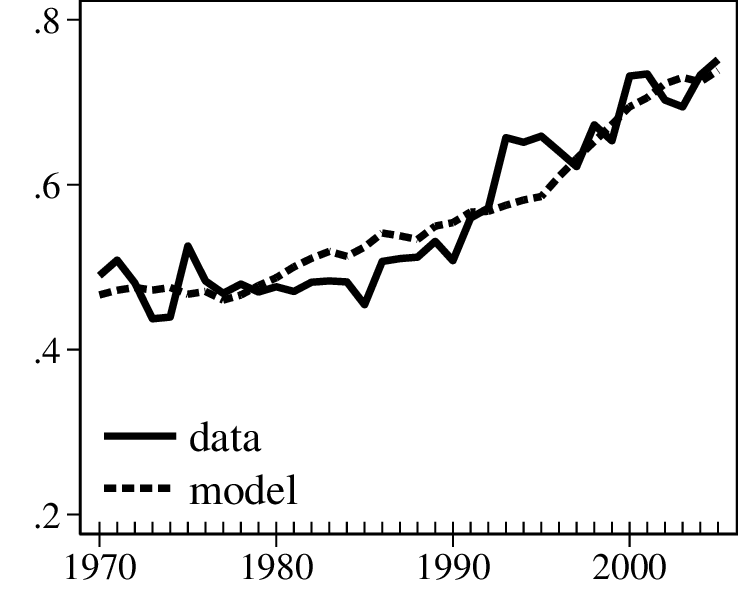}}\subfloat[{\small{}(b) }Australia]{
\centering{}\includegraphics[scale=0.35]{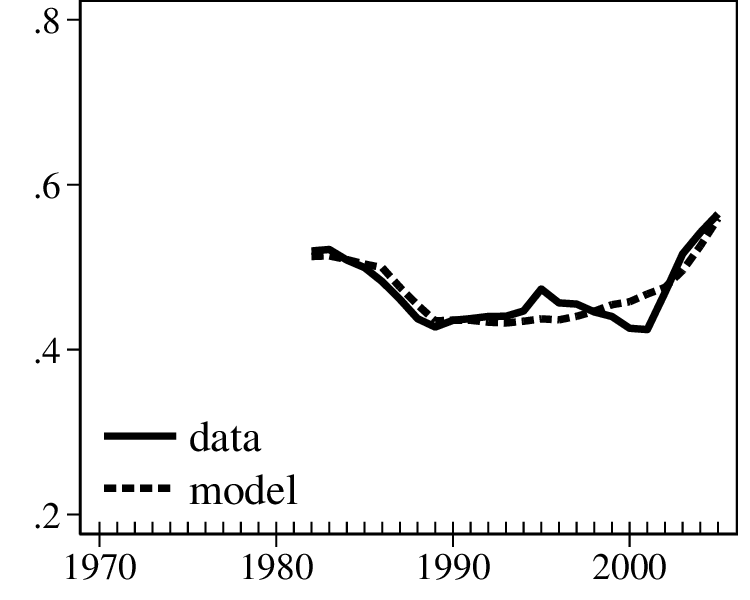}}\subfloat[{\small{}(c) }Austria]{
\centering{}\includegraphics[scale=0.35]{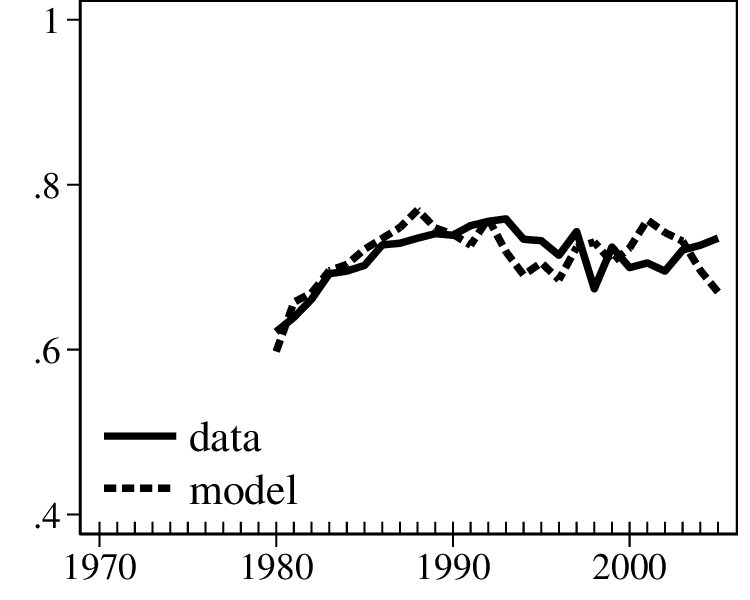}}
\par\end{centering}
\begin{centering}
\subfloat[{\small{}(d) }Czech Republic]{
\centering{}\includegraphics[scale=0.35]{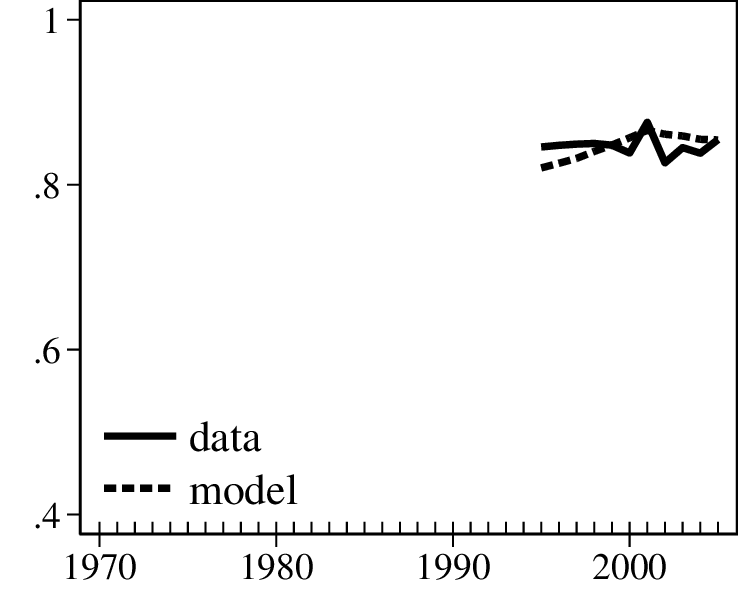}}\subfloat[{\small{}(e) }Denmark]{
\centering{}\includegraphics[scale=0.35]{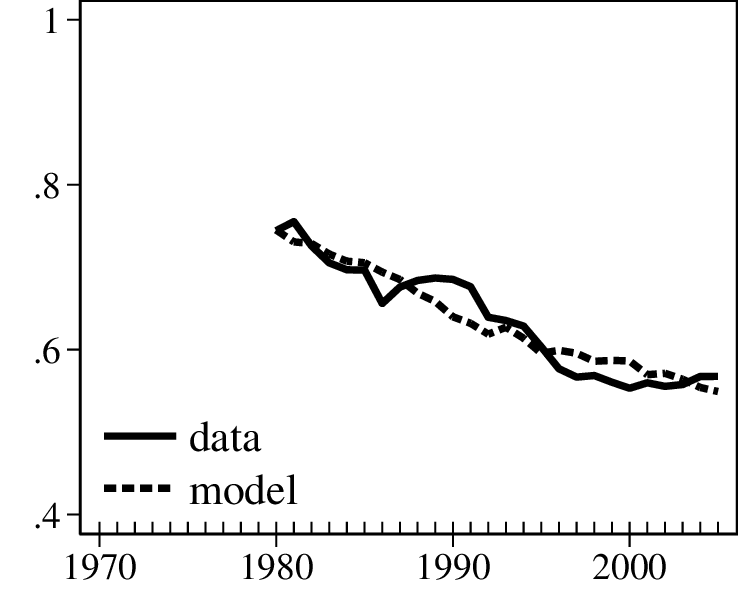}}\subfloat[{\small{}(f) }Finland]{
\centering{}\includegraphics[scale=0.35]{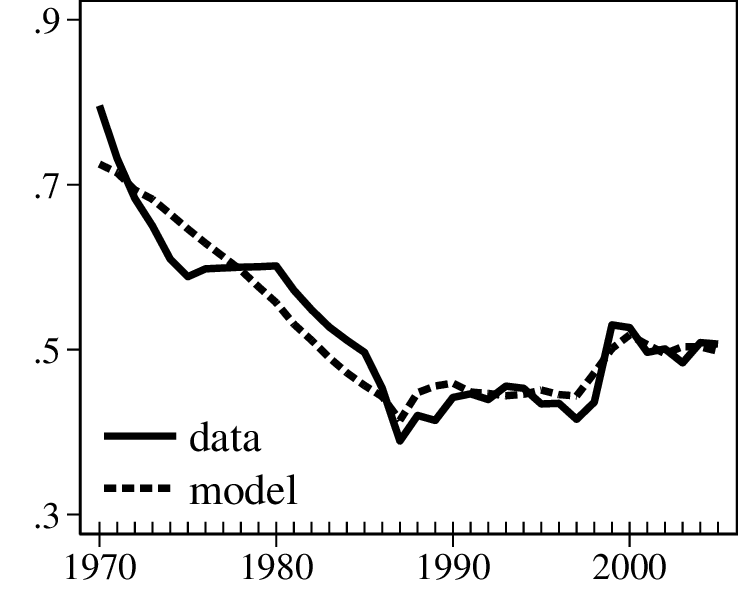}}
\par\end{centering}
\begin{centering}
\subfloat[{\small{}(g) }Germany]{
\centering{}\includegraphics[scale=0.35]{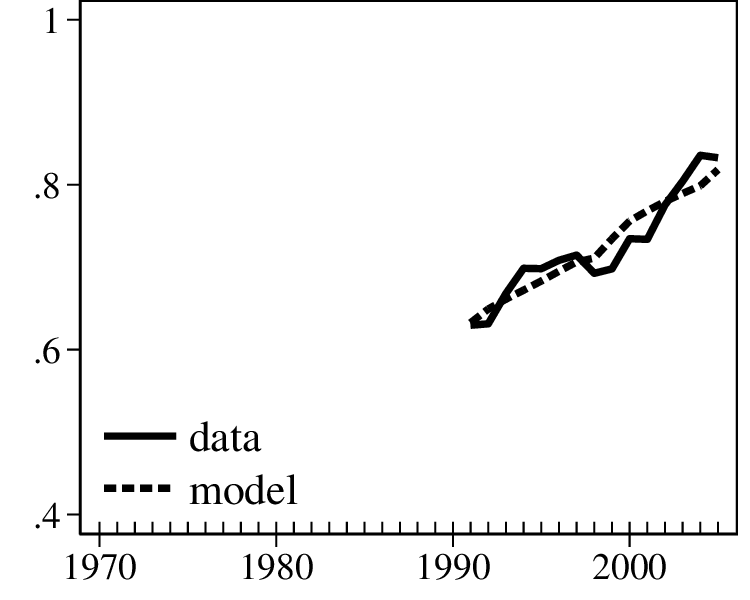}}\subfloat[{\small{}(h) }Italy]{
\centering{}\includegraphics[scale=0.35]{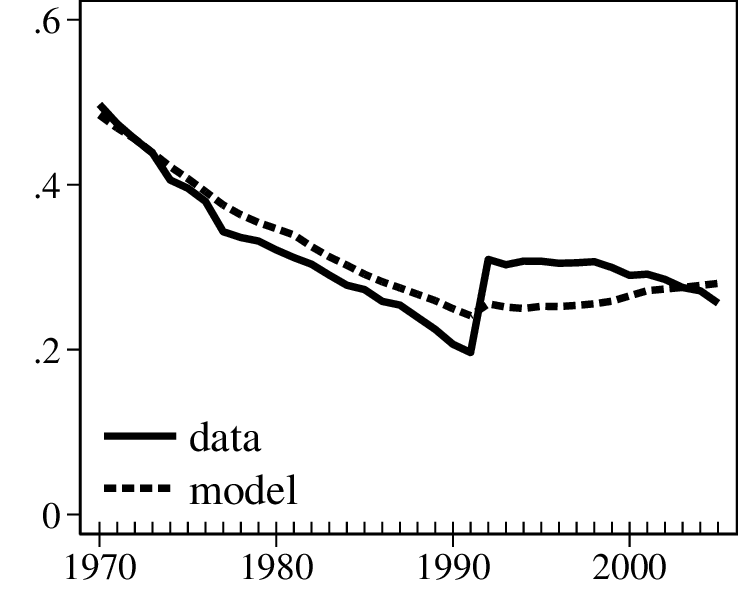}}\subfloat[{\small{}(i) }Japan]{
\centering{}\includegraphics[scale=0.35]{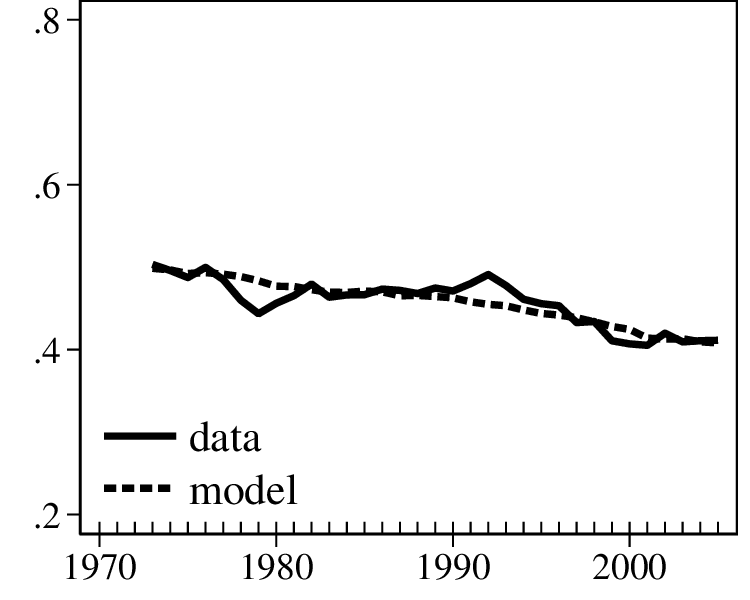}}
\par\end{centering}
\begin{centering}
\subfloat[{\small{}(j) }Netherlands]{
\centering{}\includegraphics[scale=0.35]{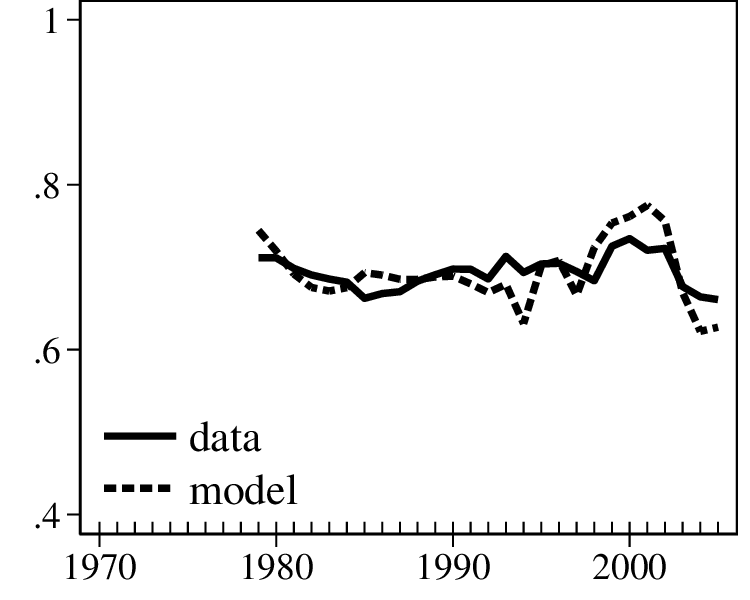}}\subfloat[{\small{}(k) }Portugal]{
\centering{}\includegraphics[scale=0.35]{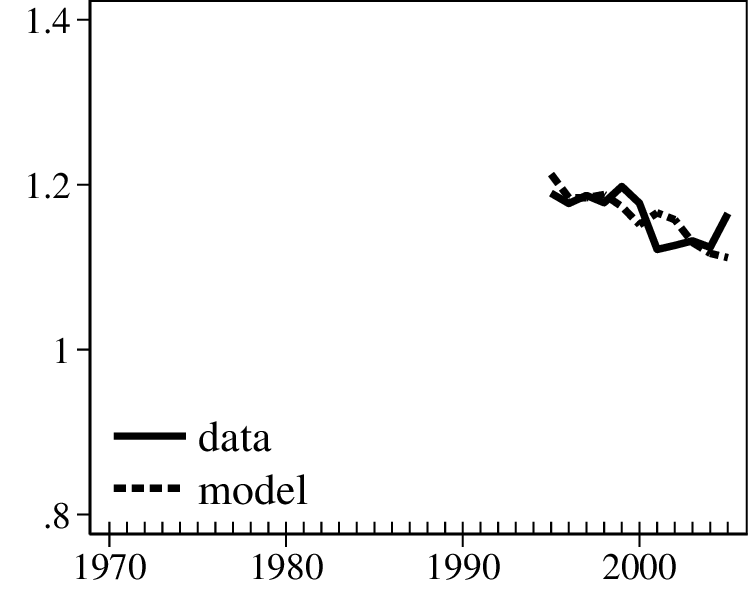}}\subfloat[{\small{}(l) }Slovenia]{
\centering{}\includegraphics[scale=0.35]{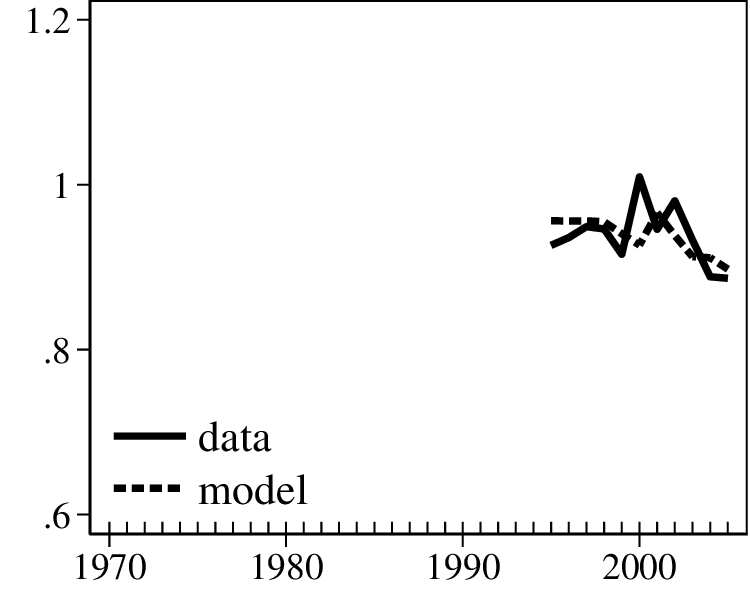}}
\par\end{centering}
\begin{centering}
\subfloat[{\small{}(m) }Sweden]{
\centering{}\includegraphics[scale=0.35]{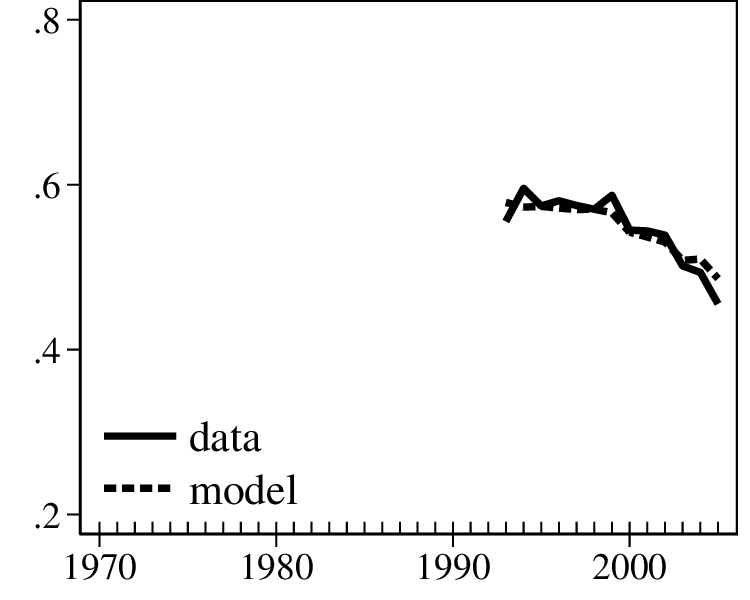}}\subfloat[{\small{}(n) }United Kingdom]{
\centering{}\includegraphics[scale=0.35]{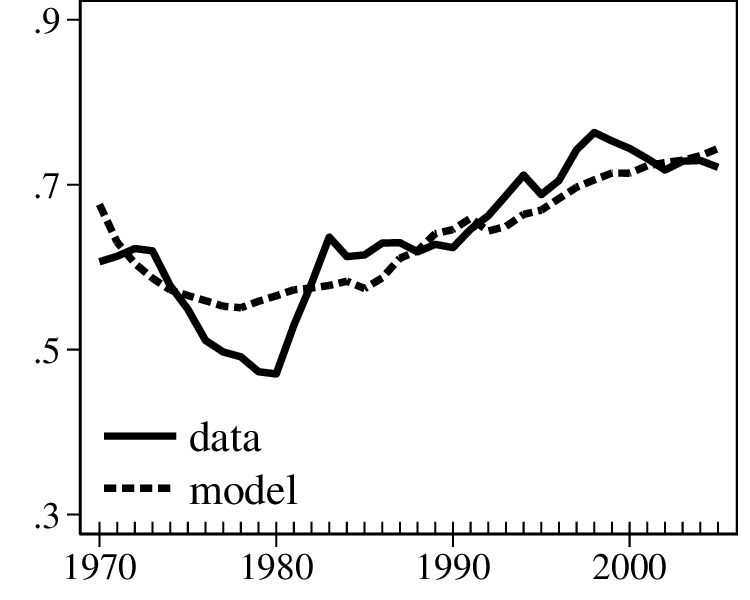}}
\par\end{centering}
\textit{\footnotesize{}Notes}{\footnotesize{}: The bold and dashed
lines indicate the actual and predicted values, respectively. All
series are logged.}{\footnotesize\par}
\end{figure}

\begin{figure}[H]
\caption{Actual and predicted skill premium ($w_{h}/w_{u}$) in the service
sector\label{fig: Wh/Wu_4_oecd_service}}

\begin{centering}
\subfloat[{\small{}(a) }United States]{
\centering{}\includegraphics[scale=0.35]{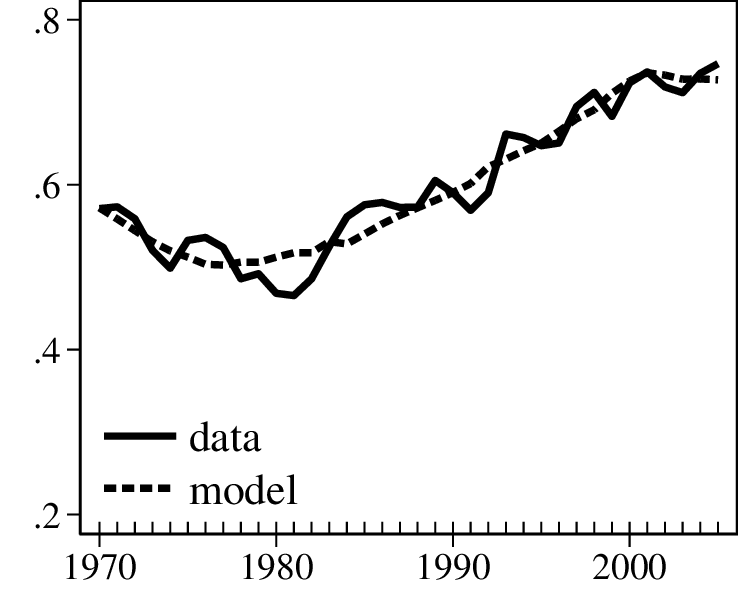}}\subfloat[{\small{}(b) }Australia]{
\centering{}\includegraphics[scale=0.35]{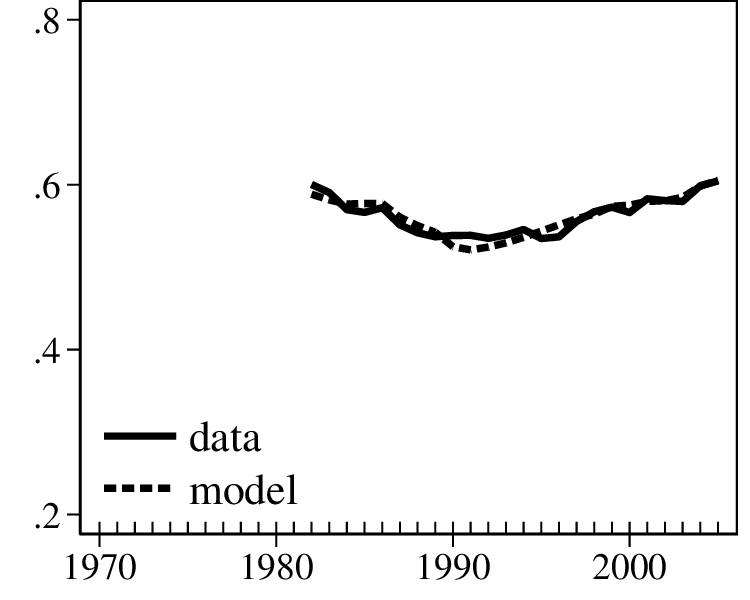}}\subfloat[{\small{}(c) }Austria]{
\centering{}\includegraphics[scale=0.35]{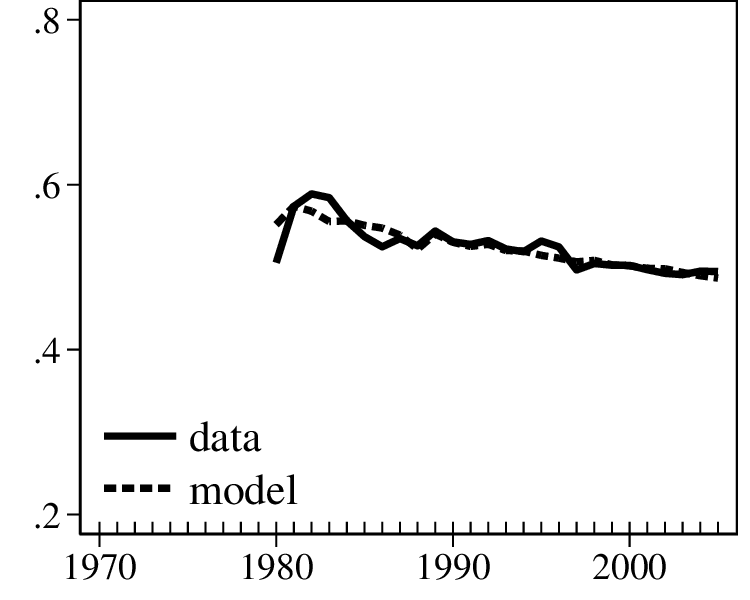}}
\par\end{centering}
\begin{centering}
\subfloat[{\small{}(d) }Czech Republic]{
\centering{}\includegraphics[scale=0.35]{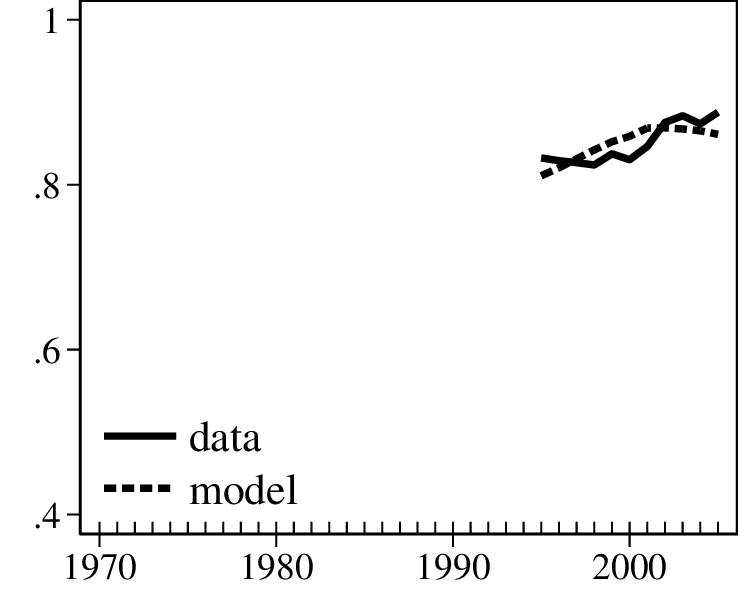}}\subfloat[{\small{}(e) }Denmark]{
\centering{}\includegraphics[scale=0.35]{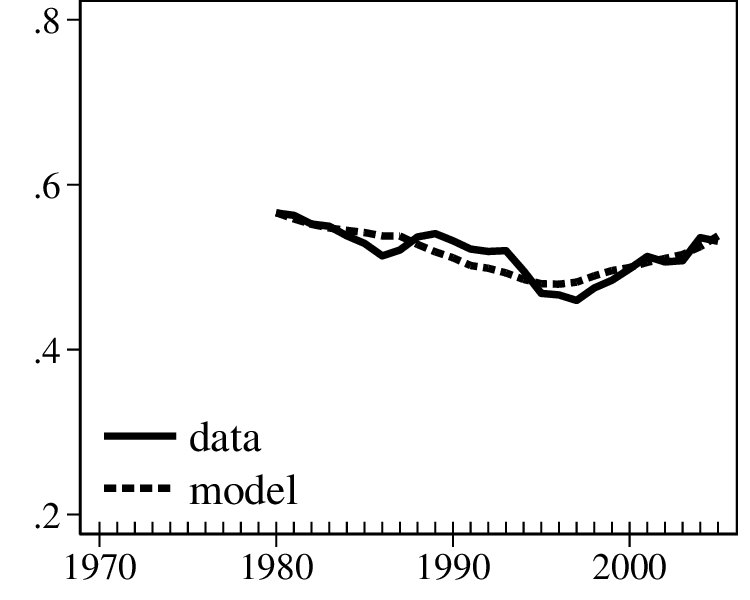}}\subfloat[{\small{}(f) }Finland]{
\centering{}\includegraphics[scale=0.35]{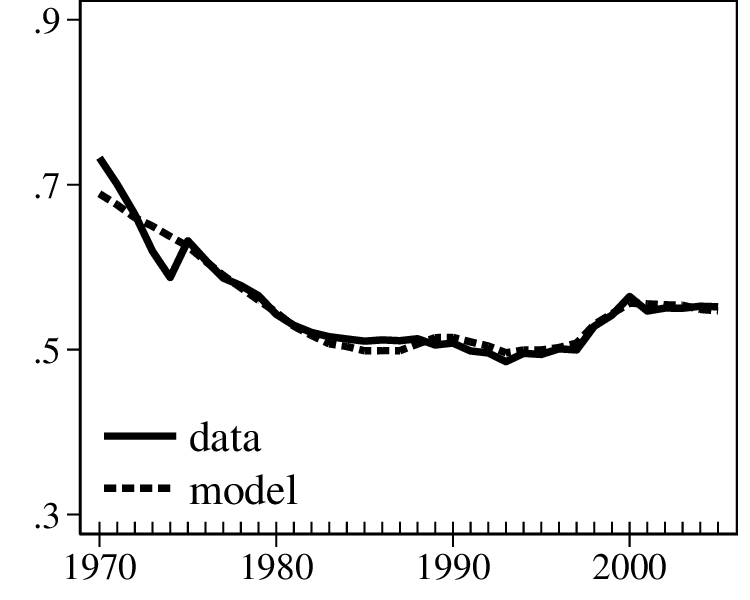}}
\par\end{centering}
\begin{centering}
\subfloat[{\small{}(g) }Germany]{
\centering{}\includegraphics[scale=0.35]{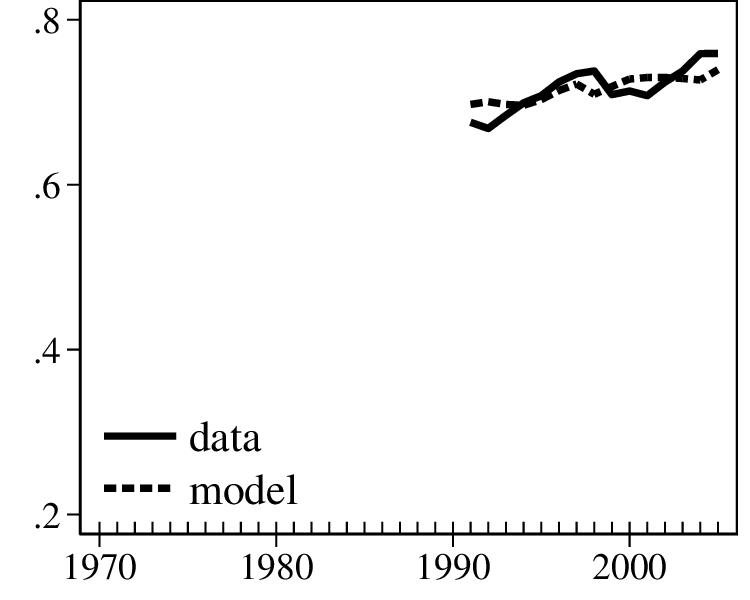}}\subfloat[{\small{}(h) }Italy]{
\centering{}\includegraphics[scale=0.35]{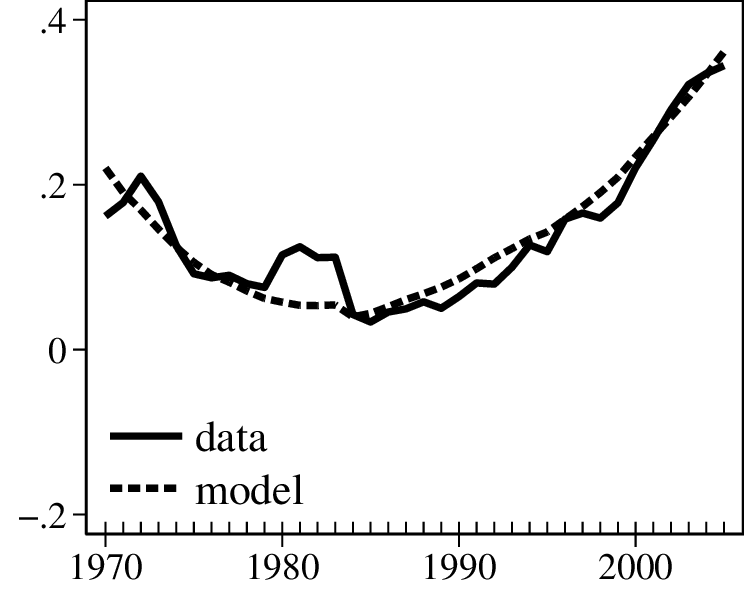}}\subfloat[{\small{}(i) }Japan]{
\centering{}\includegraphics[scale=0.35]{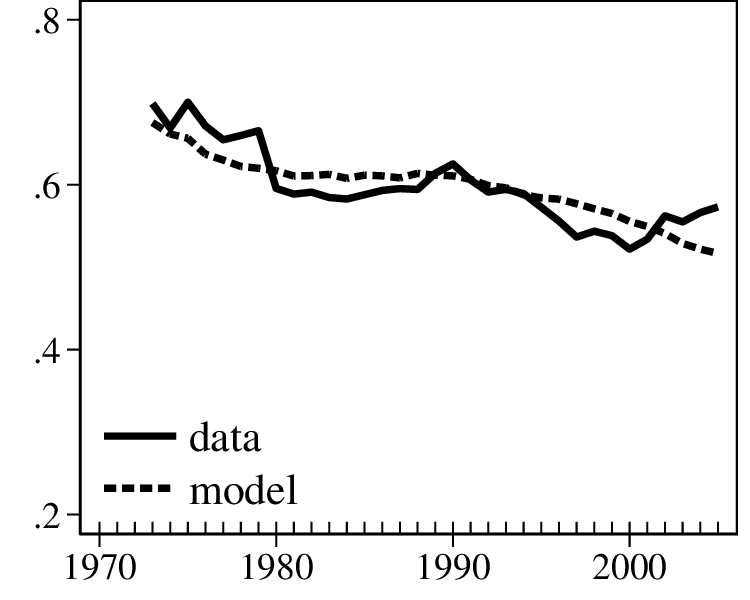}}
\par\end{centering}
\begin{centering}
\subfloat[{\small{}(j) }Netherlands]{
\centering{}\includegraphics[scale=0.35]{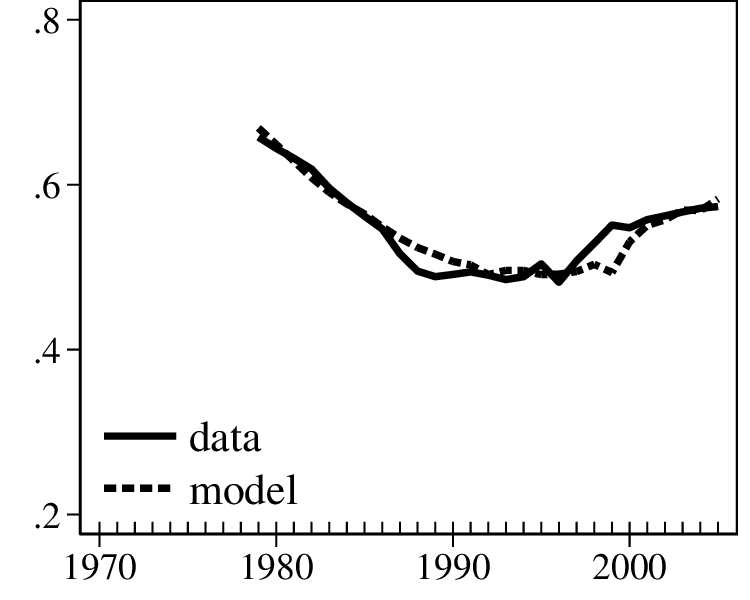}}\subfloat[{\small{}(k) }Portugal]{
\centering{}\includegraphics[scale=0.35]{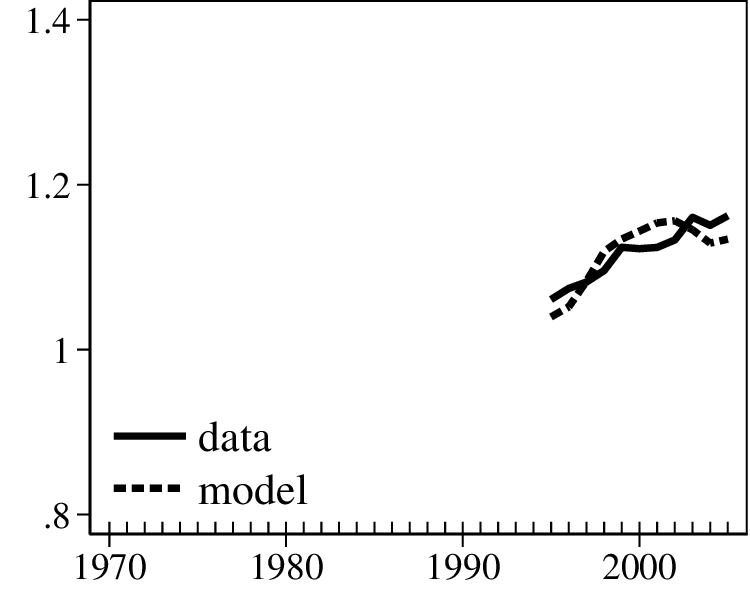}}\subfloat[{\small{}(l) }Slovenia]{
\centering{}\includegraphics[scale=0.35]{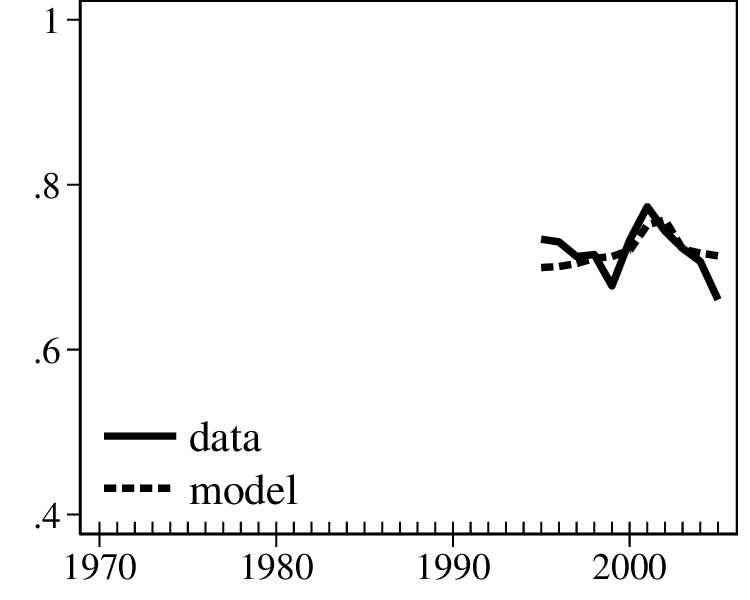}}
\par\end{centering}
\begin{centering}
\subfloat[{\small{}(m) }Sweden]{
\centering{}\includegraphics[scale=0.35]{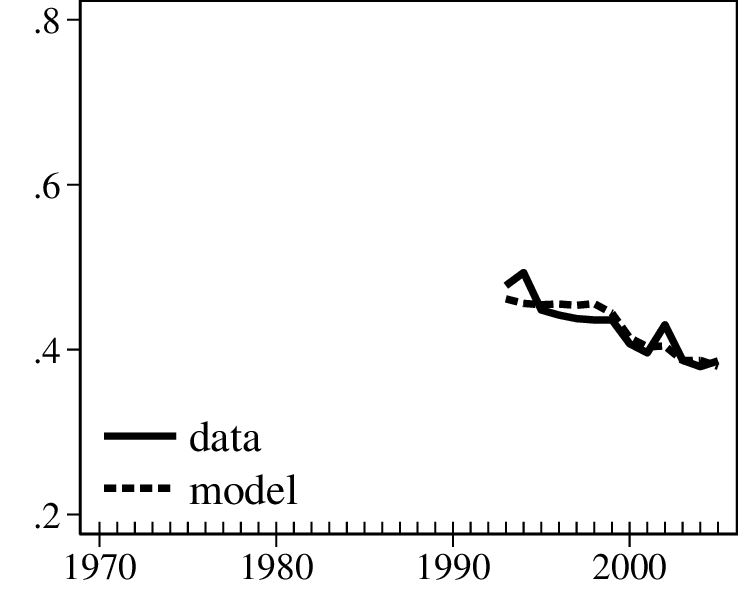}}\subfloat[{\small{}(n) }United Kingdom]{
\centering{}\includegraphics[scale=0.35]{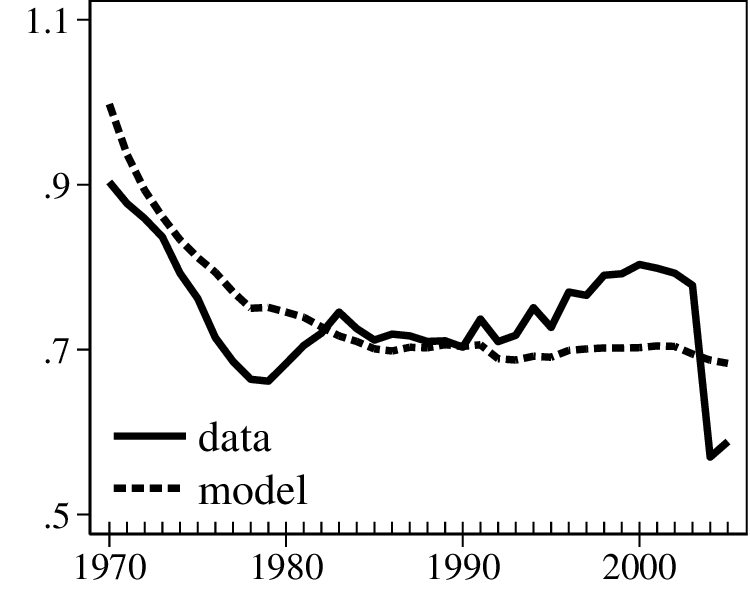}}
\par\end{centering}
\textit{\footnotesize{}Notes}{\footnotesize{}: The bold and dashed
lines indicate the actual and predicted values, respectively. All
series are logged.}{\footnotesize\par}
\end{figure}

\begin{figure}[H]
\caption{Actual and predicted ratio of the wages of skilled labor to the rental
price of ICT capital ($w_{h}/r_{i}$) in the goods sector \label{fig: Wh/Ri_oecd_goods}}

\begin{centering}
\subfloat[{\small{}(a) }United States]{
\centering{}\includegraphics[scale=0.35]{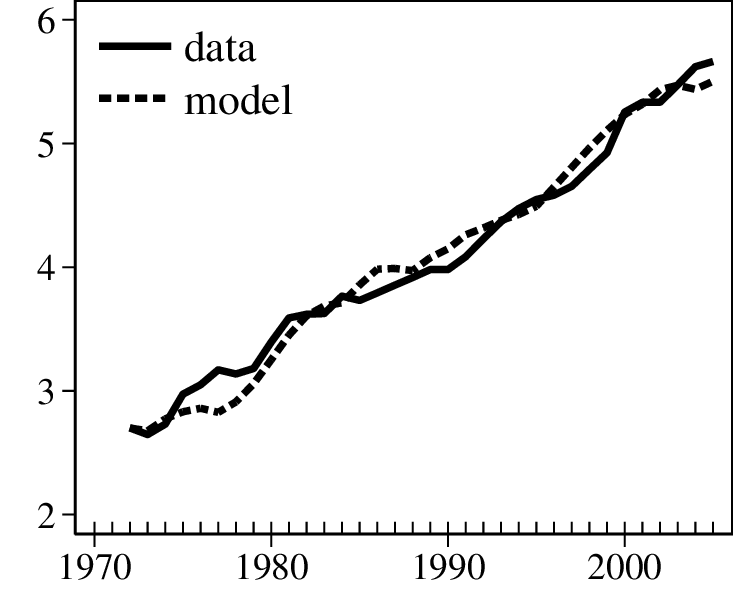}}\subfloat[{\small{}(b) }Australia]{
\centering{}\includegraphics[scale=0.35]{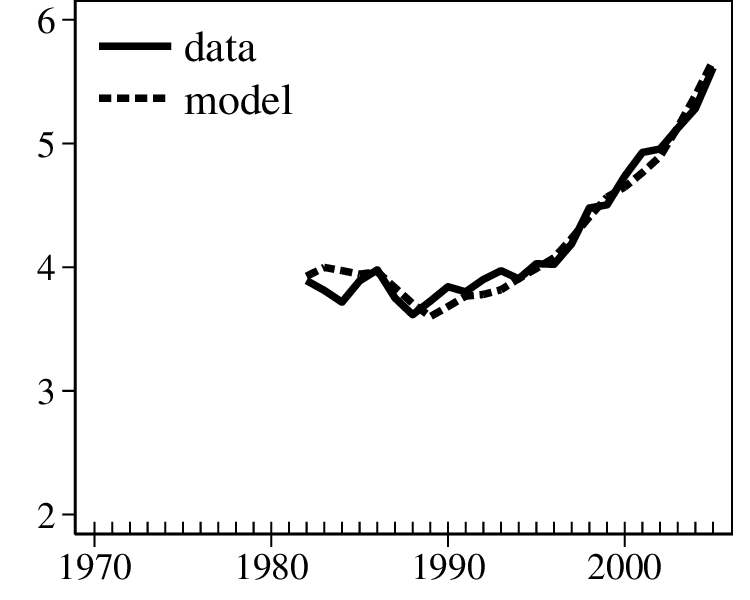}}\subfloat[{\small{}(c) }Austria]{
\centering{}\includegraphics[scale=0.35]{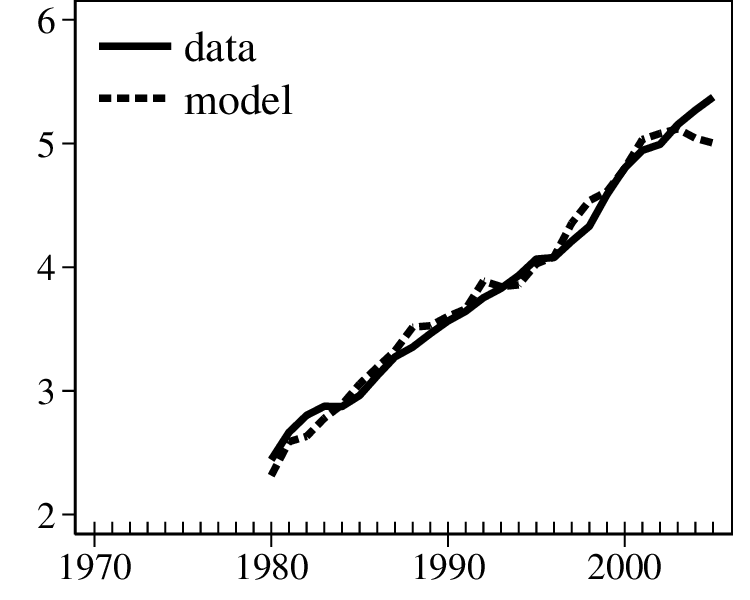}}
\par\end{centering}
\begin{centering}
\subfloat[{\small{}(d) }Czech Republic]{
\centering{}\includegraphics[scale=0.35]{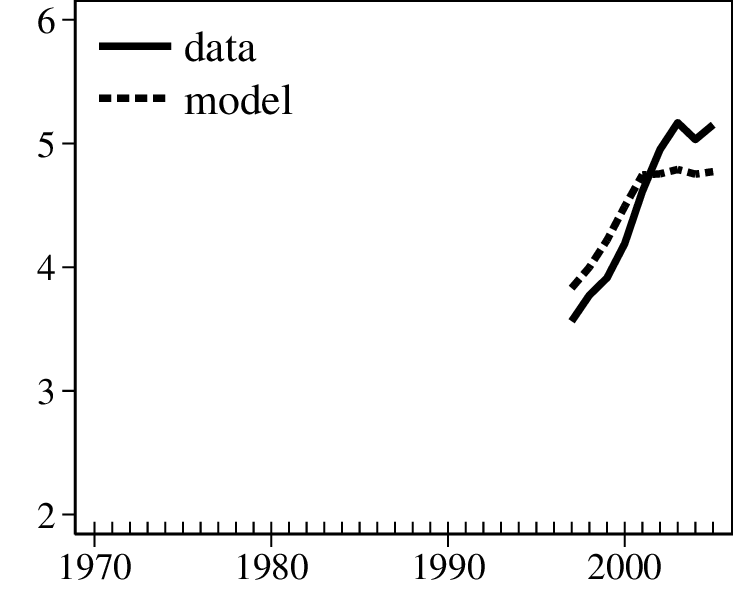}}\subfloat[{\small{}(e) }Denmark]{
\centering{}\includegraphics[scale=0.35]{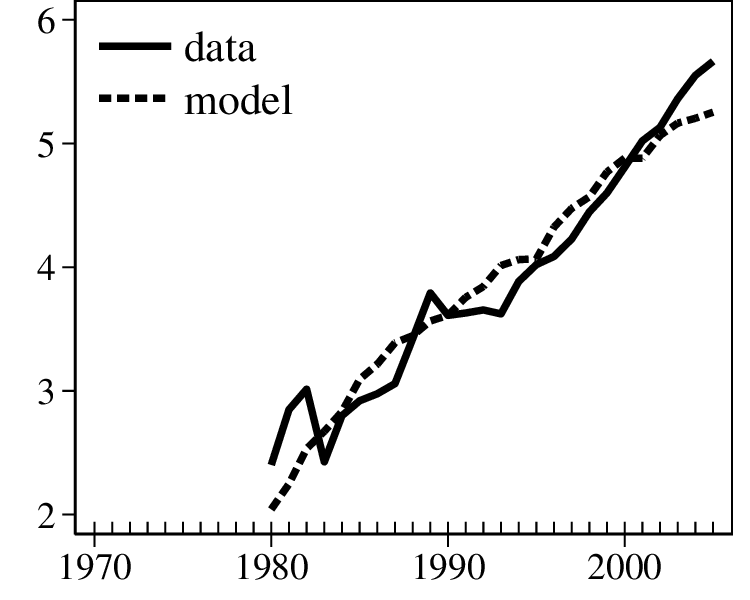}}\subfloat[{\small{}(f) }Finland]{
\centering{}\includegraphics[scale=0.35]{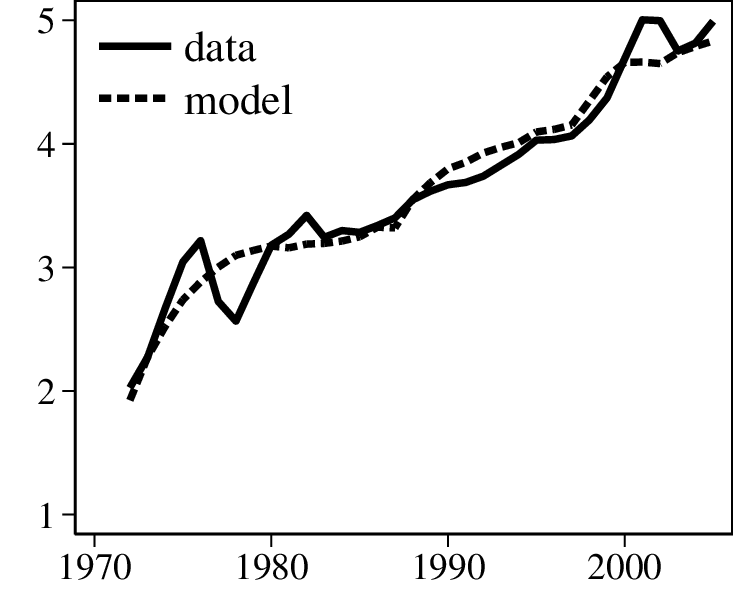}}
\par\end{centering}
\begin{centering}
\subfloat[{\small{}(g) }Germany]{
\centering{}\includegraphics[scale=0.35]{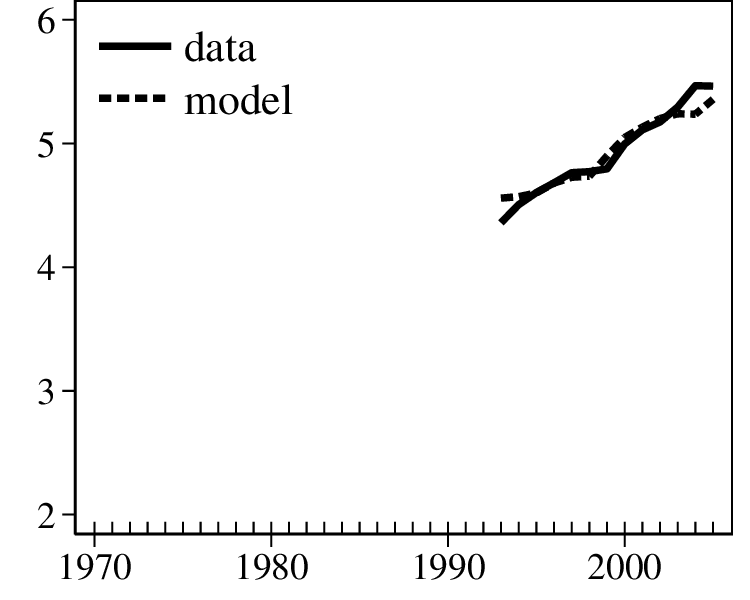}}\subfloat[{\small{}(h) }Italy]{
\centering{}\includegraphics[scale=0.35]{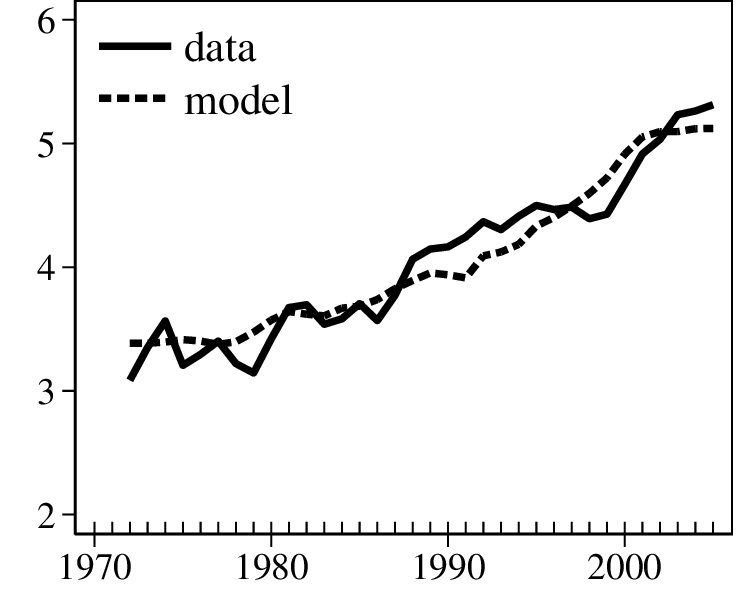}}\subfloat[{\small{}(i) }Japan]{
\centering{}\includegraphics[scale=0.35]{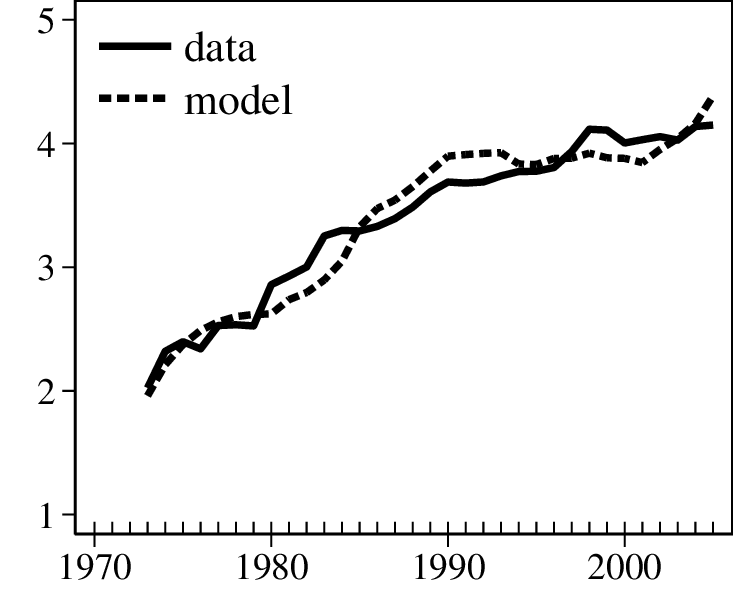}}
\par\end{centering}
\begin{centering}
\subfloat[{\small{}(j) }Netherlands]{
\centering{}\includegraphics[scale=0.35]{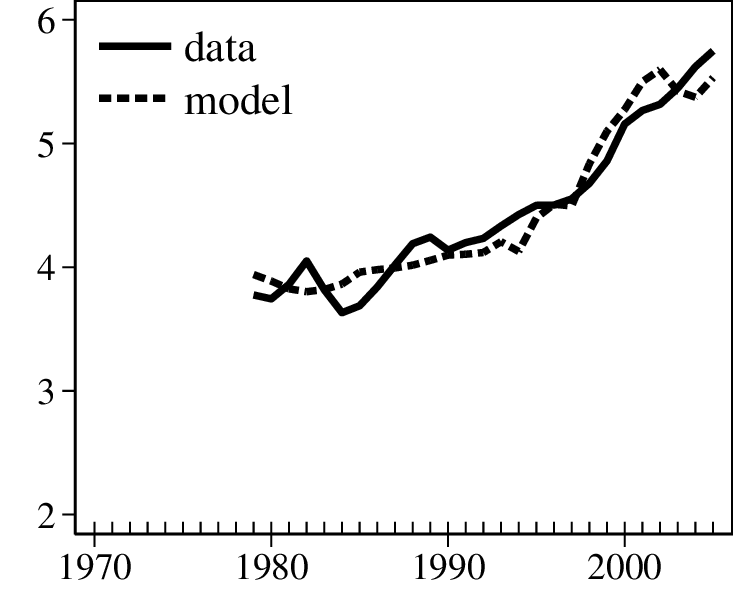}}\subfloat[{\small{}(k) }Portugal]{
\centering{}\includegraphics[scale=0.35]{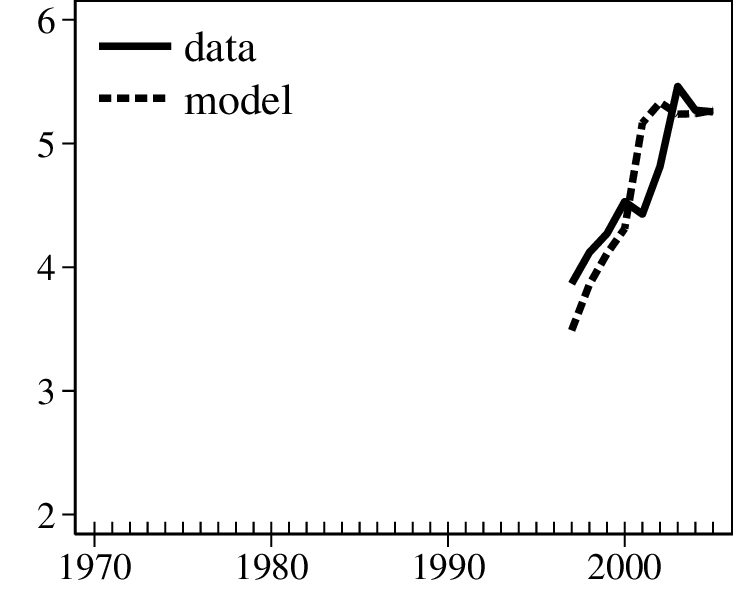}}\subfloat[{\small{}(l) }Slovenia]{
\centering{}\includegraphics[scale=0.35]{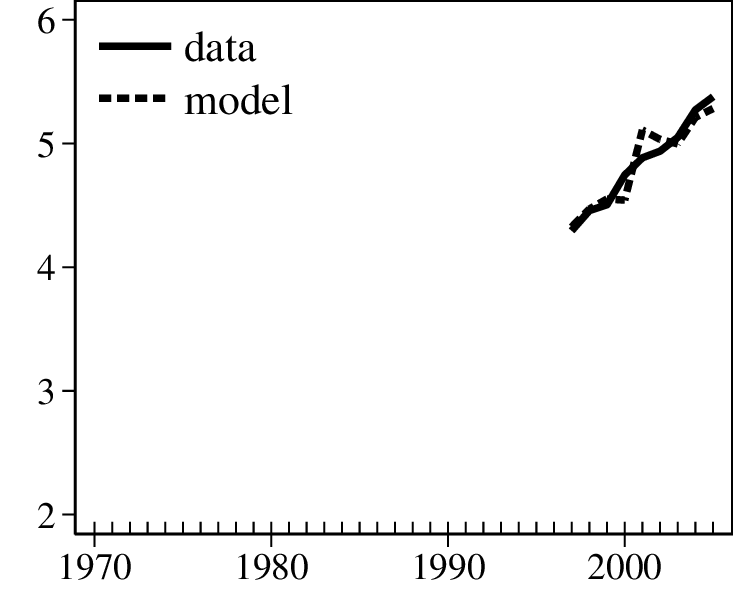}}
\par\end{centering}
\begin{centering}
\subfloat[{\small{}(m) }Sweden]{
\centering{}\includegraphics[scale=0.35]{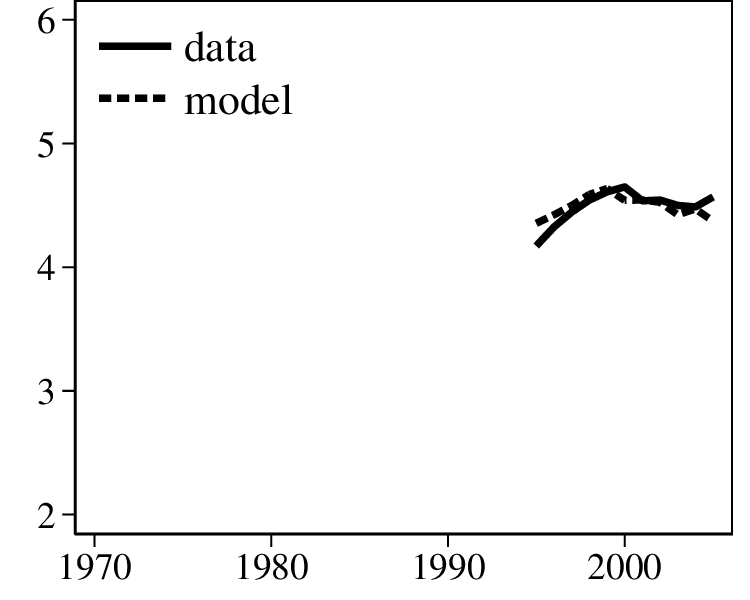}}\subfloat[{\small{}(n) }United Kingdom]{
\centering{}\includegraphics[scale=0.35]{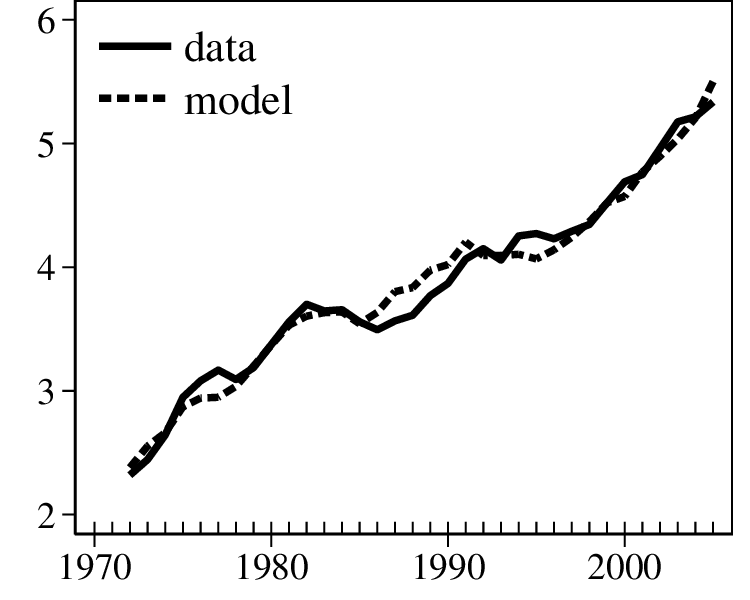}}
\par\end{centering}
\textit{\footnotesize{}Notes}{\footnotesize{}: The bold and dashed
lines indicate the actual and predicted values, respectively. All
series are logged.}{\footnotesize\par}
\end{figure}

\begin{figure}[H]
\caption{Actual and predicted ratio of the wages of skilled labor to the rental
price of ICT capital ($w_{h}/r_{i}$) in the service sector\label{fig: Wh/Ri_oecd_service}}

\begin{centering}
\subfloat[{\small{}(a) }United States]{
\centering{}\includegraphics[scale=0.35]{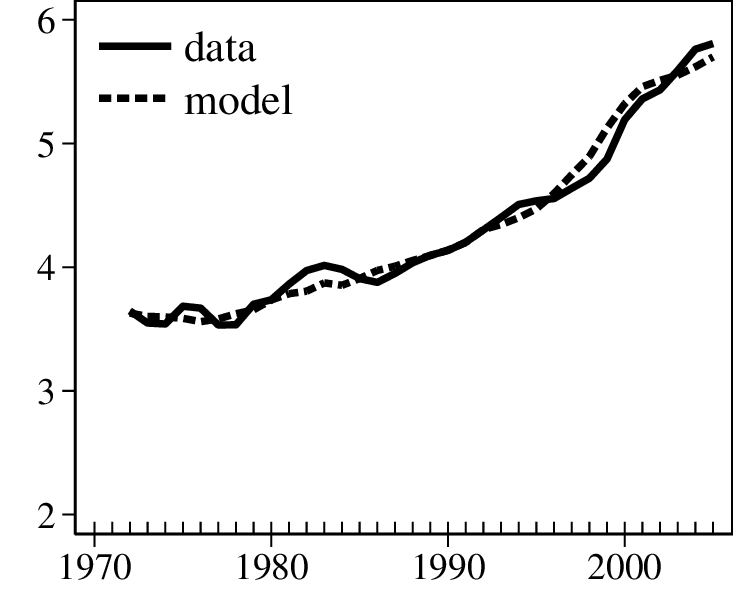}}\subfloat[{\small{}(b) }Australia]{
\centering{}\includegraphics[scale=0.35]{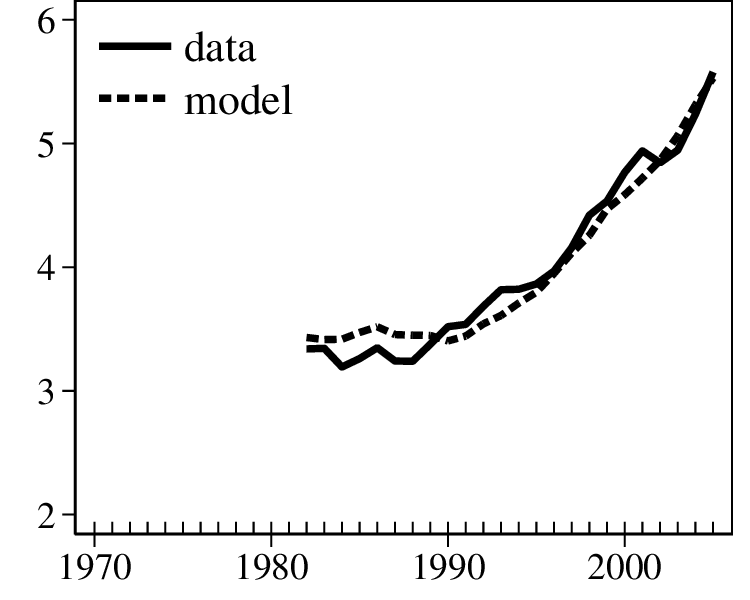}}\subfloat[{\small{}(c) }Austria]{
\centering{}\includegraphics[scale=0.35]{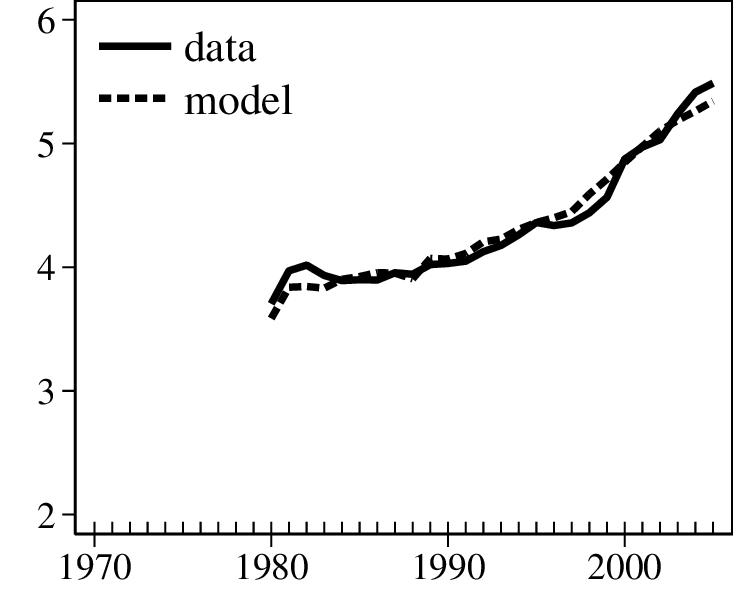}}
\par\end{centering}
\begin{centering}
\subfloat[{\small{}(d) }Czech Republic]{
\centering{}\includegraphics[scale=0.35]{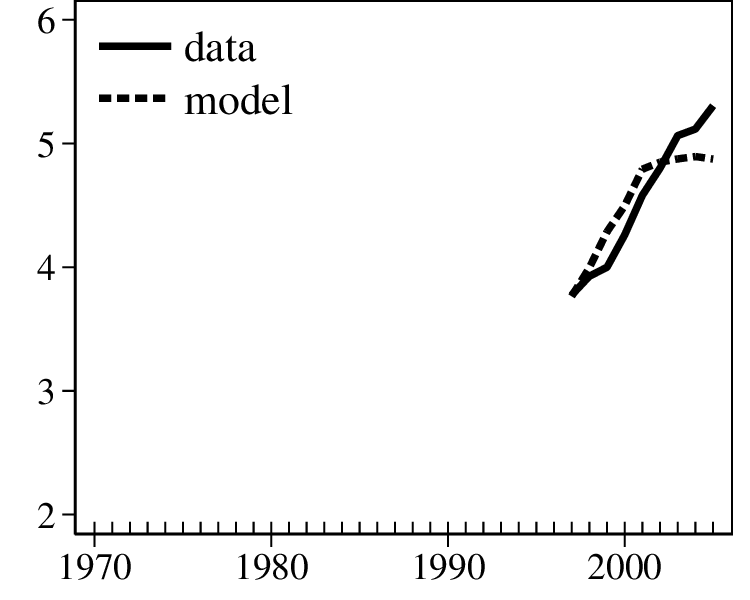}}\subfloat[{\small{}(e) }Denmark]{
\centering{}\includegraphics[scale=0.35]{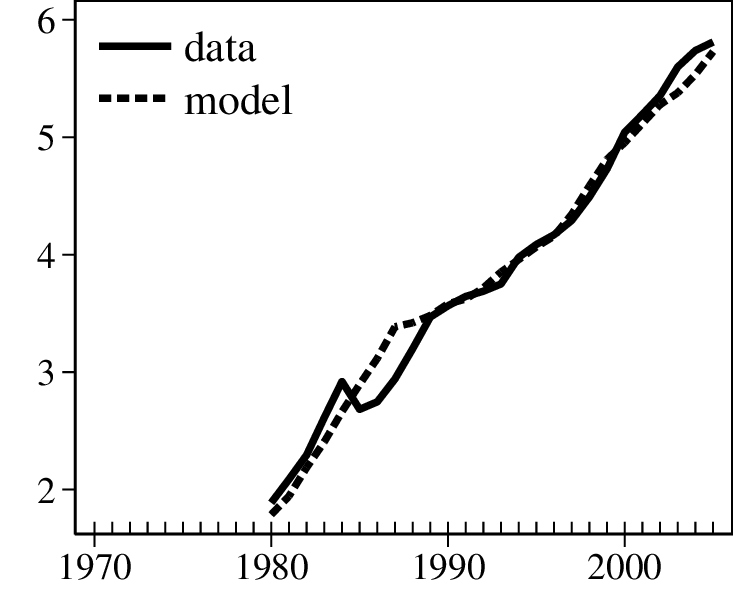}}\subfloat[{\small{}(f) }Finland]{
\centering{}\includegraphics[scale=0.35]{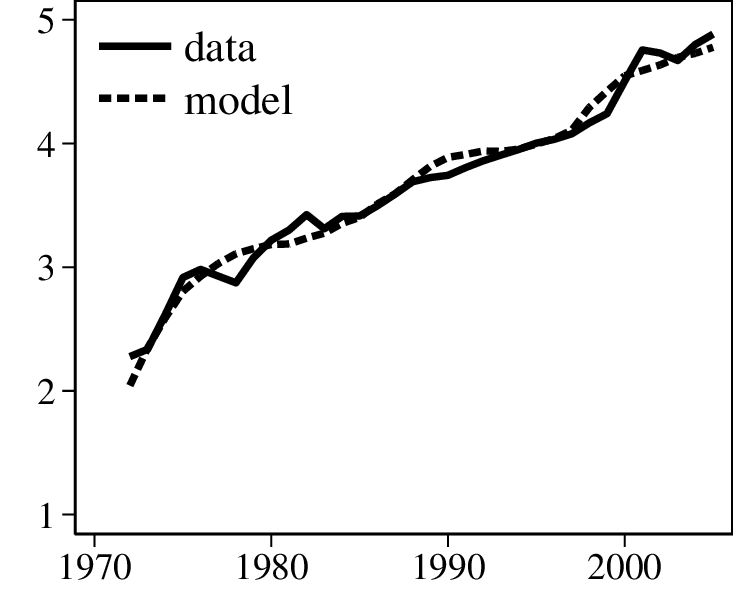}}
\par\end{centering}
\begin{centering}
\subfloat[{\small{}(g) }Germany]{
\centering{}\includegraphics[scale=0.35]{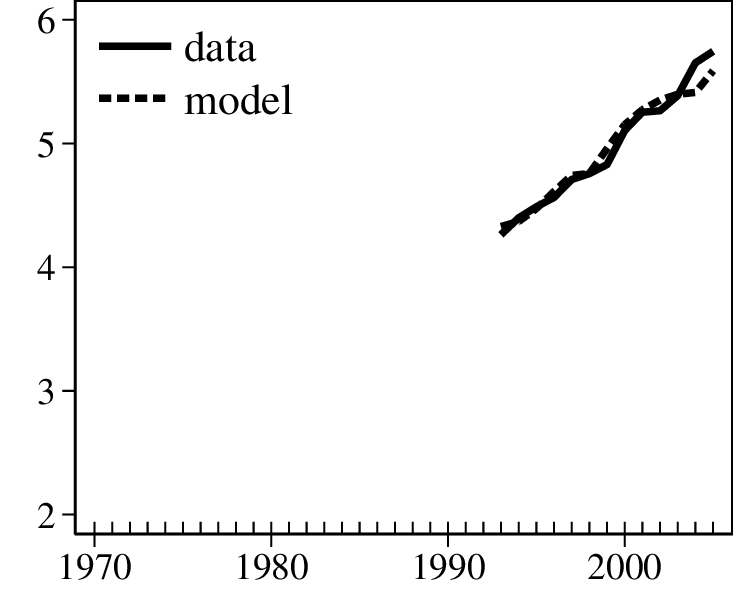}}\subfloat[{\small{}(h) }Italy]{
\centering{}\includegraphics[scale=0.35]{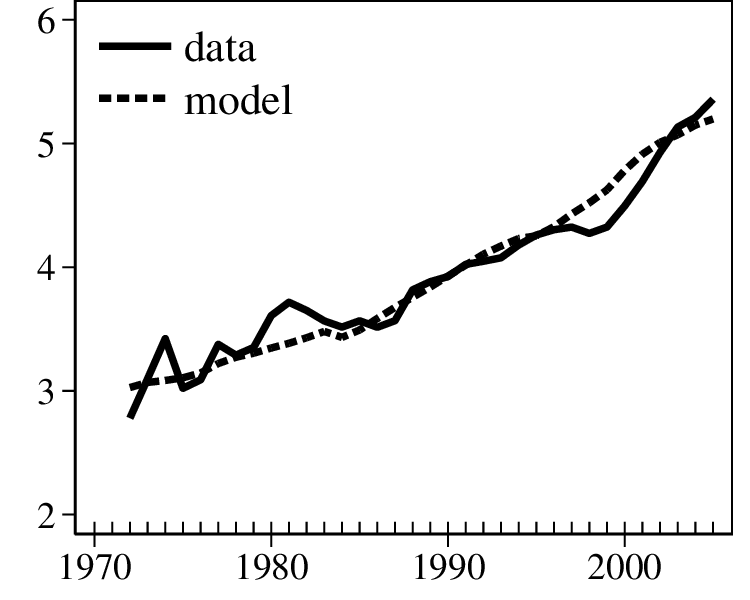}}\subfloat[{\small{}(i) }Japan]{
\centering{}\includegraphics[scale=0.35]{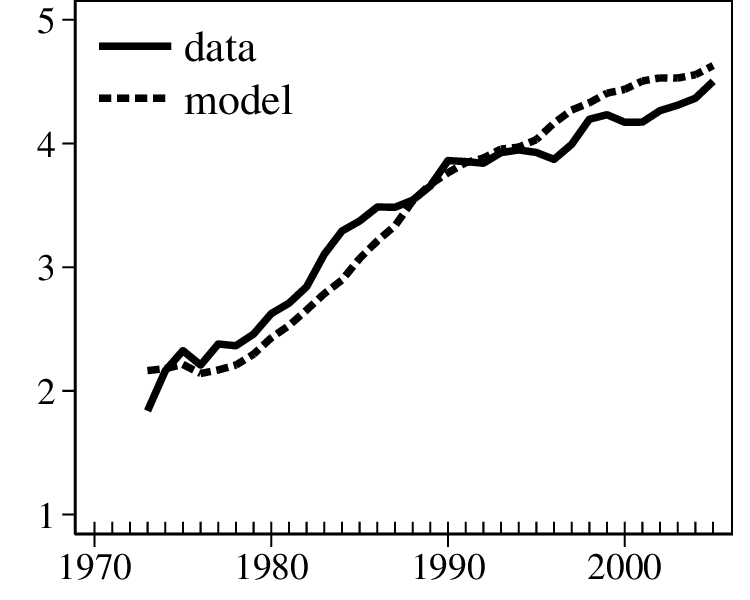}}
\par\end{centering}
\begin{centering}
\subfloat[{\small{}(j) }Netherlands]{
\centering{}\includegraphics[scale=0.35]{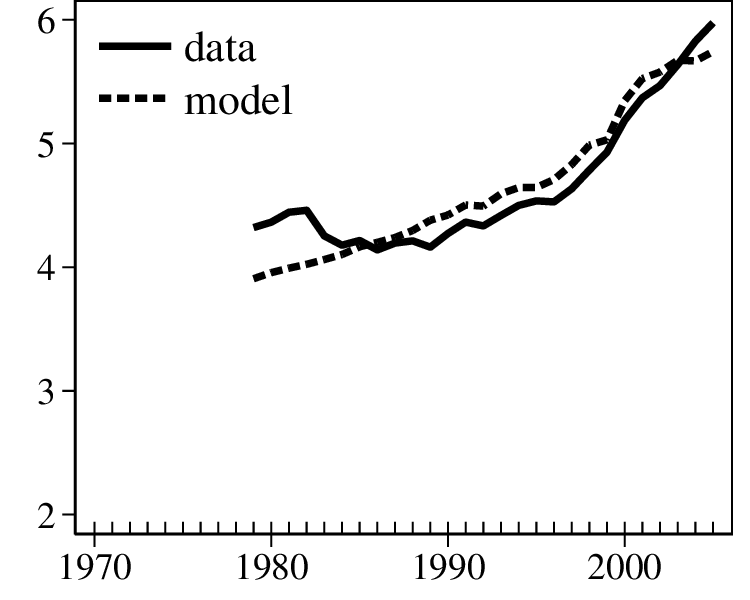}}\subfloat[{\small{}(k) }Portugal]{
\centering{}\includegraphics[scale=0.35]{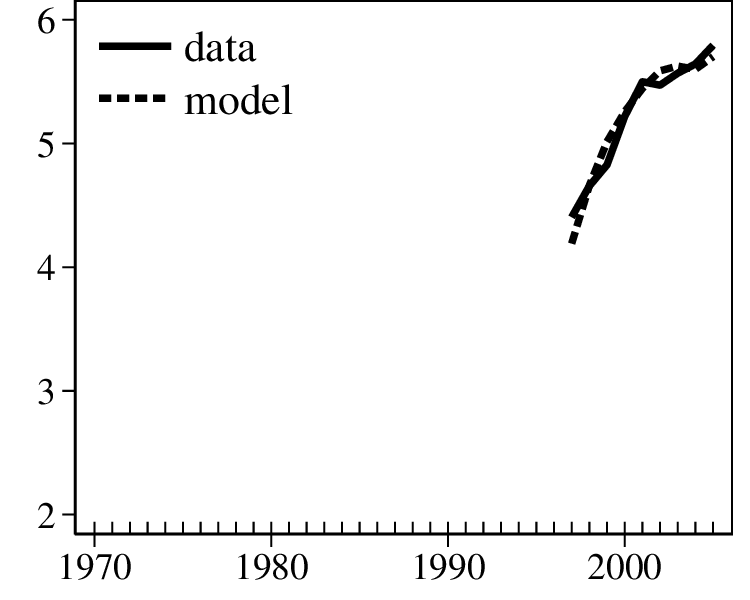}}\subfloat[{\small{}(l) }Slovenia]{
\centering{}\includegraphics[scale=0.35]{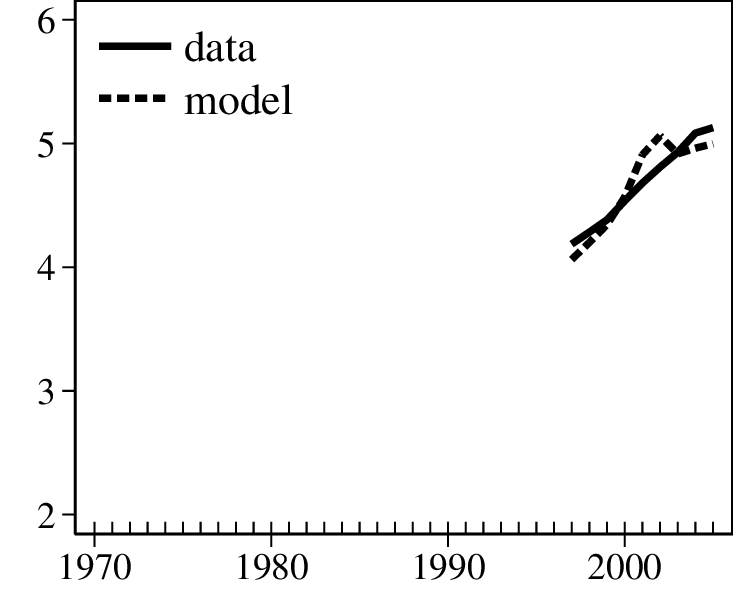}}
\par\end{centering}
\begin{centering}
\subfloat[{\small{}(m) }Sweden]{
\centering{}\includegraphics[scale=0.35]{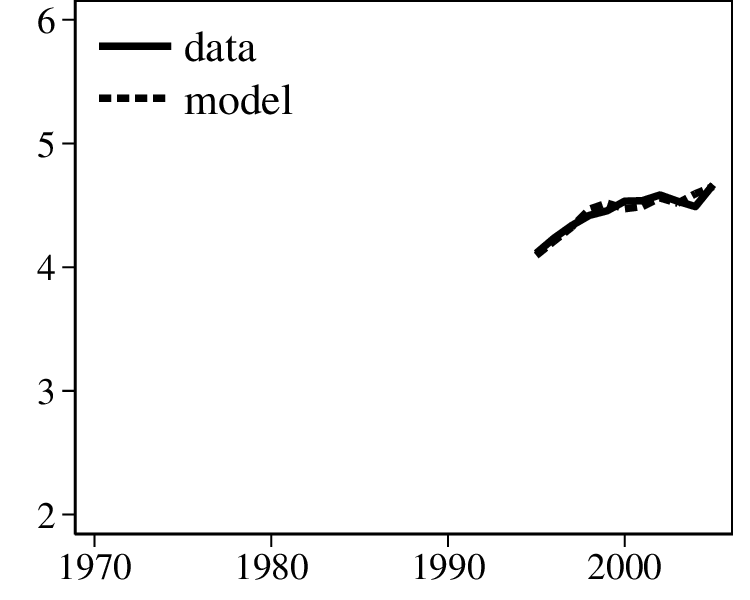}}\subfloat[{\small{}(n) }United Kingdom]{
\centering{}\includegraphics[scale=0.35]{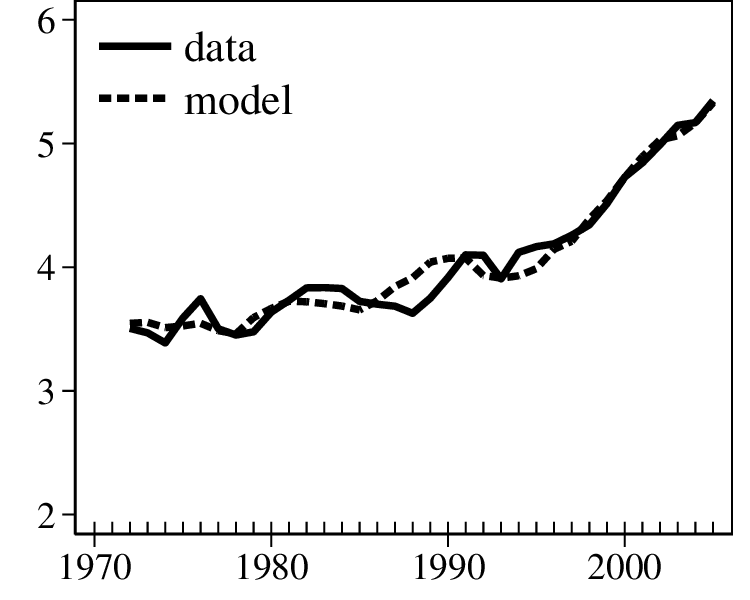}}
\par\end{centering}
\textit{\footnotesize{}Notes}{\footnotesize{}: The bold and dashed
lines indicate the actual and predicted values, respectively. All
series are logged.}{\footnotesize\par}
\end{figure}

\subsection{Skill premium}

\subsubsection{Accounting for changes in the skill premium}

Changes in the skill premium over a few decades can be explained well
by the production function with capital\textendash skill complementarity
and factor-augmenting technology for all countries and sectors. Figures
\ref{fig: Wh/Wu_4_oecd_goods} and \ref{fig: Wh/Wu_4_oecd_service}
show that the actual values of the skill premium ($w_{h}/w_{u}$)
almost overlap the predicted values from equation \eqref{eq: Wh/Wu}
in the goods and service sectors of all countries. Figures \ref{fig: Wh/Ri_oecd_goods}
and \ref{fig: Wh/Ri_oecd_service} show that the actual values of
the ratio of skilled wages to the rental price of ICT capital ($w_{h}/r_{i}$)
almost overlap the predicted values from equation \eqref{eq: Wh/Ri}
in the goods and service sectors of all countries. The orders of the
trend polynomials in equation \eqref{eq: lambda=000026mu} are chosen
to fit the data for each country and sector.\footnote{The results presented here are obtained when we include in $\lambda$
cubic trends for the goods (service) sector of Australia, Austria,
and Finland (Australia, Austria, Finland, and the United States);
quadratic trends for the goods (service) sector of Italy, Japan, the
Netherlands, Portugal, and the United States (Denmark, Italy, and
the Netherlands); linear trends for the goods (service) sector of
Denmark, Germany, and the United Kingdom (Japan and Sweden); and no
trend for the goods (service) sector of the Czech Republic, Slovenia,
and Sweden (the Czech Republic, Germany, Portugal, Slovenia, and the
United Kingdom). We also include in $\mu$ cubic trends for the goods
(service) sector of Japan and the United Kingdom (the United Kingdom);
quadratic trends for the goods (service) sector of the Netherlands
(Italy); linear trends for the goods (service) sector of Austria,
Denmark, Finland, Slovenia, and the United States (Finland and the
United States); and no trend for the goods (service) sector of Australia,
the Czech Republic, Germany, Italy, Portugal, and Sweden (Australia,
Austria, the Czech Republic, Denmark, Germany, Japan, the Netherlands,
Portugal, Slovenia, and Sweden).}

\subsubsection{Decomposition of changes in the skill premium}

\begin{figure}[H]
\caption{Decomposition of changes in the skill premium in the goods sector\label{fig: Wh/Wu_decomp_4_goods}}

\begin{centering}
\subfloat[{\small{}(a) }United States]{
\centering{}\includegraphics[scale=0.35]{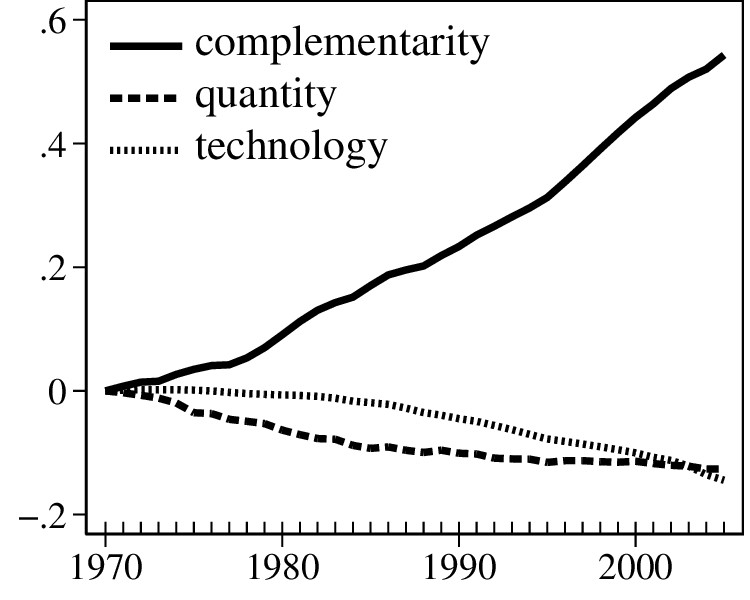}}\subfloat[{\small{}(b) }Australia]{
\centering{}\includegraphics[scale=0.35]{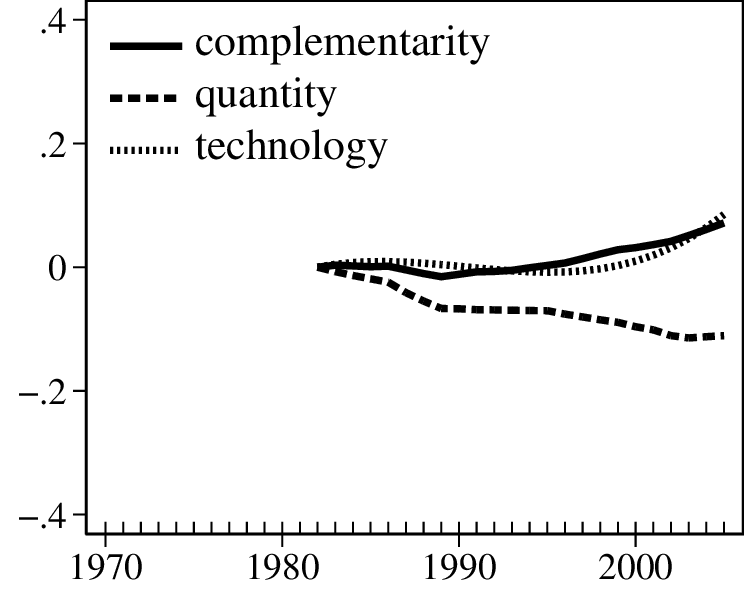}}\subfloat[{\small{}(c) }Austria]{
\centering{}\includegraphics[scale=0.35]{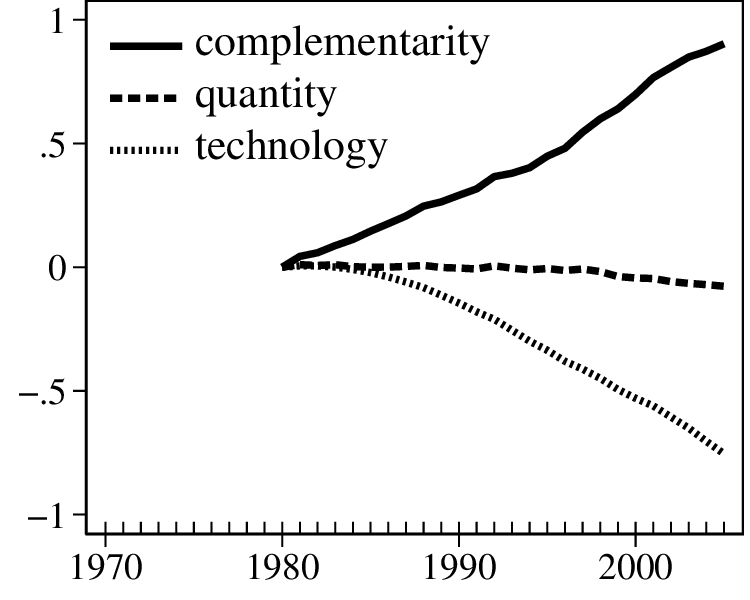}}
\par\end{centering}
\begin{centering}
\subfloat[{\small{}(d) }Czech Republic]{
\centering{}\includegraphics[scale=0.35]{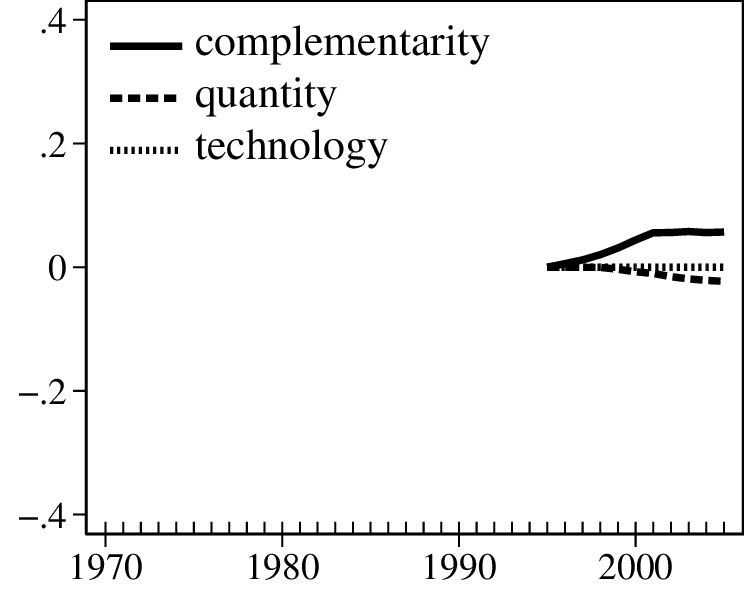}}\subfloat[{\small{}(e) }Denmark]{
\centering{}\includegraphics[scale=0.35]{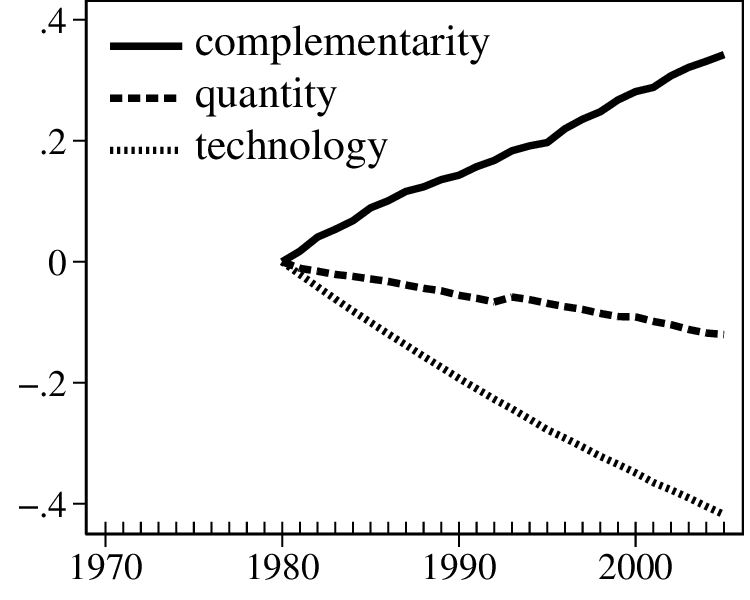}}\subfloat[{\small{}(f) }Finland]{
\centering{}\includegraphics[scale=0.35]{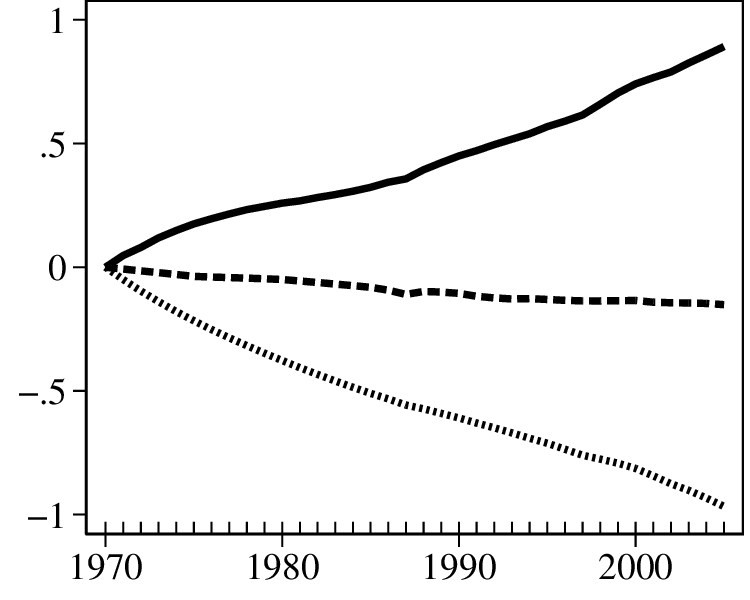}}
\par\end{centering}
\begin{centering}
\subfloat[{\small{}(g) }Germany]{
\centering{}\includegraphics[scale=0.35]{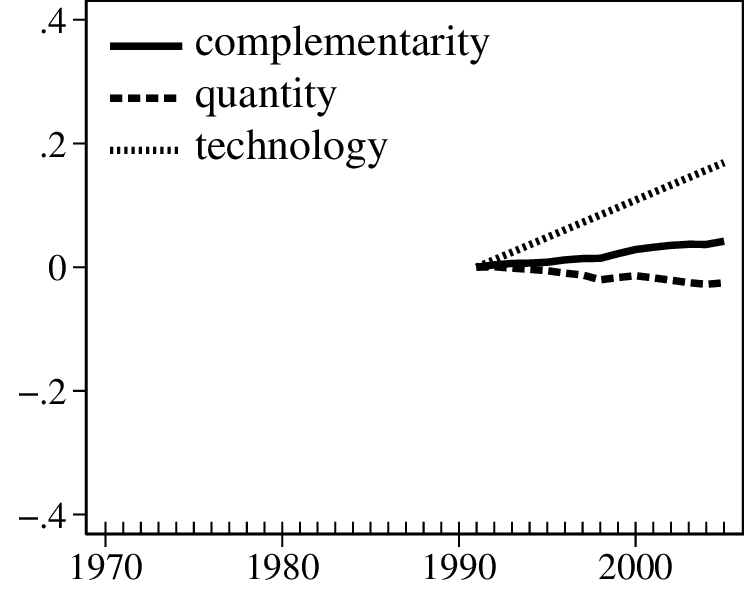}}\subfloat[{\small{}(h) }Italy]{
\centering{}\includegraphics[scale=0.35]{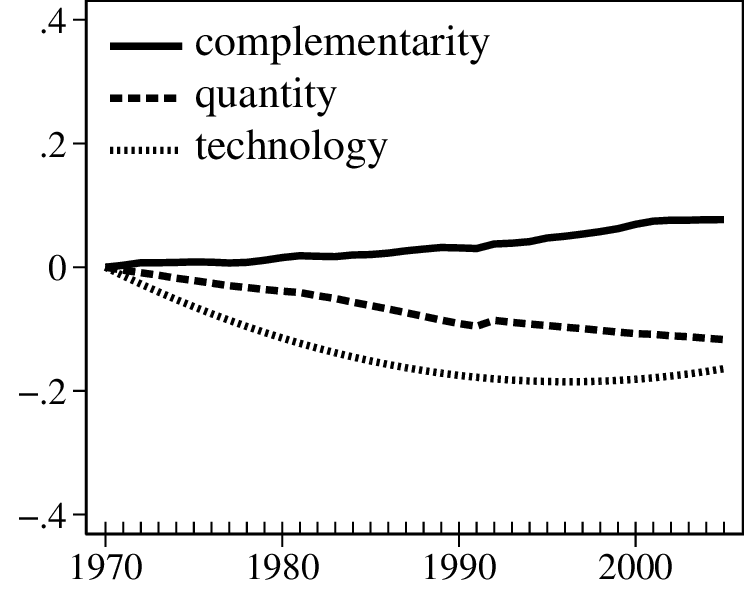}}\subfloat[{\small{}(i) }Japan]{
\centering{}\includegraphics[scale=0.35]{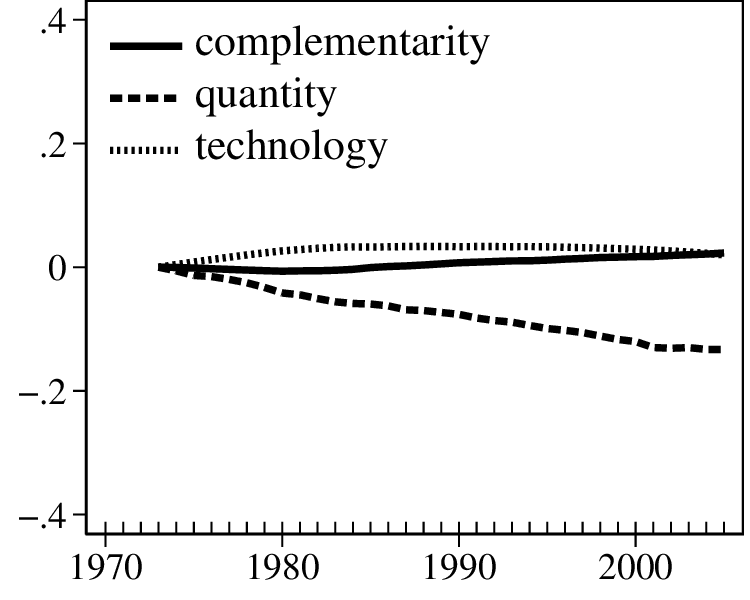}}
\par\end{centering}
\begin{centering}
\subfloat[{\small{}(j) }Netherlands]{
\centering{}\includegraphics[scale=0.35]{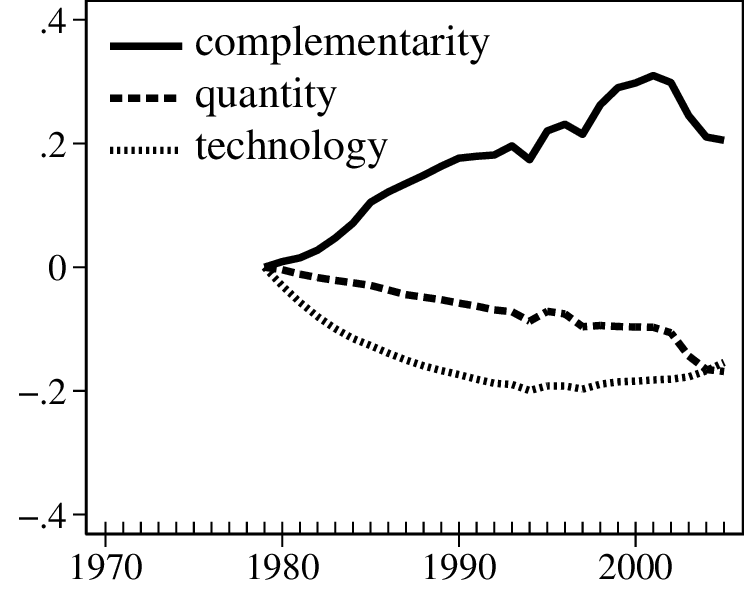}}\subfloat[{\small{}(k) }Portugal]{
\centering{}\includegraphics[scale=0.35]{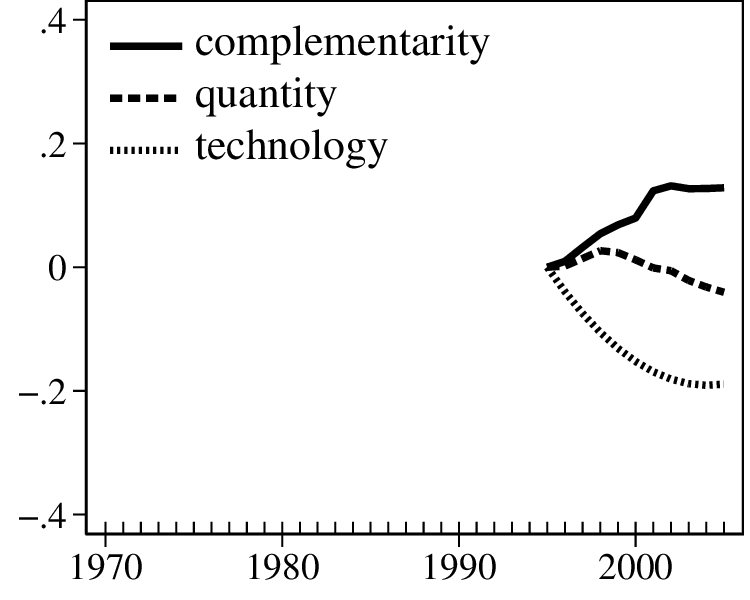}}\subfloat[{\small{}(l) }Slovenia]{
\centering{}\includegraphics[scale=0.35]{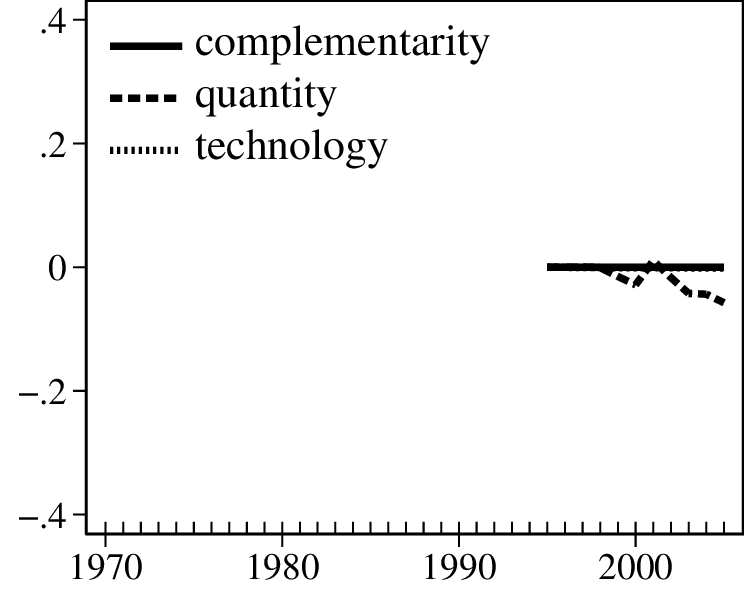}}
\par\end{centering}
\begin{centering}
\subfloat[{\small{}(m) }Sweden]{
\centering{}\includegraphics[scale=0.35]{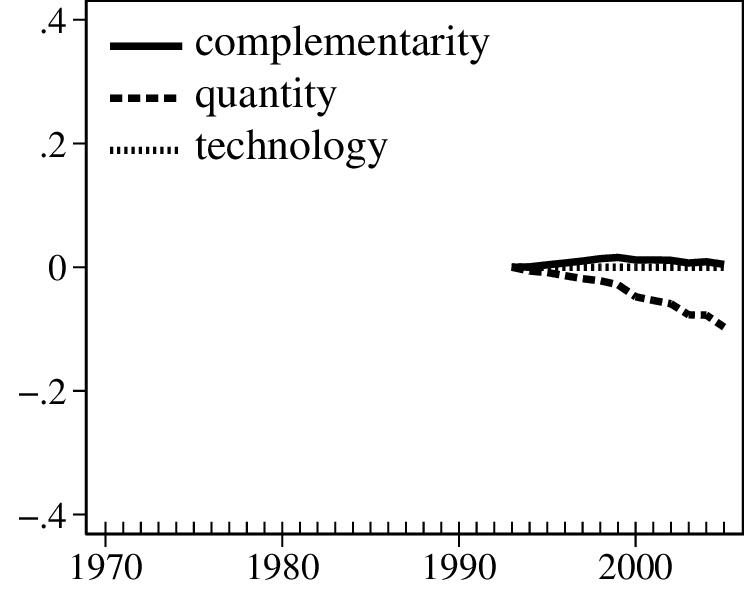}}\subfloat[{\small{}(n) }United Kingdom]{
\centering{}\includegraphics[scale=0.35]{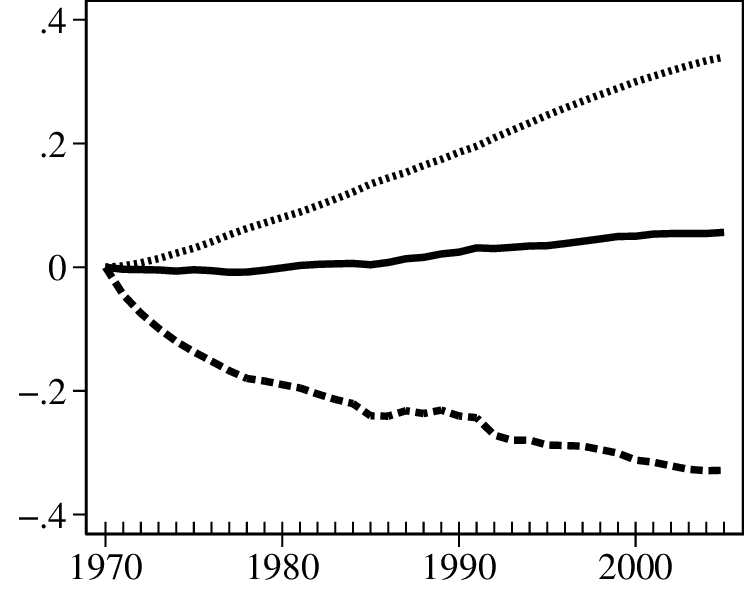}}
\par\end{centering}
\textit{\footnotesize{}Notes}{\footnotesize{}: The bold, dashed, and
dotted lines indicate the capital\textendash skill complementarity
effect, relative labor quantity effect, and relative factor-augmenting
technology effect, respectively. All series are logged and normalized
to zero in the initial year.}{\footnotesize\par}
\end{figure}

\begin{figure}[H]
\caption{Decomposition of changes in the skill premium in the service sector\label{fig: Wh/Wu_decomp_4_service}}

\begin{centering}
\subfloat[{\small{}(a) }United States]{
\centering{}\includegraphics[scale=0.35]{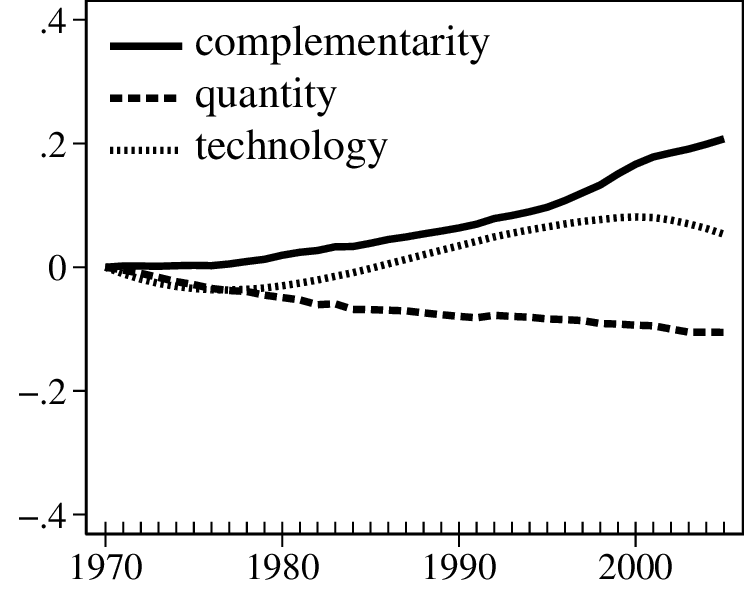}}\subfloat[{\small{}(b) }Australia]{
\centering{}\includegraphics[scale=0.35]{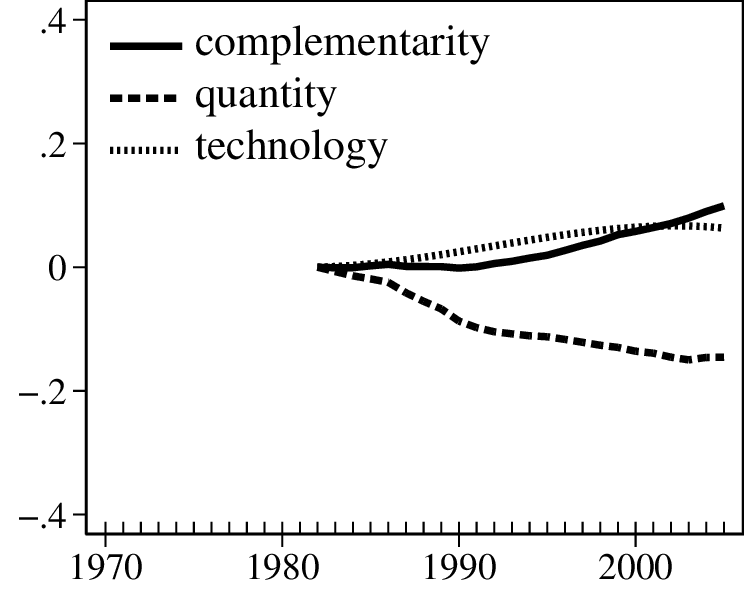}}\subfloat[{\small{}(c) }Austria]{
\centering{}\includegraphics[scale=0.35]{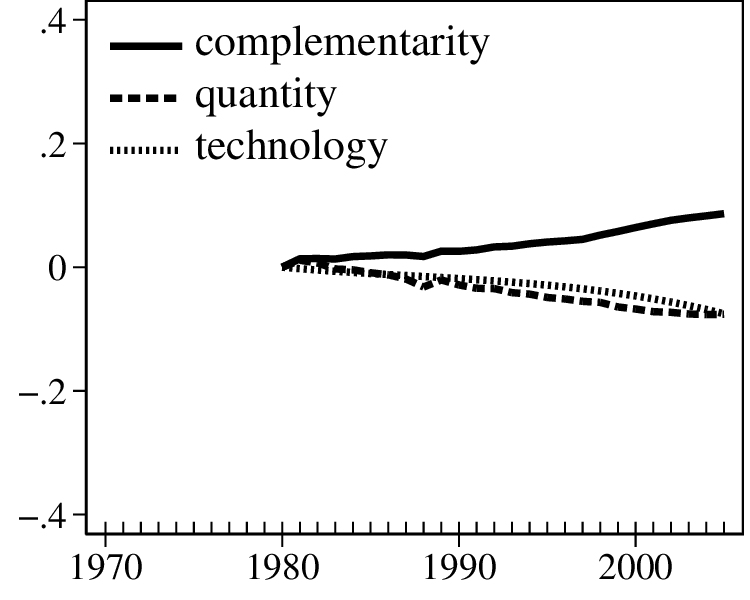}}
\par\end{centering}
\begin{centering}
\subfloat[{\small{}(d) }Czech Republic]{
\centering{}\includegraphics[scale=0.35]{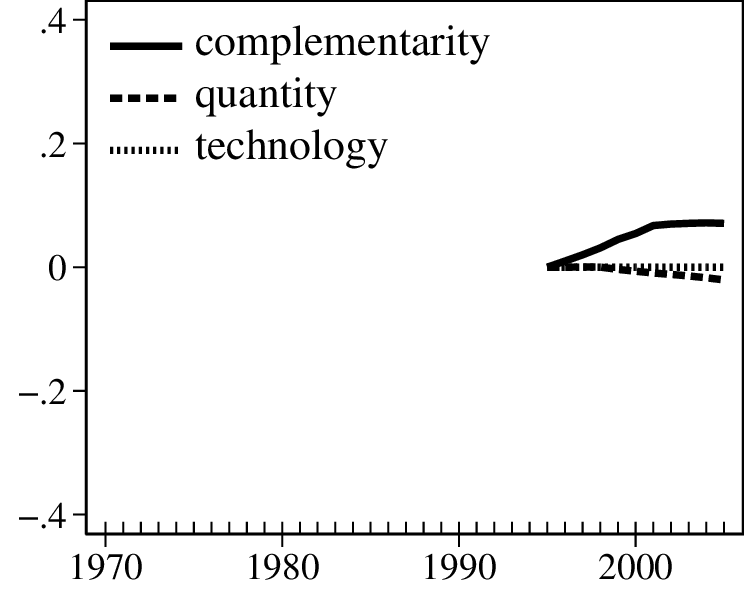}}\subfloat[{\small{}(e) }Denmark]{
\centering{}\includegraphics[scale=0.35]{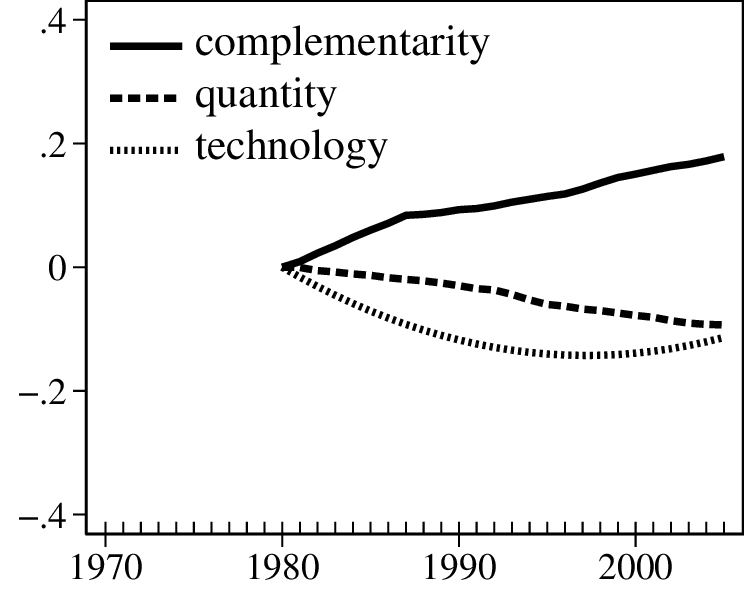}}\subfloat[{\small{}(f) }Finland]{
\centering{}\includegraphics[scale=0.35]{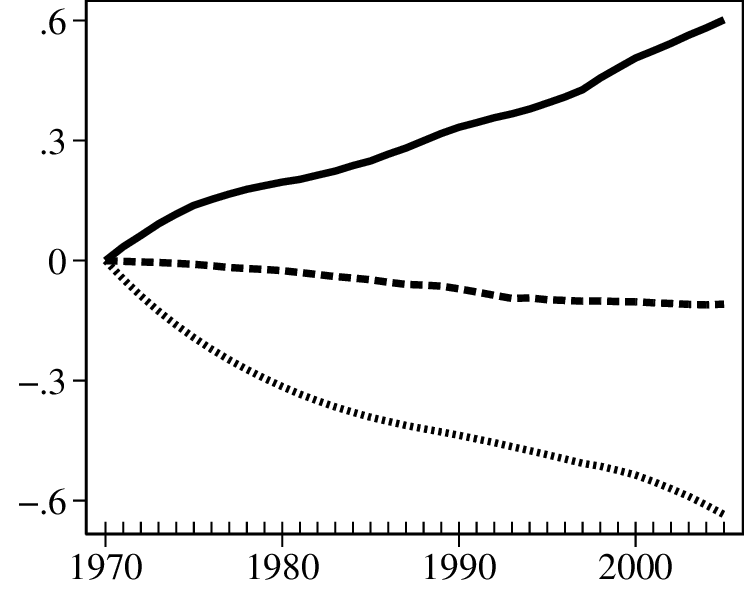}}
\par\end{centering}
\begin{centering}
\subfloat[{\small{}(g) }Germany]{
\centering{}\includegraphics[scale=0.35]{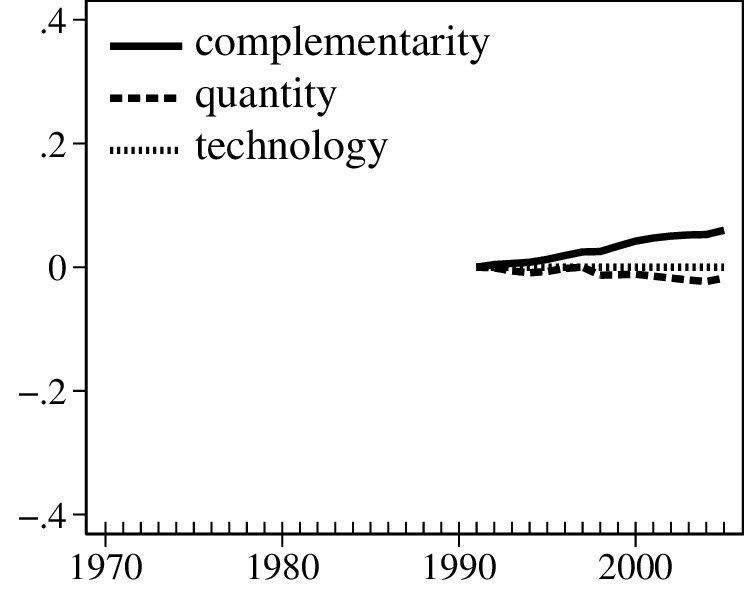}}\subfloat[{\small{}(h) }Italy]{
\centering{}\includegraphics[scale=0.35]{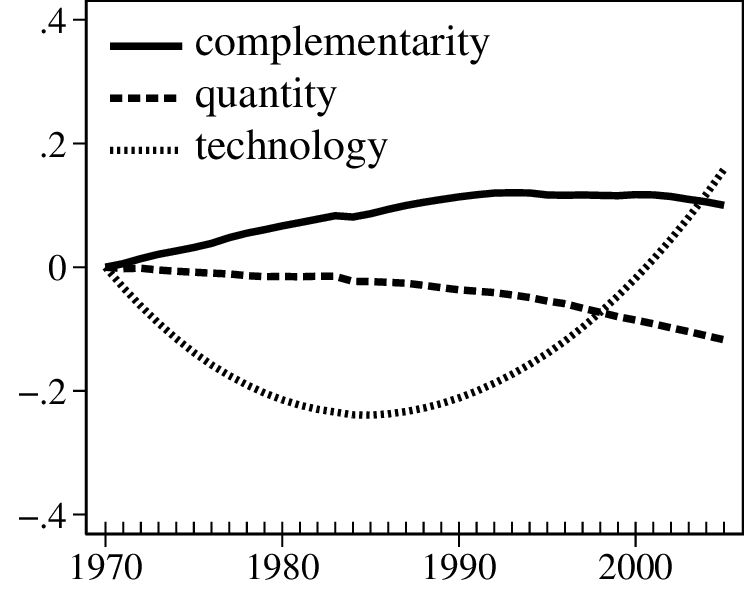}}\subfloat[{\small{}(i) }Japan]{
\centering{}\includegraphics[scale=0.35]{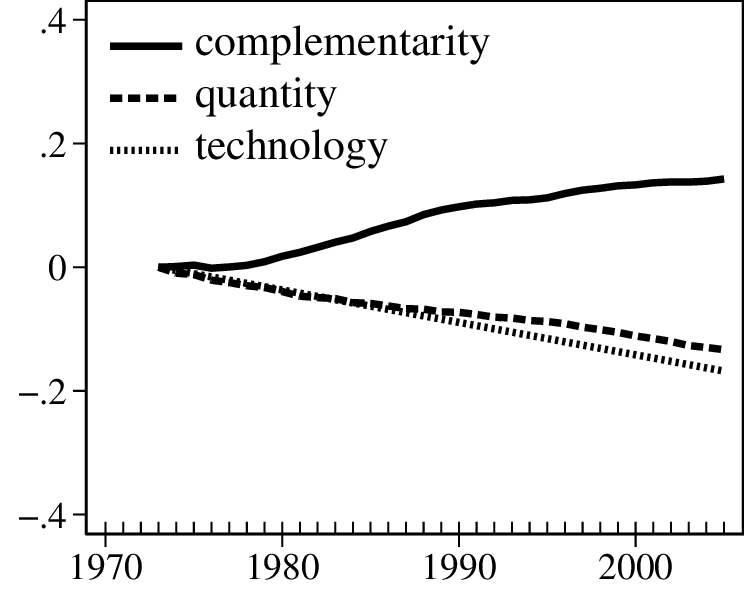}}
\par\end{centering}
\begin{centering}
\subfloat[{\small{}(j) }Netherlands]{
\centering{}\includegraphics[scale=0.35]{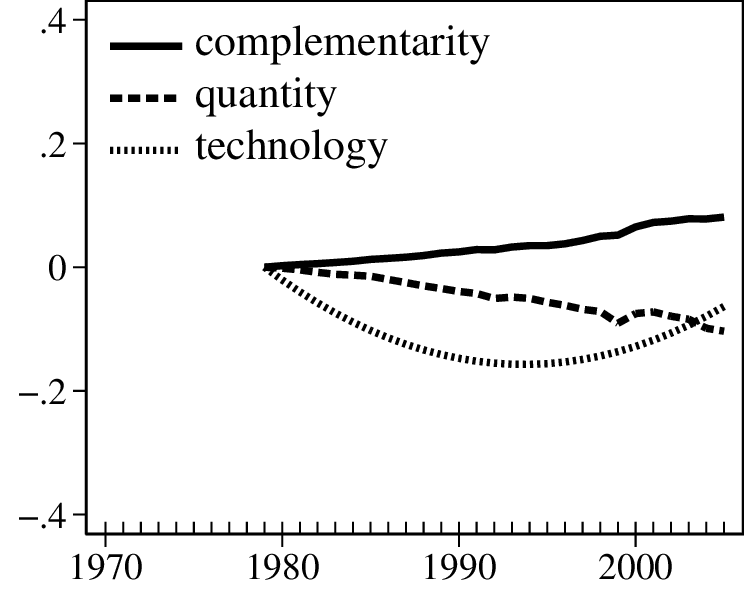}}\subfloat[{\small{}(k) }Portugal]{
\centering{}\includegraphics[scale=0.35]{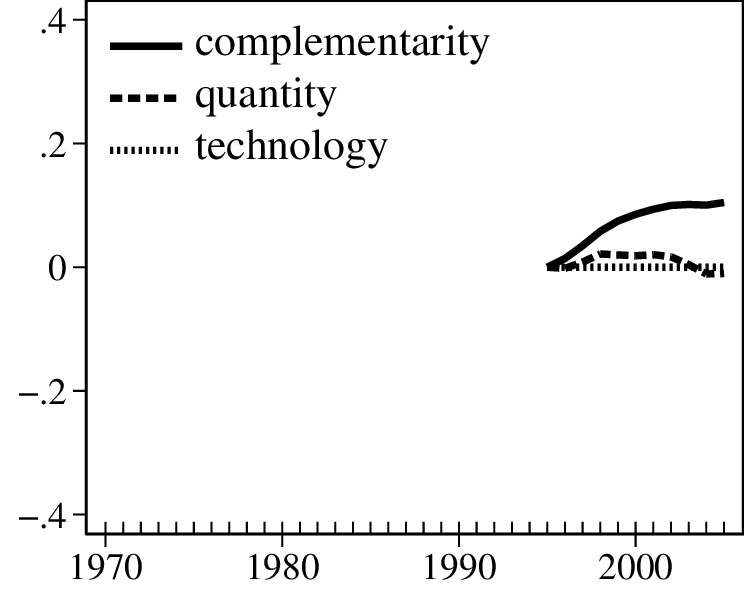}}\subfloat[{\small{}(l) }Slovenia]{
\centering{}\includegraphics[scale=0.35]{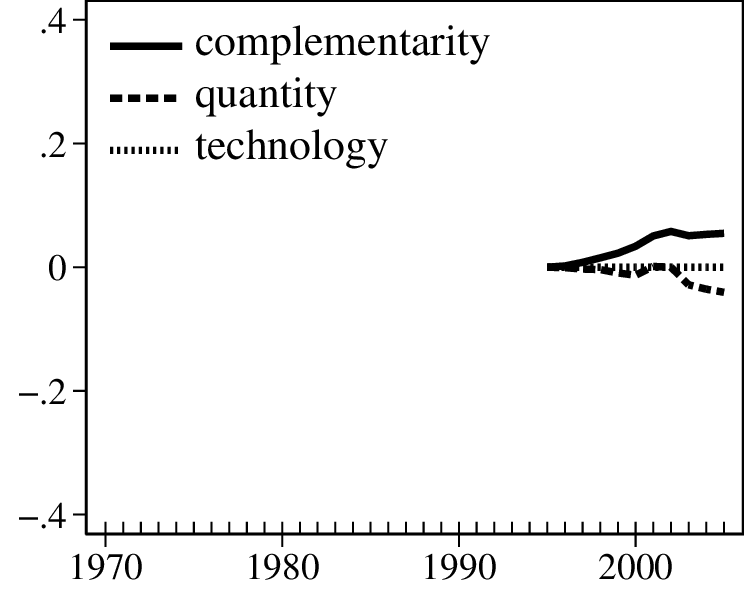}}
\par\end{centering}
\begin{centering}
\subfloat[{\small{}(m) }Sweden]{
\centering{}\includegraphics[scale=0.35]{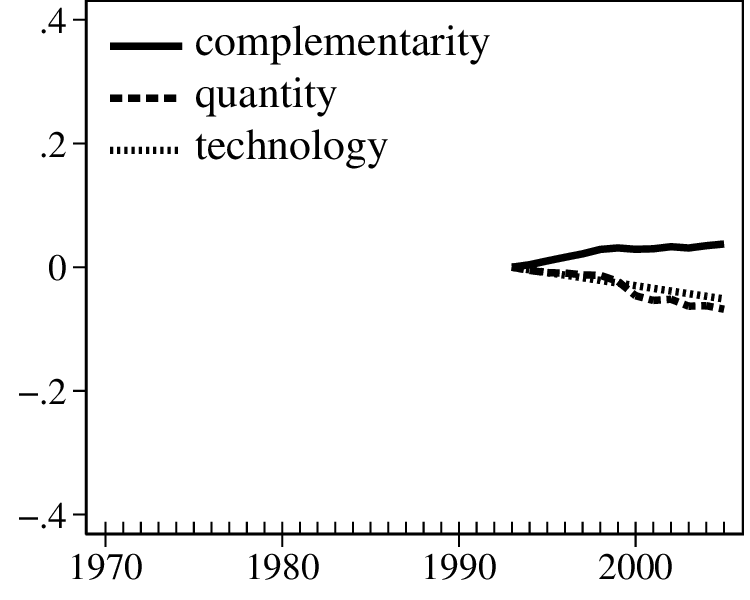}}\subfloat[{\small{}(n) }United Kingdom]{
\centering{}\includegraphics[scale=0.35]{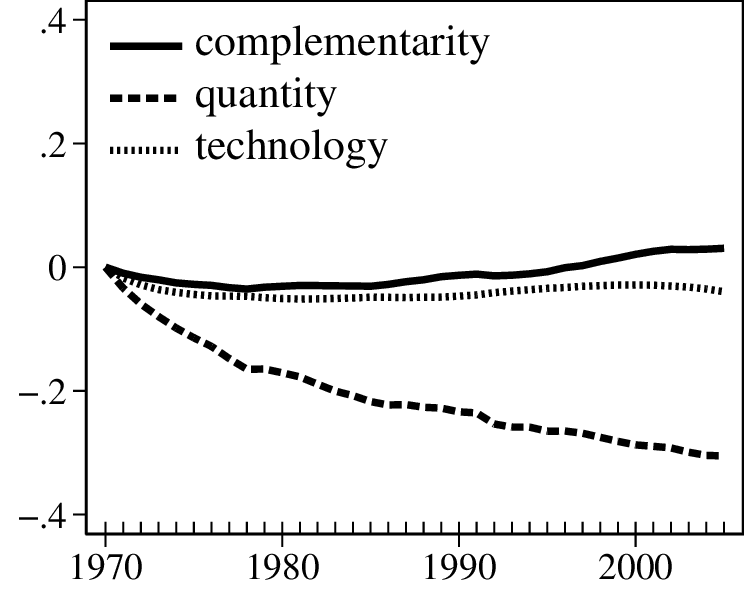}}
\par\end{centering}
\textit{\footnotesize{}Notes}{\footnotesize{}: The bold, dashed, and
dotted lines indicate the capital\textendash skill complementarity
effect, relative labor quantity effect, and relative factor-augmenting
technology effect, respectively. All series are logged and normalized
to zero in the initial year.}{\footnotesize\par}
\end{figure}

We decompose changes in the skill premium into the capital\textendash skill
complementarity effect, relative labor quantity effect, and relative
factor-augmenting technology effect. Figures \ref{fig: Wh/Wu_decomp_4_goods}
and \ref{fig: Wh/Wu_decomp_4_service} show the direction and magnitude
of the three effects in the goods and service sectors of 14 OECD countries.
The rise in the skill premium from the 1980s to the 2000s in the goods
and service sectors of the United States is attributable mostly to
the capital\textendash skill complementarity effect, while the decline
in the skill premium in the 1970s in the goods and service sectors
of the United States is attributable to the relative labor quantity
effect and relative factor-augmenting technology effect. Similarly,
the increase in the skill premium in the service sector of Germany,
the Czech Republic, and Portugal is attributable entirely to the capital\textendash skill
complementarity effect. Perhaps surprisingly, the relative factor-augmenting
technology effect is not needed to account for the increasing trend
in the skill premium in these countries. The capital\textendash skill
complementarity effect associated with a rise in ICT equipment is
large enough to account for a rise in the skill premium in almost
all countries. The increase in the skill premium in the goods sector
of Germany and the United Kingdom is, however, attributable mainly
to the relative factor-augmenting technology effect.

\begin{figure}[H]
\caption{Trends in the skill premium attributable to observed factors in the
goods sector\label{fig: Wh/Wu_decomp_KiKo4_observed_goods}}

\begin{centering}
\subfloat[{\small{}(a) }United States]{
\centering{}\includegraphics[scale=0.35]{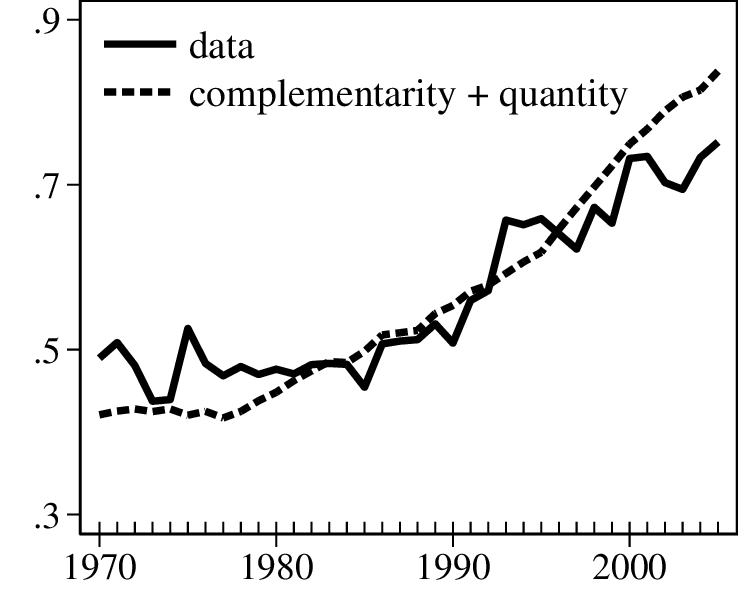}}\subfloat[{\small{}(b) }Australia]{
\centering{}\includegraphics[scale=0.35]{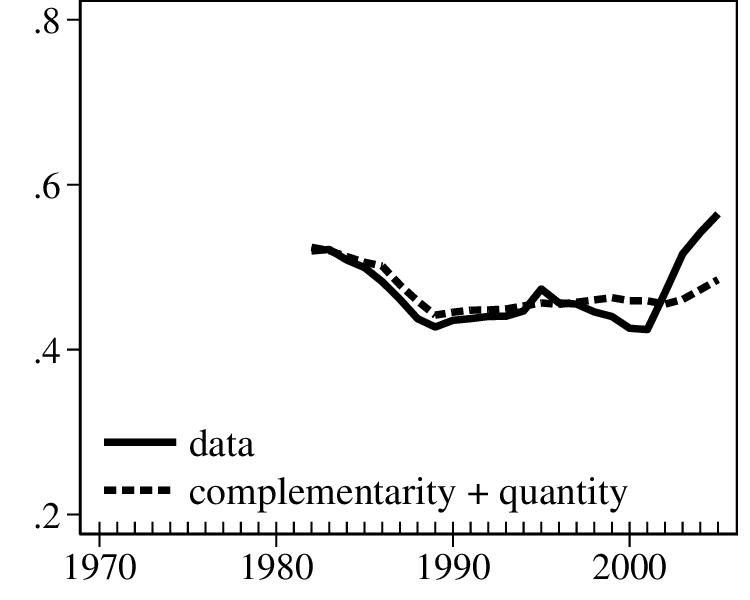}}\subfloat[{\small{}(c) }Austria]{
\centering{}\includegraphics[scale=0.35]{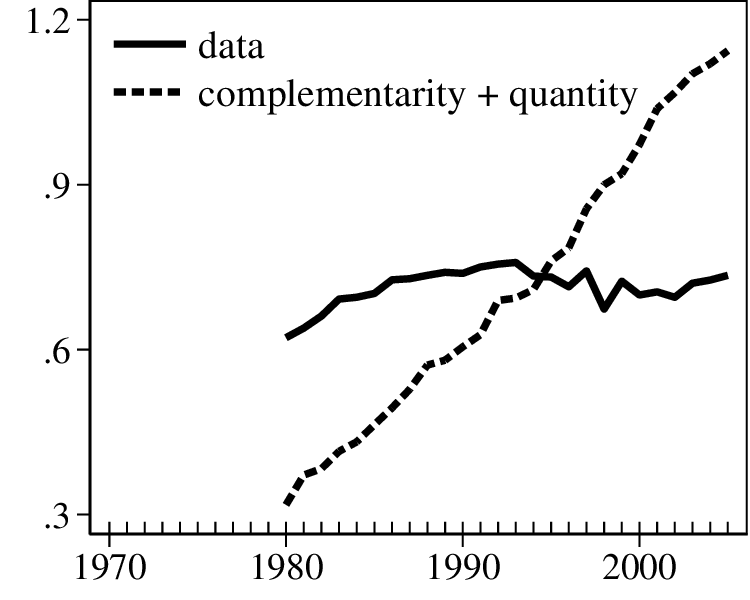}}
\par\end{centering}
\begin{centering}
\subfloat[{\small{}(d) }Czech Republic]{
\centering{}\includegraphics[scale=0.35]{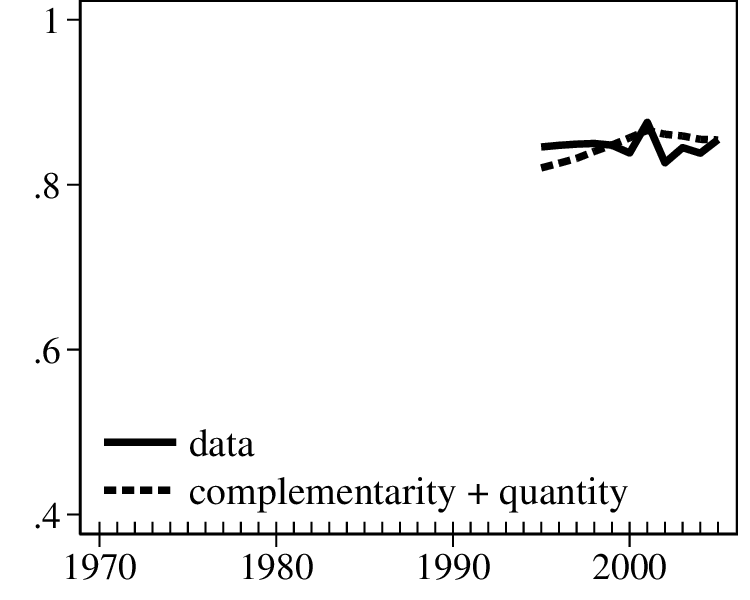}}\subfloat[{\small{}(e) }Denmark]{
\centering{}\includegraphics[scale=0.35]{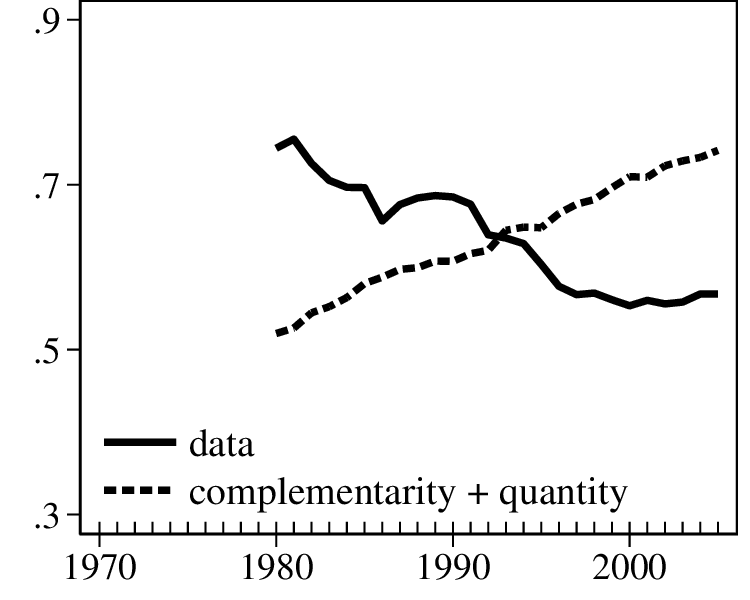}}\subfloat[{\small{}(f) }Finland]{
\centering{}\includegraphics[scale=0.35]{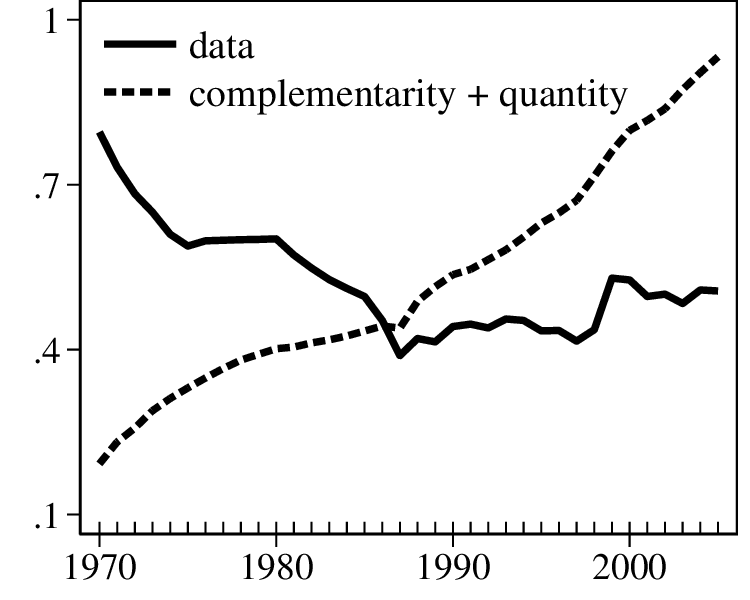}}
\par\end{centering}
\begin{centering}
\subfloat[{\small{}(g) }Germany]{
\centering{}\includegraphics[scale=0.35]{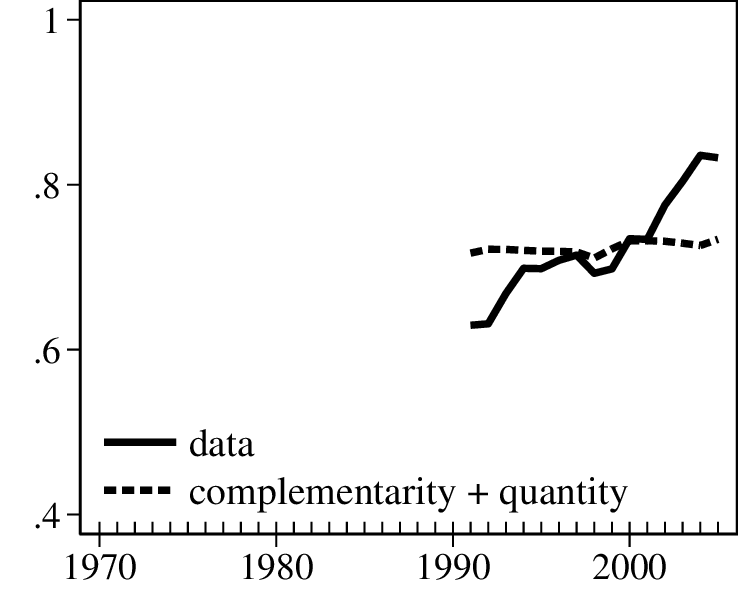}}\subfloat[{\small{}(h) }Italy]{
\centering{}\includegraphics[scale=0.35]{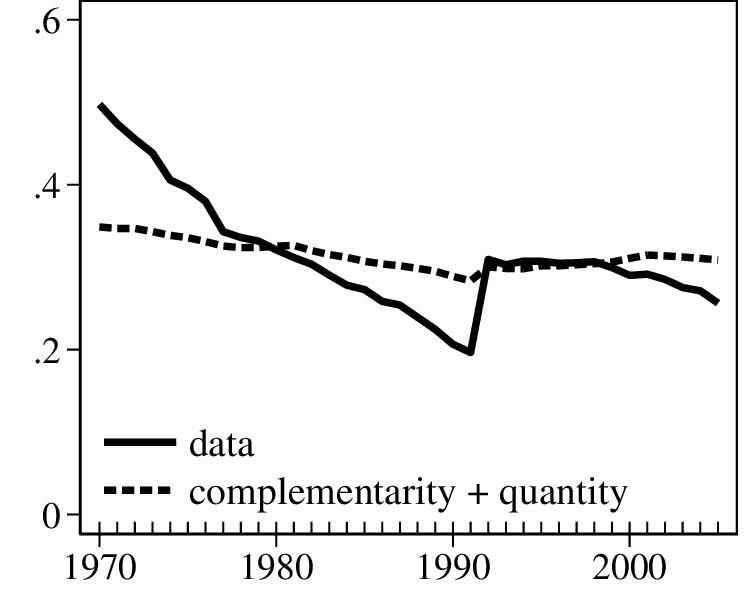}}\subfloat[{\small{}(i) }Japan]{
\centering{}\includegraphics[scale=0.35]{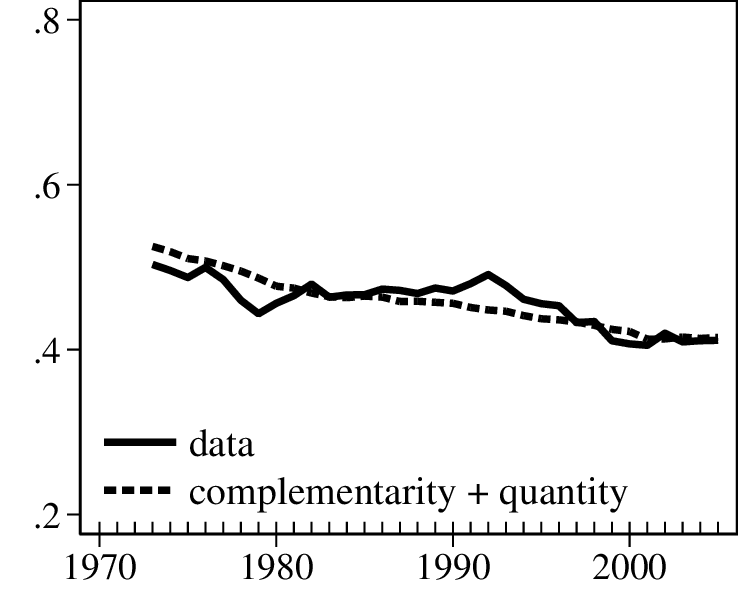}}
\par\end{centering}
\begin{centering}
\subfloat[{\small{}(j) }Netherlands]{
\centering{}\includegraphics[scale=0.35]{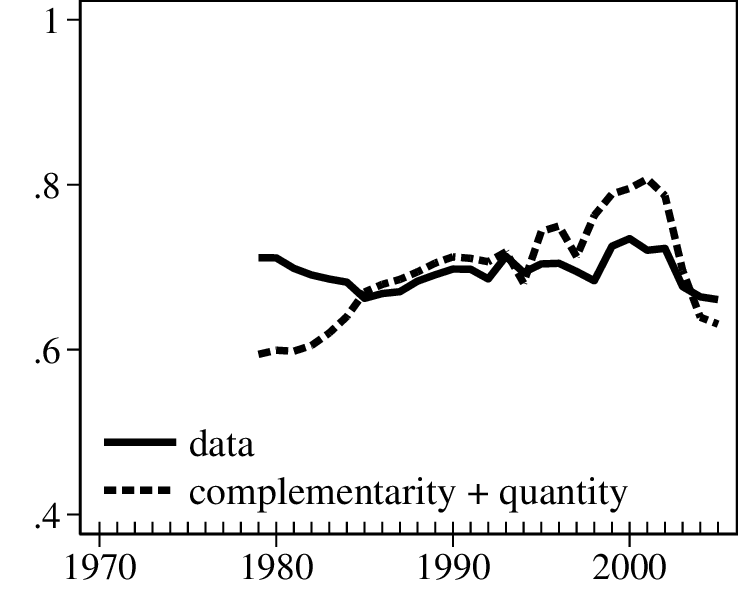}}\subfloat[{\small{}(k) }Portugal]{
\centering{}\includegraphics[scale=0.35]{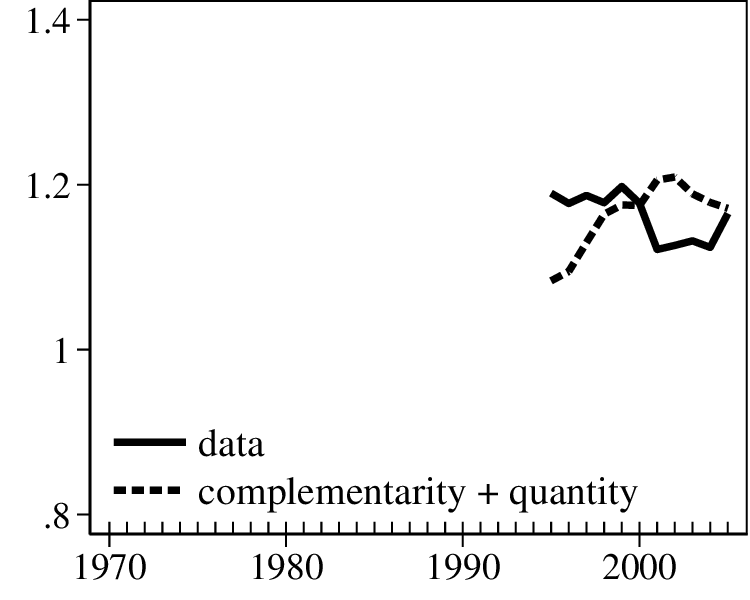}}\subfloat[{\small{}(l) }Slovenia]{
\centering{}\includegraphics[scale=0.35]{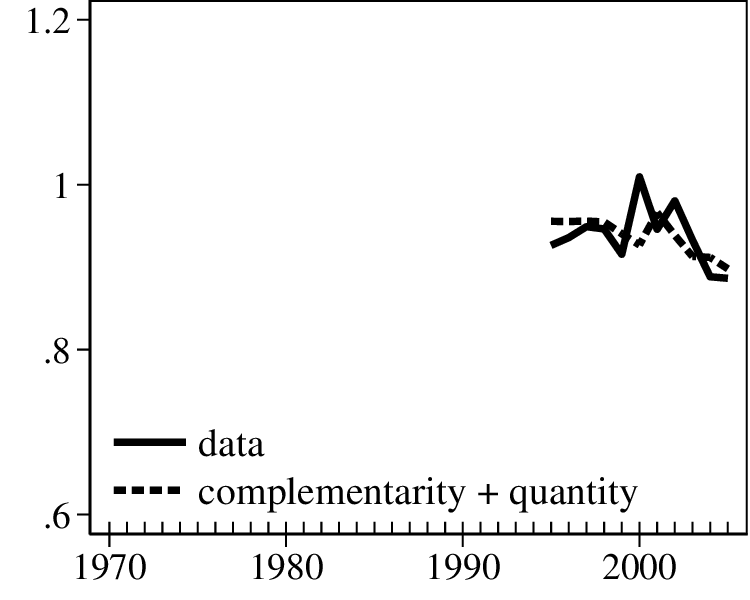}}
\par\end{centering}
\begin{centering}
\subfloat[{\small{}(m) }Sweden]{
\centering{}\includegraphics[scale=0.35]{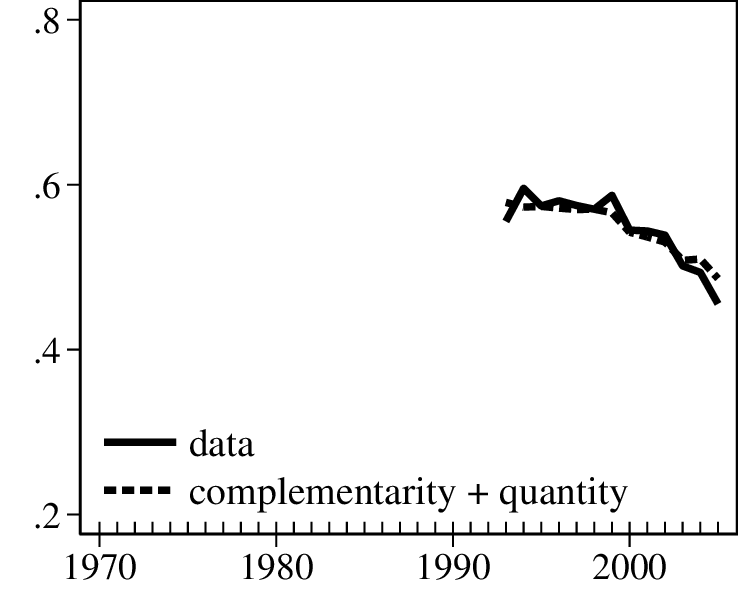}}\subfloat[{\small{}(n) }United Kingdom]{
\centering{}\includegraphics[scale=0.35]{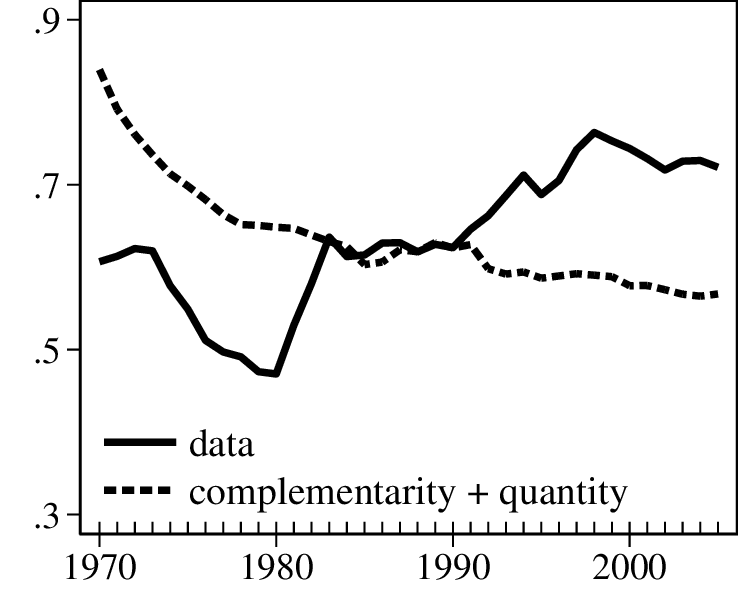}}
\par\end{centering}
\textit{\footnotesize{}Notes}{\footnotesize{}: The bold line indicates
the actual values. The dashed line indicates the sum of the capital\textendash skill
complementarity effect and relative labor quantity effect. The relative
factor-augmenting technology effect is held constant at the sample
mean.}{\footnotesize\par}
\end{figure}

\begin{figure}[H]
\caption{Trends in the skill premium attributable to observed factors the service
sector\label{fig: Wh/Wu_decomp_KiKo4_observed_service}}

\begin{centering}
\subfloat[{\small{}(a) }United States]{
\centering{}\includegraphics[scale=0.35]{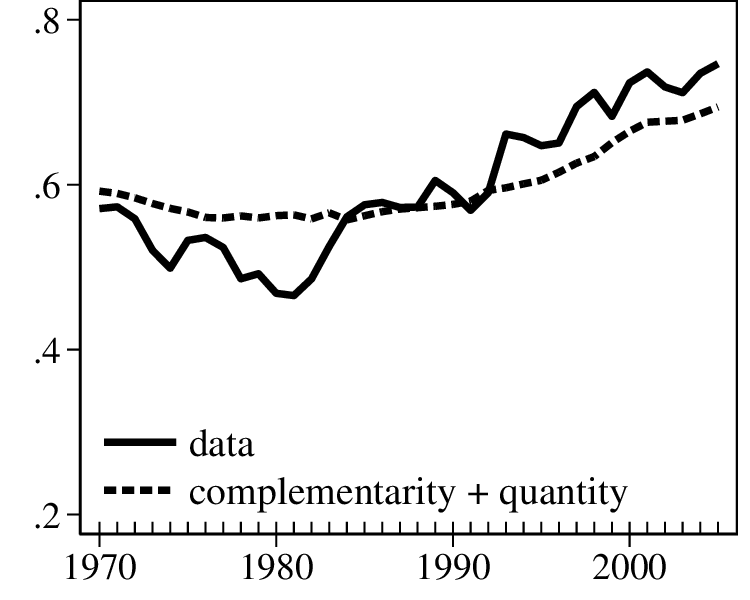}}\subfloat[{\small{}(b) }Australia]{
\centering{}\includegraphics[scale=0.35]{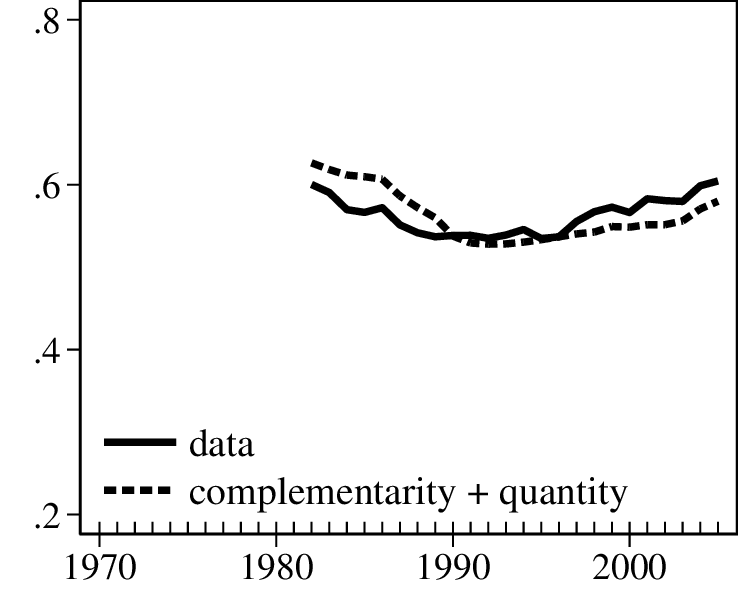}}\subfloat[{\small{}(c) }Austria]{
\centering{}\includegraphics[scale=0.35]{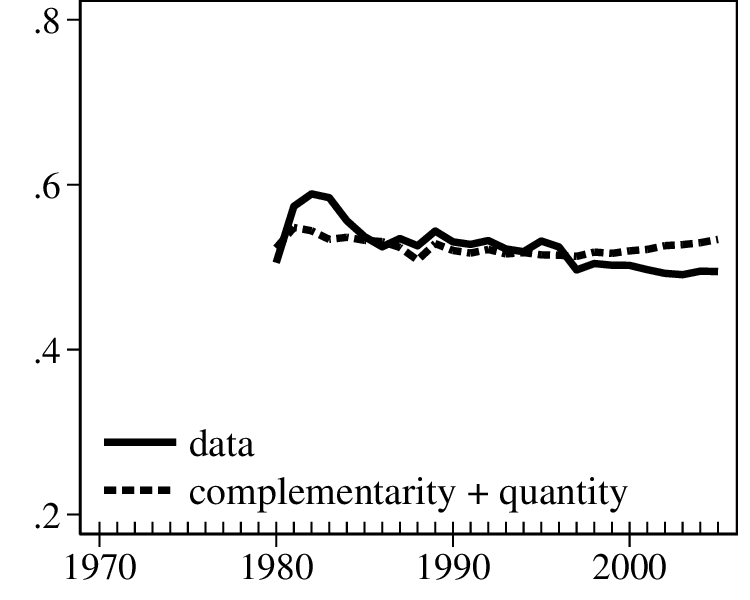}}
\par\end{centering}
\begin{centering}
\subfloat[{\small{}(d) }Czech Republic]{
\centering{}\includegraphics[scale=0.35]{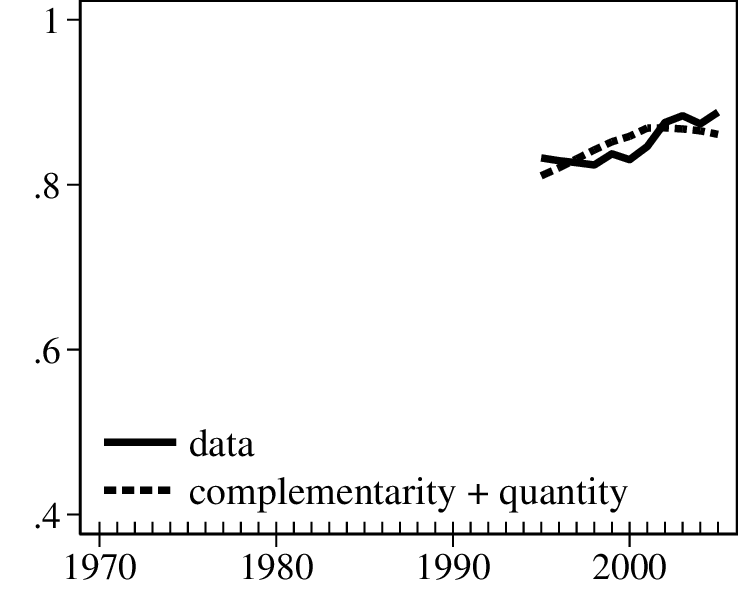}}\subfloat[{\small{}(e) }Denmark]{
\centering{}\includegraphics[scale=0.35]{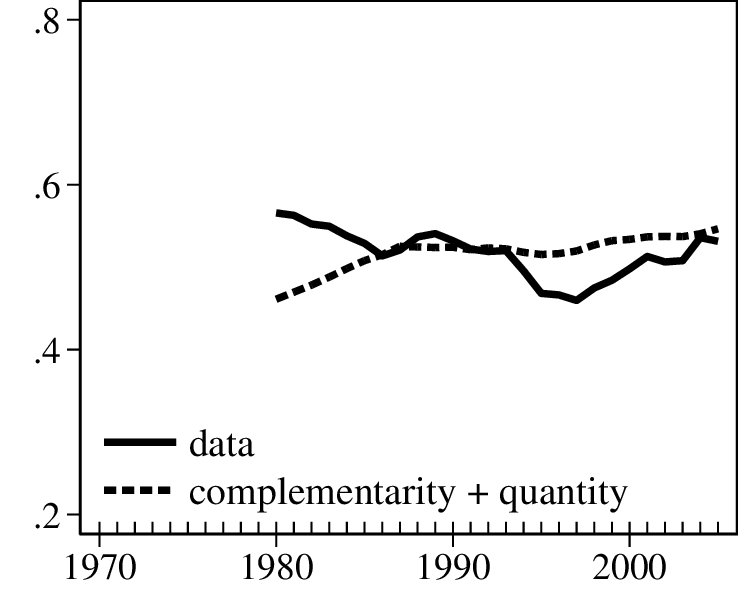}}\subfloat[{\small{}(f) }Finland]{
\centering{}\includegraphics[scale=0.35]{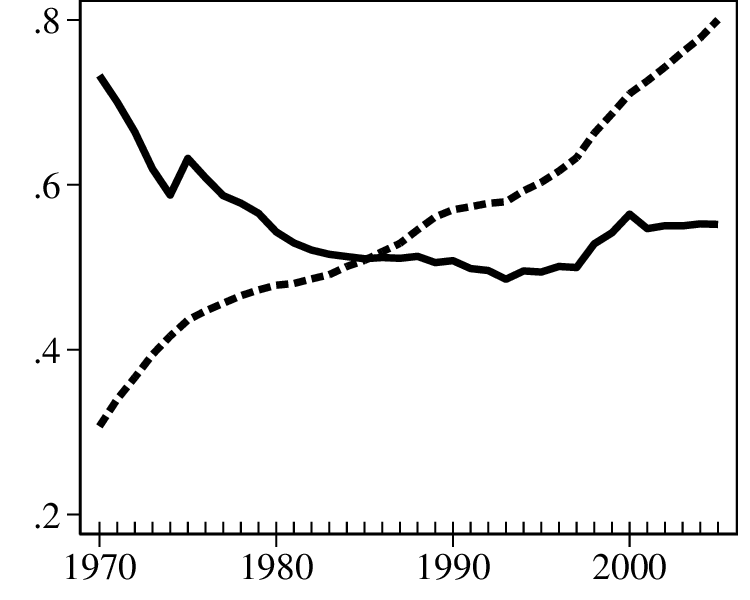}}
\par\end{centering}
\begin{centering}
\subfloat[{\small{}(g) }Germany]{
\centering{}\includegraphics[scale=0.35]{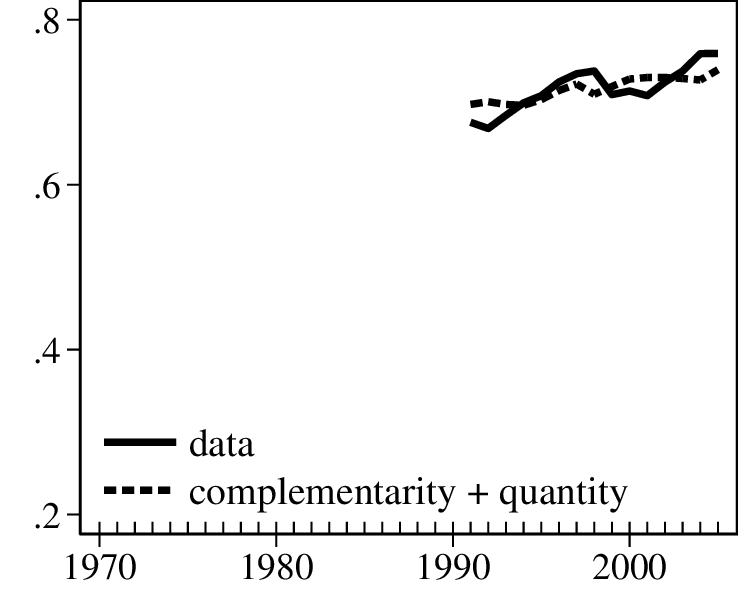}}\subfloat[{\small{}(h) }Italy]{
\centering{}\includegraphics[scale=0.35]{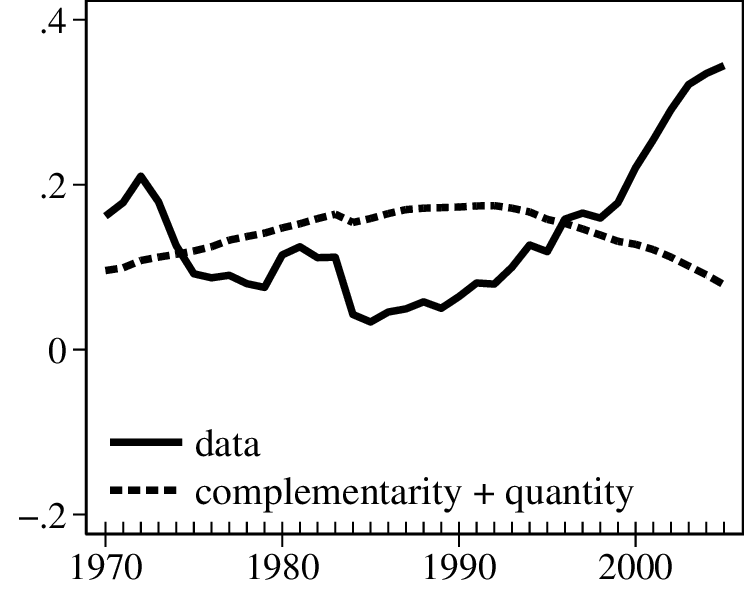}}\subfloat[{\small{}(i) }Japan]{
\centering{}\includegraphics[scale=0.35]{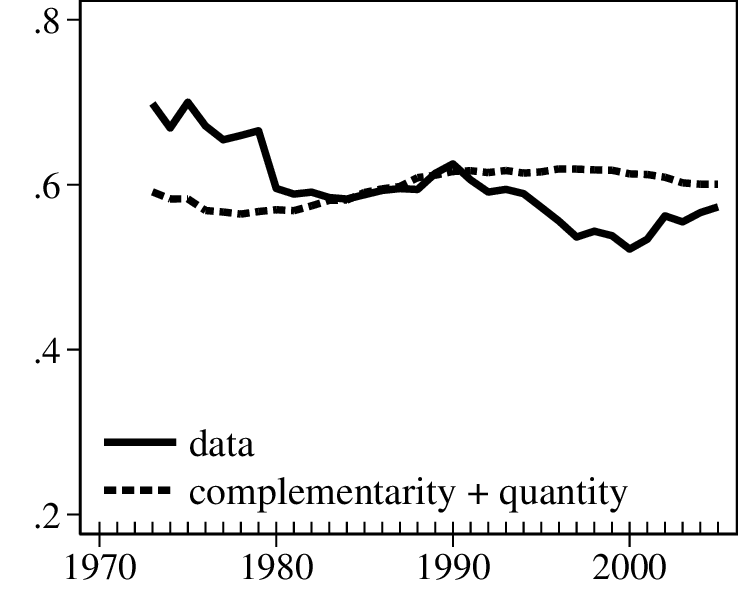}}
\par\end{centering}
\begin{centering}
\subfloat[{\small{}(j) }Netherlands]{
\centering{}\includegraphics[scale=0.35]{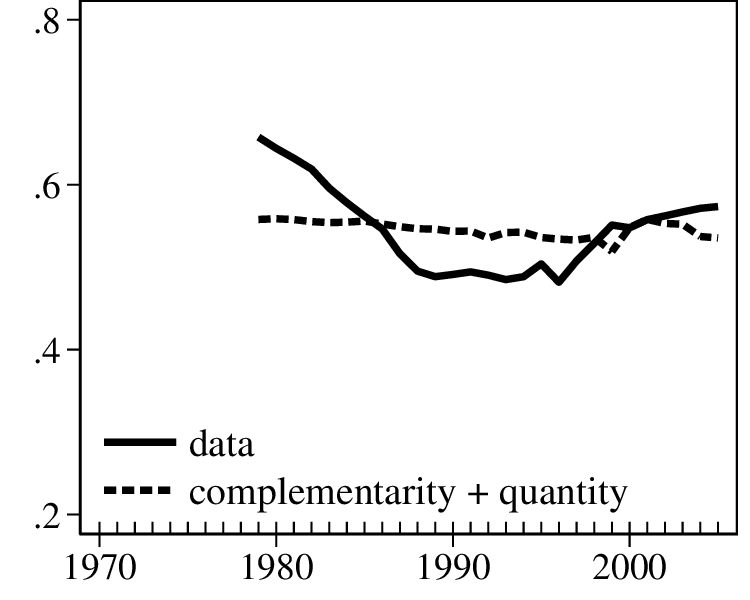}}\subfloat[{\small{}(k) }Portugal]{
\centering{}\includegraphics[scale=0.35]{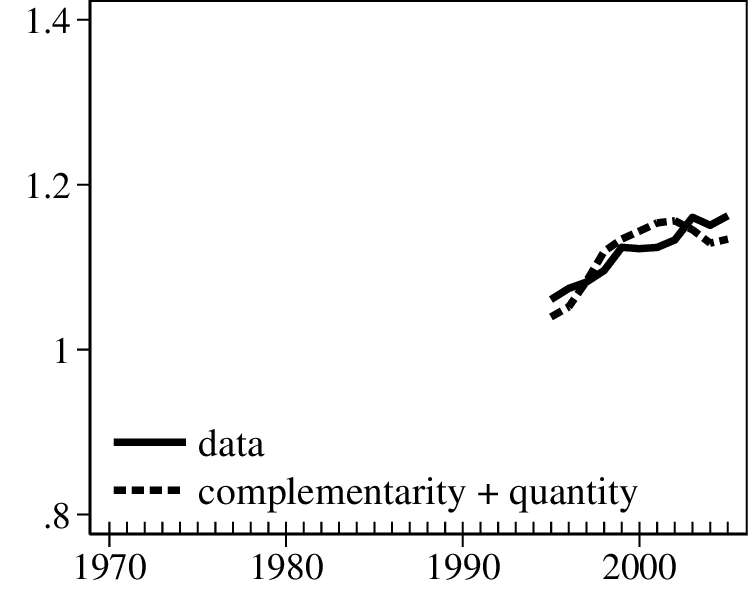}}\subfloat[{\small{}(l) }Slovenia]{
\centering{}\includegraphics[scale=0.35]{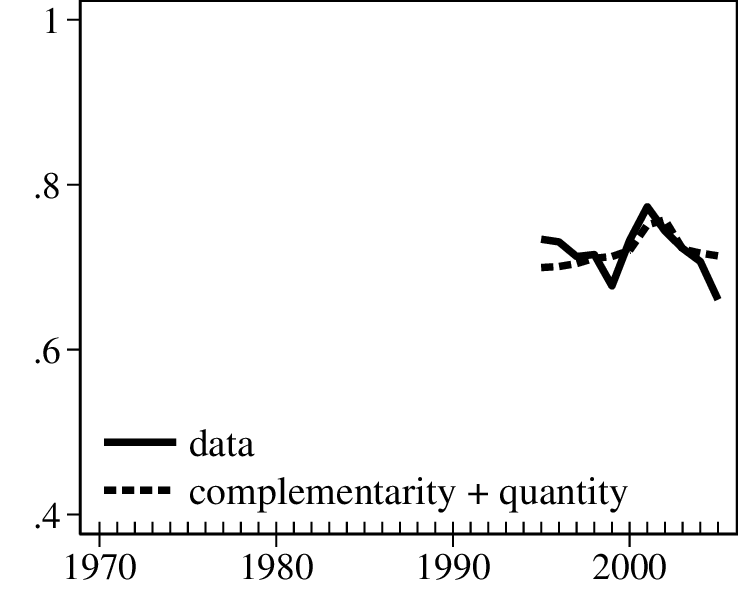}}
\par\end{centering}
\begin{centering}
\subfloat[{\small{}(m) }Sweden]{
\centering{}\includegraphics[scale=0.35]{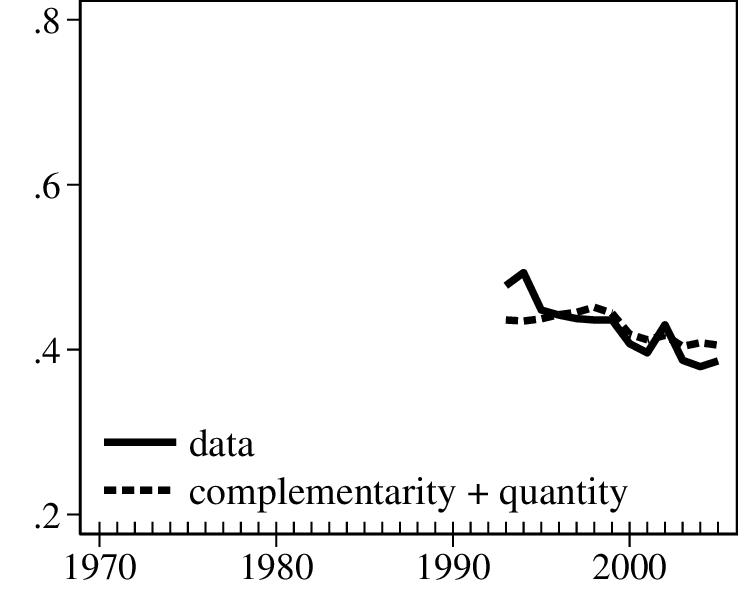}}\subfloat[{\small{}(n) }United Kingdom]{
\centering{}\includegraphics[scale=0.35]{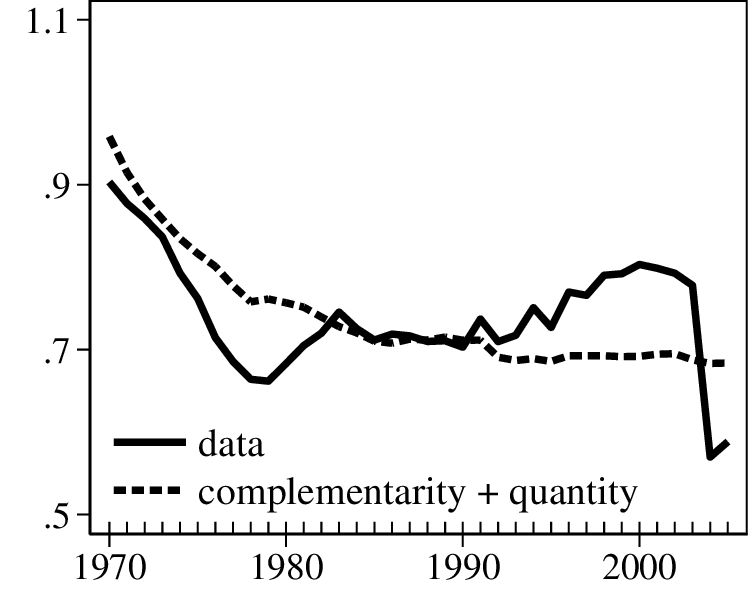}}
\par\end{centering}
\textit{\footnotesize{}Notes}{\footnotesize{}: The bold line indicates
the actual values. The dashed line indicates the sum of the capital\textendash skill
complementarity effect and relative labor quantity effect. The relative
factor-augmenting technology effect is held constant at the sample
mean.}{\footnotesize\par}
\end{figure}

It is common in all countries that the capital\textendash skill complementarity
effect contributes to widening the skill premium, while the relative
labor quantity effect contributes to narrowing it. The magnitude of
the two effects differs across countries and years, however. The capital\textendash skill
complementarity effect was particularly large in Denmark, Finland,
and the United States, while the relative labor quantity effect was
particularly large in the United Kingdom. The capital\textendash skill
complementarity effect was greater for the goods sector than the service
sector in the majority of countries, while the relative labor quantity
effect was similar for the goods and service sectors in most countries.
The relative factor-augmenting technology effect differs in direction,
as well as magnitude, across countries, sectors, and years.

Trends in the skill premium can be attributed to observed factors
in the majority of countries. Figures \ref{fig: Wh/Wu_decomp_KiKo4_observed_goods}
and \ref{fig: Wh/Wu_decomp_KiKo4_observed_service} show the trends
in the skill premium attributable to observed factors, which are calculated
as the sum of the capital\textendash skill complementarity effect
and relative labor quantity effect, along with the actual trends in
the goods and service sectors. The skill premium trends attributable
to observed factors line up with the actual trends in the goods sector
of eight countries (Australia, the Czech Republic, Italy, Japan, the
Netherlands, Slovenia, Sweden, and the United States) and the service
sector of nine countries (Australia, Austria, the Czech Republic,
Germany, Portugal, Slovenia, Sweden, the United Kingdom, and the United
States). Among these countries, the skill premium has an increasing
trend in the goods (service) sector of the United States (the Czech
Republic, Germany, Portugal, and the United States); no clear trend
in the goods (service) sector of Australia, the Czech Republic, the
Netherlands, and Slovenia (Australia, Austria, Slovenia, and the United
Kingdom); and a decreasing trend in the goods (service) sector of
Italy, Japan, and Sweden (Sweden). Trends in the skill premium attributable
to observed factors also align partially with the actual trends in
the service sector of three countries (Denmark, Japan, and the Netherlands).
These results indicate that changes in the skill premium can be interpreted
as a consequence of the capital\textendash skill complementarity effect
and relative labor quantity effect. Meanwhile, the skill premium trends
attributable to observed factors do not align with the actual trends
in the goods sector of six countries (Austria, Denmark, Finland, Germany,
Portugal, and the United Kingdom) and the service sector of two countries
(Finland and Italy). Unobserved factors cannot be disregarded as determinants
of changes in the skill premium in these countries.

\subsubsection{Decomposition of international differences in changes in the skill
premium}

We examine the extent to which international differences in changes
in the skill premium are attributed to specific observed or unobserved
factors. Table \ref{tab: Wh/Wu_decomp_4} presents the decomposition
results of the differences in changes in the skill premium compared
to the United States. The first two columns report the actual and
predicted differences in changes in the skill premium from the first
year of observation to the year 2005 between the United States and
other countries. Almost all the values in the first column are positive,
meaning that the rate of increase in the skill premium was greater
in the United States than in other countries. The next five columns
report the parts attributable to ICT equipment, skilled labor, unskilled
labor, the ratio of skilled to unskilled labor-augmenting technology,
and the ratio of ICT capital- to skilled labor-augmenting technology,
respectively.

\begin{table}[H]
\caption{Decomposition of the differences in changes in the skill premium compared
to the United States\label{tab: Wh/Wu_decomp_4}}

\begin{centering}
{\small{}}\subfloat[{\small{}Goods sector}]{{\small{}}{\small\par}
\centering{}{\small{}}%
\begin{tabular}{llr@{\extracolsep{0pt}.}lr@{\extracolsep{0pt}.}lr@{\extracolsep{0pt}.}lr@{\extracolsep{0pt}.}lr@{\extracolsep{0pt}.}lr@{\extracolsep{0pt}.}lr@{\extracolsep{0pt}.}lr@{\extracolsep{0pt}.}l}
\hline 
Countries &  & \multicolumn{4}{c}{$\ln(w_{h}/w_{u})$} & \multicolumn{2}{c}{} & \multicolumn{2}{c}{$k_{i}$} & \multicolumn{2}{c}{$\ell_{h}$} & \multicolumn{2}{c}{$\ell_{u}$} & \multicolumn{2}{c}{$A_{h}/A_{u}$} & \multicolumn{2}{c}{$A_{i}/A_{h}$}\tabularnewline
\cline{1-1} \cline{3-6} \cline{5-6} \cline{9-18} \cline{11-18} \cline{13-18} \cline{15-18} \cline{17-18} 
 &  & \multicolumn{2}{c}{Data} & \multicolumn{2}{c}{Model} & \multicolumn{2}{c}{} & \multicolumn{2}{c}{} & \multicolumn{2}{c}{} & \multicolumn{2}{c}{} & \multicolumn{2}{c}{} & \multicolumn{2}{c}{}\tabularnewline
\enskip{}AUS &  & 0&225 & 0&182 & \multicolumn{2}{c}{} & 0&303 & 0&050 & \textendash 0&005 & 0&617 & \textendash 0&783\tabularnewline
\enskip{}AUT &  & 0&162 & 0&179 & \multicolumn{2}{c}{} & \textendash 0&501 & \textendash 0&002 & 0&030 & \textendash 0&666 & 1&318\tabularnewline
\enskip{}CZE &  & 0&084 & 0&119 & \multicolumn{2}{c}{} & 0&141 & \textendash 0&006 & 0&013 & 0&348 & \textendash 0&377\tabularnewline
\enskip{}DEN &  & 0&452 & 0&447 & \multicolumn{2}{c}{} & 0&062 & 0&050 & 0&018 & 0&727 & \textendash 0&410\tabularnewline
\enskip{}FIN &  & 0&551 & 0&499 & \multicolumn{2}{c}{} & \textendash 0&318 & \textendash 0&080 & 0&074 & 0&179 & 0&643\tabularnewline
\enskip{}GER &  & \textendash 0&011 & \textendash 0&014 & \multicolumn{2}{c}{} & 0&237 & \textendash 0&080 & 0&041 & 0&297 & \textendash 0&509\tabularnewline
\enskip{}ITA &  & 0&503 & 0&476 & \multicolumn{2}{c}{} & 0&588 & \textendash 0&169 & 0&038 & 1&145 & \textendash 1&125\tabularnewline
\enskip{}JPN &  & 0&406 & 0&357 & \multicolumn{2}{c}{} & 0&624 & \textendash 0&155 & 0&049 & 0&927 & \textendash 1&089\tabularnewline
\enskip{}NED &  & 0&332 & 0&378 & \multicolumn{2}{c}{} & 0&076 & 0&261 & 0&001 & 1&166 & \textendash 1&126\tabularnewline
\enskip{}POR &  & 0&117 & 0&254 & \multicolumn{2}{c}{} & 0&052 & 0&040 & 0&002 & 0&537 & \textendash 0&377\tabularnewline
\enskip{}SLO &  & 0&133 & 0&212 & \multicolumn{2}{c}{} & 0&201 & 0&020 & 0&019 & 0&351 & \textendash 0&379\tabularnewline
\enskip{}SWE &  & 0&195 & 0&256 & \multicolumn{2}{c}{} & 0&191 & 0&097 & 0&004 & 0&408 & \textendash 0&444\tabularnewline
\enskip{}UK &  & 0&147 & 0&204 & \multicolumn{2}{c}{} & 0&566 & 0&049 & 0&073 & 0&824 & \textendash 1&309\tabularnewline
\hline 
\end{tabular}{\small\par}}{\small\par}
\par\end{centering}
\begin{centering}
{\small{}}\subfloat[{\small{}Service sector}]{{\small{}}{\small\par}
\centering{}{\small{}}%
\begin{tabular}{llr@{\extracolsep{0pt}.}lr@{\extracolsep{0pt}.}lr@{\extracolsep{0pt}.}lr@{\extracolsep{0pt}.}lr@{\extracolsep{0pt}.}lr@{\extracolsep{0pt}.}lr@{\extracolsep{0pt}.}lr@{\extracolsep{0pt}.}l}
\hline 
Countries &  & \multicolumn{4}{c}{$\ln(w_{h}/w_{u})$} & \multicolumn{2}{c}{} & \multicolumn{2}{c}{$k_{i}$} & \multicolumn{2}{c}{$\ell_{h}$} & \multicolumn{2}{c}{$\ell_{u}$} & \multicolumn{2}{c}{$A_{h}/A_{u}$} & \multicolumn{2}{c}{$A_{i}/A_{h}$}\tabularnewline
\cline{1-1} \cline{3-6} \cline{5-6} \cline{9-18} \cline{11-18} \cline{13-18} \cline{15-18} \cline{17-18} 
 &  & \multicolumn{2}{c}{Data} & \multicolumn{2}{c}{Model} & \multicolumn{2}{c}{} & \multicolumn{2}{c}{} & \multicolumn{2}{c}{} & \multicolumn{2}{c}{} & \multicolumn{2}{c}{} & \multicolumn{2}{c}{}\tabularnewline
\enskip{}AUS &  & 0&257 & 0&193 & \multicolumn{2}{c}{} & 0&056 & 0&133 & \textendash 0&004 & 0&254 & \textendash 0&246\tabularnewline
\enskip{}AUT &  & 0&289 & 0&280 & \multicolumn{2}{c}{} & 0&124 & \textendash 0&004 & 0&005 & 0&421 & \textendash 0&265\tabularnewline
\enskip{}CZE &  & 0&044 & 0&027 & \multicolumn{2}{c}{} & 0&059 & \textendash 0&023 & 0&007 & 0&095 & \textendash 0&111\tabularnewline
\enskip{}DEN &  & 0&313 & 0&243 & \multicolumn{2}{c}{} & 0&032 & \textendash 0&001 & 0&018 & 0&459 & \textendash 0&265\tabularnewline
\enskip{}FIN &  & 0&356 & 0&297 & \multicolumn{2}{c}{} & \textendash 0&458 & 0&047 & 0&020 & \textendash 0&013 & 0&701\tabularnewline
\enskip{}GER &  & 0&094 & 0&084 & \multicolumn{2}{c}{} & 0&107 & \textendash 0&042 & 0&011 & 0&162 & \textendash 0&154\tabularnewline
\enskip{}ITA &  & \textendash 0&007 & 0&014 & \multicolumn{2}{c}{} & 0&129 & 0&002 & \textendash 0&012 & 0&335 & \textendash 0&440\tabularnewline
\enskip{}JPN &  & 0&352 & 0&355 & \multicolumn{2}{c}{} & 0&082 & \textendash 0&008 & 0&035 & 0&575 & \textendash 0&329\tabularnewline
\enskip{}NED &  & 0&339 & 0&307 & \multicolumn{2}{c}{} & 0&128 & 0&037 & \textendash 0&005 & 0&422 & \textendash 0&276\tabularnewline
\enskip{}POR &  & \textendash 0&002 & \textendash 0&017 & \multicolumn{2}{c}{} & 0&021 & \textendash 0&011 & \textendash 0&011 & 0&095 & \textendash 0&111\tabularnewline
\enskip{}SLO &  & 0&173 & 0&063 & \multicolumn{2}{c}{} & 0&061 & 0&015 & 0&003 & 0&095 & \textendash 0&111\tabularnewline
\enskip{}SWE &  & 0&177 & 0&178 & \multicolumn{2}{c}{} & 0&089 & 0&027 & 0&016 & 0&178 & \textendash 0&133\tabularnewline
\enskip{}UK &  & 0&491 & 0&470 & \multicolumn{2}{c}{} & 0&183 & 0&187 & 0&007 & 0&549 & \textendash 0&456\tabularnewline
\hline 
\end{tabular}{\small\par}}{\small\par}
\par\end{centering}
\textit{\footnotesize{}Notes}{\footnotesize{}: The differences in
changes in the skill premium between the United States and other countries
are decomposed into the parts attributable to ICT capital ($k_{i}$),
skilled labor ($\ell_{h}$), unskilled labor ($\ell_{u}$), the ratio
of skilled to unskilled labor-augmenting technology ($A_{h}/A_{u}$),
and the ratio of ICT capital- to skilled labor-augmenting technology
($A_{i}/A_{h}$).}{\footnotesize\par}
\end{table}

Cross-country differences in changes in the skill premium can be explained
by specific capital and labor quantities in the majority of countries.
The main factors accounting for the differences are ICT equipment
and skilled labor in the goods (service) sector of the Netherlands,
Portugal, and Sweden (Australia and the United Kingdom), and ICT equipment
in the goods (service) sector of Australia, Italy, Japan, and Slovenia
(Austria, the Czech Republic, Germany, Japan, the Netherlands, Slovenia,
and Sweden). The differences in changes in the skill premium compared
to the United States are attributable at least in part to ICT equipment
in almost all countries. This result reflects the fact that the contribution
of ICT equipment to changes in the skill premium was large in the
United States for two reasons. First, the rate of increase in ICT
equipment ($k_{i}$) was large in the United States. Second, the rate
of increase in the ratio of ICT capital- to skilled labor-augmenting
technology ($A_{i}/A_{h}$) was small in the United States. The contribution
of ICT equipment to changes in the skill premium increases with a
decline in $A_{i}/A_{h}$ as the cross-derivative of the skill premium
with respect to $k_{i}$ and $A_{i}/A_{h}$ is negative for $\sigma>\rho$
and $\rho<0$ in equation \eqref{eq: Wh/Wu}.

Not surprisingly, however, international differences in changes in
the skill premium cannot be explained solely by observed factors.
Especially in Finland, where the capital\textendash skill complementarity
effect is greater than or at least comparable to that in the United
States, the differences in changes in the skill premium compared to
the United States are attributed mainly to the relative factor-augmenting
technology effect. The relative factor-augmenting technology effect
can be decomposed into two parts, one attributable to $A_{h}/A_{u}$
and the other attributable to $A_{i}/A_{h}$, the former of which
is typically referred to as skill-biased technological change when
it increases over time. The sixth and seventh columns show that the
two effects work in opposite directions in all countries. The relative
factor-augmenting technology effect is negligible in the United States
because the two effects are canceled out. Interestingly, skill-biased
technological change is greater in the United States than in other
countries.

\subsubsection{Decomposition of changes in relative labor demand}

We measure the contribution of factor prices to changes in the relative
demand for skilled to unskilled labor. Table \ref{tab: Lh/Lu_decomp_4}
presents the decomposition results of changes in relative labor demand
into factor prices and relative factor-augmenting technology. The
first two columns report the actual and predicted changes in relative
labor quantity from the first year of observation till the year 2005
for each country. The changes predicted from the factor demand function
fit reasonably well with the actual data, even though the parameters
are not chosen to account for changes in the relative demand for skilled
to unskilled labor.

The next five columns report the parts attributable to the rental
price of ICT capital, the wages of skilled and unskilled labor, and
two types of relative factor-augmenting technology, respectively.
As seen in Figures \ref{fig: LhLuKiKe_us}, \ref{fig: LhLuKiKe_oecd_goods},
and \ref{fig: LhLuKiKe_oecd_service}, the relative quantity of skilled
to unskilled labor increased in all countries. On the one hand, as
the relative demand for skilled labor should decline with a rise in
the relative wages of skilled labor, a rise in relative labor quantity
is not attributable to relative wages in countries such as the United
States that saw a rise in the skill premium. On the other hand, as
the relative demand for skilled labor should increase with a fall
in the rental price of ICT capital in the presence of capital\textendash skill
complementarity, a rise in relative labor quantity may be attributable
to the rental price. As seen in Figures \ref{fig: Ri/Ro_us}, \ref{fig: Ri/Ro_oecd_goods},
and \ref{fig: Ri/Ro_oecd_service}, the rental price of ICT capital
fell in all countries. An important question here is the extent to
which a rise in relative labor quantity is attributable to a fall
in the rental price of ICT capital. The results indicate that a fall
in the rental price of ICT capital is large enough to account for
a rise in relative labor quantity in the goods and service sectors
of the United States. The same applies to the goods or service sector
of almost all countries.

\begin{table}[H]
\caption{Decomposition of changes in relative labor demand\label{tab: Lh/Lu_decomp_4}}

\begin{centering}
{\small{}}\subfloat[{\small{}Goods sector}]{{\small{}}{\small\par}
\centering{}{\small{}}%
\begin{tabular}{llr@{\extracolsep{0pt}.}lr@{\extracolsep{0pt}.}lr@{\extracolsep{0pt}.}lr@{\extracolsep{0pt}.}lr@{\extracolsep{0pt}.}lr@{\extracolsep{0pt}.}lr@{\extracolsep{0pt}.}lr@{\extracolsep{0pt}.}l}
\hline 
Countries &  & \multicolumn{4}{c}{$\ln(\ell_{h}/\ell_{u})$} & \multicolumn{2}{c}{} & \multicolumn{2}{c}{$r_{i}$} & \multicolumn{2}{c}{$w_{h}$} & \multicolumn{2}{c}{$w_{u}$} & \multicolumn{2}{c}{$A_{h}/A_{u}$} & \multicolumn{2}{c}{$A_{h}/A_{i}$}\tabularnewline
\cline{1-1} \cline{3-6} \cline{5-6} \cline{9-18} \cline{11-18} \cline{13-18} \cline{15-18} \cline{17-18} 
 &  & \multicolumn{2}{c}{Data} & \multicolumn{2}{c}{Model} & \multicolumn{2}{c}{} & \multicolumn{2}{c}{} & \multicolumn{2}{c}{} & \multicolumn{2}{c}{} & \multicolumn{2}{c}{} & \multicolumn{2}{c}{}\tabularnewline
\enskip{}US &  & 1&066 & 1&207 & \multicolumn{2}{c}{} & 2&538 & \textendash 4&600 & 2&795 & 8&371 & \textendash 7&897\tabularnewline
\enskip{}AUS &  & 0&990 & 0&997 & \multicolumn{2}{c}{} & 0&591 & \textendash 1&043 & 0&687 & 0&762 & 0&000\tabularnewline
\enskip{}AUT &  & 0&688 & 1&041 & \multicolumn{2}{c}{} & 4&053 & \textendash 4&906 & 5&122 & 12&668 & \textendash 15&896\tabularnewline
\enskip{}CZE &  & 0&202 & 0&626 & \multicolumn{2}{c}{} & 0&535 & \textendash 2&801 & 2&892 & 0&000 & 0&000\tabularnewline
\enskip{}DEN &  & 1&076 & 0&982 & \multicolumn{2}{c}{} & 2&222 & \textendash 1&444 & 3&150 & 0&215 & \textendash 3&161\tabularnewline
\enskip{}FIN &  & 1&222 & 1&250 & \multicolumn{2}{c}{} & 2&846 & \textendash 6&969 & 9&826 & 7&464 & \textendash 11&916\tabularnewline
\enskip{}GER &  & 0&212 & 0&270 & \multicolumn{2}{c}{} & 0&302 & \textendash 3&051 & 1&724 & 1&295 & 0&000\tabularnewline
\enskip{}ITA &  & 0&966 & 1&361 & \multicolumn{2}{c}{} & 0&576 & \textendash 5&448 & 7&457 & \textendash 1&224 & 0&000\tabularnewline
\enskip{}JPN &  & 1&192 & 1&180 & \multicolumn{2}{c}{} & 0&104 & \textendash 9&961 & 10&901 & \textendash 0&126 & 0&262\tabularnewline
\enskip{}NED &  & 1&508 & 1&418 & \multicolumn{2}{c}{} & 2&313 & \textendash 1&931 & 2&728 & \textendash 3&498 & 1&806\tabularnewline
\enskip{}POR &  & 0&489 & \textendash 0&169 & \multicolumn{2}{c}{} & 0&633 & \textendash 0&408 & 0&627 & \textendash 1&021 & 0&000\tabularnewline
\enskip{}SLO &  & 0&517 & 0&552 & \multicolumn{2}{c}{} & 0&001 & \textendash 1&544 & 2&104 & \textendash 0&020 & 0&011\tabularnewline
\enskip{}SWE &  & 0&789 & 1&214 & \multicolumn{2}{c}{} & 0&001 & \textendash 3&356 & 4&568 & 0&000 & 0&000\tabularnewline
\enskip{}UK &  & 2&269 & 2&632 & \multicolumn{2}{c}{} & 0&366 & \textendash 7&785 & 7&058 & 2&413 & 0&580\tabularnewline
\hline 
\end{tabular}{\small\par}}{\small\par}
\par\end{centering}
\begin{centering}
{\small{}}\subfloat[{\small{}Service sector}]{{\small{}}{\small\par}
\centering{}{\small{}}%
\begin{tabular}{llr@{\extracolsep{0pt}.}lr@{\extracolsep{0pt}.}lr@{\extracolsep{0pt}.}lr@{\extracolsep{0pt}.}lr@{\extracolsep{0pt}.}lr@{\extracolsep{0pt}.}lr@{\extracolsep{0pt}.}lr@{\extracolsep{0pt}.}l}
\hline 
Countries &  & \multicolumn{4}{c}{$\ln(\ell_{h}/\ell_{u})$} & \multicolumn{2}{c}{} & \multicolumn{2}{c}{$r_{i}$} & \multicolumn{2}{c}{$w_{h}$} & \multicolumn{2}{c}{$w_{u}$} & \multicolumn{2}{c}{$A_{h}/A_{u}$} & \multicolumn{2}{c}{$A_{h}/A_{i}$}\tabularnewline
\cline{1-1} \cline{3-6} \cline{5-6} \cline{9-18} \cline{11-18} \cline{13-18} \cline{15-18} \cline{17-18} 
 &  & \multicolumn{2}{c}{Data} & \multicolumn{2}{c}{Model} & \multicolumn{2}{c}{} & \multicolumn{2}{c}{} & \multicolumn{2}{c}{} & \multicolumn{2}{c}{} & \multicolumn{2}{c}{} & \multicolumn{2}{c}{}\tabularnewline
\enskip{}US &  & 0&854 & 0&875 & \multicolumn{2}{c}{} & 1&156 & \textendash 3&228 & 1&800 & 3&656 & \textendash 2&509\tabularnewline
\enskip{}AUS &  & 1&302 & 1&477 & \multicolumn{2}{c}{} & 0&870 & \textendash 1&584 & 1&625 & 0&566 & 0&000\tabularnewline
\enskip{}AUT &  & 0&681 & 0&196 & \multicolumn{2}{c}{} & 0&745 & \textendash 0&530 & 0&653 & \textendash 0&673 & 0&000\tabularnewline
\enskip{}CZE &  & 0&188 & 0&062 & \multicolumn{2}{c}{} & 0&523 & \textendash 1&852 & 1&391 & 0&000 & 0&000\tabularnewline
\enskip{}DEN &  & 0&834 & 0&863 & \multicolumn{2}{c}{} & 1&484 & \textendash 1&815 & 2&210 & \textendash 1&016 & 0&000\tabularnewline
\enskip{}FIN &  & 0&943 & 0&895 & \multicolumn{2}{c}{} & 2&732 & \textendash 0&295 & 1&338 & 4&155 & \textendash 7&036\tabularnewline
\enskip{}GER &  & 0&103 & \textendash 0&108 & \multicolumn{2}{c}{} & 0&490 & \textendash 1&620 & 1&021 & 0&000 & 0&000\tabularnewline
\enskip{}ITA &  & 1&035 & 1&764 & \multicolumn{2}{c}{} & 1&207 & \textendash 0&493 & \textendash 0&677 & 0&651 & 1&076\tabularnewline
\enskip{}JPN &  & 1&192 & 1&038 & \multicolumn{2}{c}{} & 1&145 & \textendash 4&304 & 5&699 & \textendash 1&502 & 0&000\tabularnewline
\enskip{}NED &  & 0&925 & 0&809 & \multicolumn{2}{c}{} & 0&646 & 0&475 & 0&256 & \textendash 0&569 & 0&000\tabularnewline
\enskip{}POR &  & 0&167 & \textendash 0&156 & \multicolumn{2}{c}{} & 0&557 & \textendash 0&110 & \textendash 0&604 & 0&000 & 0&000\tabularnewline
\enskip{}SLO &  & 0&336 & 0&881 & \multicolumn{2}{c}{} & 0&376 & \textendash 0&738 & 1&243 & 0&000 & 0&000\tabularnewline
\enskip{}SWE &  & 0&529 & 0&412 & \multicolumn{2}{c}{} & 0&203 & \textendash 0&765 & 1&357 & \textendash 0&382 & 0&000\tabularnewline
\enskip{}UK &  & 2&197 & 2&752 & \multicolumn{2}{c}{} & 0&399 & \textendash 0&018 & 2&438 & \textendash 0&173 & 0&106\tabularnewline
\hline 
\end{tabular}{\small\par}}{\small\par}
\par\end{centering}
\textit{\footnotesize{}Notes}{\footnotesize{}: Changes in the relative
demand for skilled to unskilled labor are decomposed into the parts
attributable to the rental price of ICT capital ($r_{i}$), the wages
of skilled labor ($w_{h}$), the wages of unskilled labor ($w_{u}$),
the ratio of skilled to unskilled labor-augmenting technology ($A_{h}/A_{u}$),
and the ratio of ICT capital- to skilled labor-augmenting technology
($A_{i}/A_{h}$).}{\footnotesize\par}
\end{table}

\subsubsection{Impact of higher education expansion}

The increased use of ICT results in a widening of the wage gap between
skilled and unskilled workers, while it would result in an increase
in production efficiency. One way to prevent a rise in the skill wage
gap is to raise the supply of skilled labor by expanding access to
tertiary education. When ICT equipment is more complementary to skilled
labor than unskilled labor, such a policy is more effective in reducing
the skill premium as it not only raises the relative labor quantity
effect but also reduces the capital\textendash skill complementarity
effect. We end our analysis by measuring the quantitative importance
of capital\textendash skill complementarity in the evaluation for
the impact of higher education expansion.

\begin{table}[H]
\caption{Contribution of an increase in skilled labor to reducing the skill
premium\label{tab: WhWu_decomp_4_Lh}}

\begin{centering}
{\small{}}\subfloat[{\small{}Goods sector}]{{\small{}}{\small\par}
\centering{}{\small{}}%
\begin{tabular}{lcr@{\extracolsep{0pt}.}lr@{\extracolsep{0pt}.}lr@{\extracolsep{0pt}.}lr@{\extracolsep{0pt}.}lc}
\hline 
Countries &  & \multicolumn{2}{c}{$\ell_{h}$} & \multicolumn{2}{c}{} & \multicolumn{2}{c}{$\ell_{h}^{CSC}$} & \multicolumn{2}{c}{$\ell_{h}^{RLQ}$} & $(\ell_{h}^{CSC}+\ell_{h}^{RLQ})/\ell_{h}^{RLQ}$\tabularnewline
\cline{1-1} \cline{3-4} \cline{7-11} \cline{9-11} \cline{11-11} 
\enskip{}US &  & \textendash 0&274 & \multicolumn{2}{c}{} & \textendash 0&154 & \textendash 0&121 & 2.27\tabularnewline
\enskip{}AUS &  & \textendash 0&170 & \multicolumn{2}{c}{} & \textendash 0&051 & \textendash 0&118 & 1.43\tabularnewline
\enskip{}AUT &  & \textendash 0&127 & \multicolumn{2}{c}{} & \textendash 0&087 & \textendash 0&040 & 3.19\tabularnewline
\enskip{}CZE &  & \textendash 0&007 & \multicolumn{2}{c}{} & \textendash 0&002 & \textendash 0&005 & 1.51\tabularnewline
\enskip{}DEN &  & \textendash 0&179 & \multicolumn{2}{c}{} & \textendash 0&085 & \textendash 0&095 & 1.90\tabularnewline
\enskip{}FIN &  & \textendash 0&195 & \multicolumn{2}{c}{} & \textendash 0&123 & \textendash 0&072 & 2.72\tabularnewline
\enskip{}GER &  & 0&025 & \multicolumn{2}{c}{} & 0&008 & 0&017 & 1.47\tabularnewline
\enskip{}ITA &  & \textendash 0&105 & \multicolumn{2}{c}{} & \textendash 0&031 & \textendash 0&074 & 1.43\tabularnewline
\enskip{}JPN &  & \textendash 0&082 & \multicolumn{2}{c}{} & \textendash 0&009 & \textendash 0&073 & 1.12\tabularnewline
\enskip{}NED &  & \textendash 0&404 & \multicolumn{2}{c}{} & \textendash 0&248 & \textendash 0&156 & 2.59\tabularnewline
\enskip{}POR &  & \textendash 0&053 & \multicolumn{2}{c}{} & \textendash 0&020 & \textendash 0&034 & 1.59\tabularnewline
\enskip{}SLO &  & \textendash 0&034 & \multicolumn{2}{c}{} & 0&000 & \textendash 0&033 & 1.00\tabularnewline
\enskip{}SWE &  & \textendash 0&133 & \multicolumn{2}{c}{} & \textendash 0&041 & \textendash 0&092 & 1.45\tabularnewline
\enskip{}UK &  & \textendash 0&323 & \multicolumn{2}{c}{} & \textendash 0&074 & \textendash 0&249 & 1.30\tabularnewline
\hline 
\end{tabular}{\small\par}}{\small\par}
\par\end{centering}
\begin{centering}
{\small{}}\subfloat[{\small{}Service sector}]{{\small{}}{\small\par}
\centering{}{\small{}}%
\begin{tabular}{lcr@{\extracolsep{0pt}.}lr@{\extracolsep{0pt}.}lr@{\extracolsep{0pt}.}lr@{\extracolsep{0pt}.}lc}
\hline 
Countries &  & \multicolumn{2}{c}{$\ell_{h}$} & \multicolumn{2}{c}{} & \multicolumn{2}{c}{$\ell_{h}^{CSC}$} & \multicolumn{2}{c}{$\ell_{h}^{RLQ}$} & $(\ell_{h}^{CSC}+\ell_{h}^{RLQ})/\ell_{h}^{RLQ}$\tabularnewline
\cline{1-1} \cline{3-4} \cline{7-11} \cline{9-11} \cline{11-11} 
\enskip{}US &  & \textendash 0&275 & \multicolumn{2}{c}{} & \textendash 0&120 & \textendash 0&155 & 1.77\tabularnewline
\enskip{}AUS &  & \textendash 0&284 & \multicolumn{2}{c}{} & \textendash 0&096 & \textendash 0&189 & 1.51\tabularnewline
\enskip{}AUT &  & \textendash 0&166 & \multicolumn{2}{c}{} & \textendash 0&057 & \textendash 0&109 & 1.52\tabularnewline
\enskip{}CZE &  & \textendash 0&033 & \multicolumn{2}{c}{} & \textendash 0&011 & \textendash 0&022 & 1.49\tabularnewline
\enskip{}DEN &  & \textendash 0&169 & \multicolumn{2}{c}{} & \textendash 0&056 & \textendash 0&113 & 1.49\tabularnewline
\enskip{}FIN &  & \textendash 0&322 & \multicolumn{2}{c}{} & \textendash 0&183 & \textendash 0&139 & 2.32\tabularnewline
\enskip{}GER &  & \textendash 0&036 & \multicolumn{2}{c}{} & \textendash 0&011 & \textendash 0&025 & 1.45\tabularnewline
\enskip{}ITA &  & \textendash 0&277 & \multicolumn{2}{c}{} & \textendash 0&098 & \textendash 0&179 & 1.55\tabularnewline
\enskip{}JPN &  & \textendash 0&233 & \multicolumn{2}{c}{} & \textendash 0&089 & \textendash 0&144 & 1.62\tabularnewline
\enskip{}NED &  & \textendash 0&213 & \multicolumn{2}{c}{} & \textendash 0&068 & \textendash 0&145 & 1.47\tabularnewline
\enskip{}POR &  & \textendash 0&045 & \multicolumn{2}{c}{} & \textendash 0&015 & \textendash 0&030 & 1.51\tabularnewline
\enskip{}SLO &  & \textendash 0&071 & \multicolumn{2}{c}{} & \textendash 0&025 & \textendash 0&047 & 1.53\tabularnewline
\enskip{}SWE &  & \textendash 0&100 & \multicolumn{2}{c}{} & \textendash 0&034 & \textendash 0&066 & 1.52\tabularnewline
\enskip{}UK &  & \textendash 0&462 & \multicolumn{2}{c}{} & \textendash 0&114 & \textendash 0&348 & 1.33\tabularnewline
\hline 
\end{tabular}{\small\par}}{\small\par}
\par\end{centering}
\textit{\footnotesize{}Notes}{\footnotesize{}: Changes in the log
of the skill premium attributable to skilled labor ($\ell_{h}$) are
decomposed into the part attributable to the capital\textendash skill
complementarity effect ($\ell_{h}^{CSC}$) and the part attributable
to the relative labor quantity effect ($\ell_{h}^{RLQ}$).}{\footnotesize\par}
\end{table}

Table \ref{tab: WhWu_decomp_4_Lh} decomposes the contribution of
skilled labor to changes in the skill premium into the contribution
through the capital\textendash skill complementarity effect and the
contribution through the relative labor quantity effect. The first
column reports the extent to which a rise in skilled labor had an
effect on changes in the log of the skill premium from the first year
of observation to the year 2005 for each country. The second and third
columns report the parts attributable to the capital\textendash skill
complementarity effect and the relative labor quantity effect, respectively.
The contribution through the capital\textendash skill complementarity
effect is not negligible in all countries and is comparable to or
greater than the contribution through the relative labor quantity
effect in a few countries such as Finland and the United States. The
fourth column reports the ratio of the impact of skilled labor on
the skill premium in the presence and absence of capital\textendash skill
complementarity. The results imply that the impact of higher education
expansion is 1.78 (1.58) times greater for the goods (service) sector
on average when the capital\textendash skill complementarity effect
is taken into account than when it is not.

\section{Conclusion\label{sec: conclusion}}

This study examines the sources and mechanisms of changes in the skill
premium for each country and sector and the differences in such changes
across countries. We estimate a sector-level production function extended
to allow for capital\textendash skill complementarity and factor-biased
technological change using cross-country and cross-industry panel
data. We then decompose changes in the skill premium into the capital\textendash skill
complementarity effect, relative labor quantity effect, and relative
factor-augmenting technology effect in the goods and service sectors
of 14 OECD countries from the 1970s to the 2000s. The capital\textendash skill
complementarity effect and relative labor quantity effect arise from
changes in the composition of capital and labor in production, while
the relative factor-augmenting technology effect arises from other
factors such as unobserved factor-biased technological change. Our
results indicate that most of the changes in the skill premium can
be accounted for by the capital\textendash skill complementarity effect
and relative labor quantity effect in the majority of countries when
ICT equipment is distinguished from non-ICT capital. A large part
of the differences in changes in the skill premium among those countries
can be explained by ICT equipment, skilled labor, and unskilled labor.

Consequently, we show that whether the skill premium would eventually
increase or decrease depends on the outcome of the race between demand
shifts driven by the expansion of ICT and supply shifts driven by
the expansion of access to higher education. On the one hand, the
skill premium can increase due to the capital\textendash skill complementarity
effect, the magnitude of which is proportional to the rate of increase
in the ratio of ICT equipment to skilled labor. On the other hand,
the skill premium can decrease due to the relative labor quantity
effect, the magnitude of which is proportional to the rate of increase
in the ratio of skilled to unskilled labor. The expansion of ICT can
result in a widening of the wage gap between skilled and unskilled
workers, although it can yield significant benefits to the economy
as a whole. Our results imply that the expansion of access to higher
education is effective in preventing a rise in the skill wage gap
not only by raising the relative labor quantity effect but also by
reducing the capital\textendash skill complementarity effect.

\clearpage{}

\bibliographystyle{ecta}
\bibliography{../../references}

\clearpage{}

\appendix

\section{Appendix}

\setcounter{table}{0} \renewcommand{\thetable}{A\arabic{table}}

\setcounter{figure}{0} \renewcommand{\thefigure}{A\arabic{figure}}

\subsection{Derivation of the first-order conditions\label{subsec: FOC}}

We consider the profit maximization problem of a representative firm
in competitive markets. Production technology is given by
\begin{equation}
y_{t}=f\left(\boldsymbol{k}_{t},\boldsymbol{\ell}_{t};A_{t}\right),
\end{equation}
where $\boldsymbol{k}_{t}$ and $\boldsymbol{\ell}_{t}$ are the vectors
of capital and labor inputs in period $t$, that is, $\boldsymbol{k}_{t}=(k_{1t},\ldots,k_{jt},\ldots,k_{J_{k}t})$
and $\boldsymbol{\ell}_{t}=(\ell_{1t},\ldots,\ell_{jt},\ldots,\ell_{J_{\ell}t})$,
respectively.

Let $\chi$ denote investment and $q$ the price of investment. The
Bellman equation of the problem can be written as
\begin{equation}
V\left(\boldsymbol{k}_{t}\right)=\max_{k_{j,t+1},\ell_{jt},\chi_{jt}}\left\{ y_{t}-\sum_{j=1}^{J_{k}}q_{jt}\chi_{jt}-\sum_{j=1}^{J_{\ell}}w_{jt}\ell_{jt}+\beta_{t+1}\E_{t}\left[V\left(\boldsymbol{k}_{t+1}\right)\right]\right\} 
\end{equation}
subject to the law of motion of capital:
\begin{equation}
k_{j,t+1}=\left(1-\delta_{j}\right)k_{jt}+\chi_{jt}.\label{eq: LoM}
\end{equation}
The discount factor is given by $\beta_{t}=1/(1+i_{t})$, where $i$
is the interest rate.

The first-order condition with respect to $k_{j,t+1}$ is
\begin{equation}
q_{jt}=\beta_{t+1}\E_{t}\left[\frac{\partial V_{t+1}}{\partial k_{j,t+1}}\right].
\end{equation}
From the envelope theorem,
\begin{equation}
\frac{\partial V_{t}}{\partial k_{jt}}=\frac{\partial f_{t}}{\partial k_{jt}}+q_{jt}\left(1-\delta_{j}\right).
\end{equation}
Assume that the firm makes its investment decisions with knowledge
of $A_{t+1}$ and $q_{j,t+1}$. The first-order condition can be rewritten
as
\begin{equation}
\frac{\partial f_{t+1}}{\partial k_{j,t+1}}=\delta_{j}q_{j,t+1}+i_{t+1}q_{jt}-\left(q_{j,t+1}-q_{jt}\right)\equiv r_{j,t+1},\label{eq: FOC_k}
\end{equation}
where $r_{jt}$ is referred to as the rental price of capital or the
user cost of capital. The first-order condition with respect to $\ell_{jt}$
is given by
\begin{equation}
\frac{\partial f_{t}}{\partial\ell_{jt}}=w_{jt}.\label{eq: FOC_l}
\end{equation}
Equations for the no-arbitrage condition for capital equipment and
structure and for the wage-bill ratio in \citet{Krusell_Ohanian_RiosRull_Violante_EM00}
can be obtained from equations \eqref{eq: FOC_k} and \eqref{eq: FOC_l},
respectively. In the main text, we allow for the deviation from the
profit-maximizing conditions in competitive markets.

\subsection{Factor demand and substitution\label{subsec: factor}}

The factor demand functions can be derived from the four-factor production
function \eqref{eq: F(Ki,Ko,Lh,Lu;Ai,Ah,Au)} as
\begin{eqnarray}
\ell_{h} & = & y\left[\frac{\left(1-\alpha\right)r_{o}}{\omega_{o}\alpha}\right]^{\alpha}\left(\frac{\omega_{h}}{w_{h}}\right)^{\frac{1}{1-\rho}}A_{h}^{\frac{\rho}{1-\rho}}B^{-\frac{\sigma-\rho}{\left(1-\rho\right)\left(1-\sigma\right)}}C^{\frac{1-\alpha+\alpha\sigma}{1-\sigma}},\label{eq: Lh_4}\\
k_{i} & = & y\left[\frac{\left(1-\alpha\right)r_{o}}{\omega_{o}\alpha}\right]^{\alpha}\left(\frac{\omega_{i}}{r_{i}}\right)^{\frac{1}{1-\rho}}A_{i}^{\frac{\rho}{1-\rho}}B^{-\frac{\sigma-\rho}{\left(1-\rho\right)\left(1-\sigma\right)}}C^{\frac{1-\alpha+\alpha\sigma}{1-\sigma}},\label{eq: Ki_4}\\
\ell_{u} & = & y\left[\frac{\left(1-\alpha\right)r_{o}}{\omega_{o}\alpha}\right]^{\alpha}\left(\frac{\omega_{u}}{w_{u}}\right)^{\frac{1}{1-\sigma}}A_{u}^{\frac{\sigma}{1-\sigma}}C^{\frac{1-\alpha+\alpha\sigma}{1-\sigma}},\label{eq: Lu_4}\\
k_{o} & = & y\left[\frac{\left(1-\alpha\right)r_{o}}{\omega_{o}\alpha}\right]^{\alpha-1}C^{1-\alpha},\label{eq: Ko_4}
\end{eqnarray}
where
\begin{eqnarray}
B & = & \left(\omega_{i}^{\frac{\rho}{1-\rho}}A_{i}^{\frac{\rho}{1-\rho}}r_{i}^{-\frac{\rho}{1-\rho}}+\omega_{h}^{\frac{\rho}{1-\rho}}A_{h}^{\frac{\rho}{1-\rho}}w_{h}^{-\frac{\rho}{1-\rho}}\right)^{-\frac{1-\rho}{\rho}},\\
C & = & \left(B^{-\frac{\sigma}{1-\sigma}}+\omega_{u}^{\frac{\sigma}{1-\sigma}}A_{u}^{\frac{\sigma}{1-\sigma}}w_{u}^{-\frac{\sigma}{1-\sigma}}\right)^{-\frac{1-\sigma}{\sigma}}.
\end{eqnarray}

The relative demand for skilled to unskilled labor can be written
in logs using equations \eqref{eq: Lh_4} and \eqref{eq: Lu_4} as
\begin{eqnarray}
\ln\left(\frac{\ell_{h}}{\ell_{u}}\right) & = & -\frac{1}{1-\rho}\ln w_{h}+\frac{1}{1-\sigma}\ln w_{u}-\frac{\sigma-\rho}{\left(1-\sigma\right)\left(1-\rho\right)}\ln D\nonumber \\
 &  & +\frac{1}{1-\sigma}\left[\sigma\ln\left(\frac{A_{h}}{A_{u}}\right)+\ln\left(\frac{\omega_{h}}{\omega_{u}}\right)\right]-\frac{\sigma-\rho}{\left(1-\sigma\right)\left(1-\rho\right)}\left[\ln\left(\frac{A_{h}}{A_{i}}\right)+\ln\left(\frac{\omega_{h}}{\omega_{i}}\right)\right],
\end{eqnarray}
where
\begin{equation}
D=\left[r_{i}^{-\frac{\rho}{1-\rho}}+\left(\frac{\omega_{h}}{\omega_{i}}\right)^{\frac{\rho}{1-\rho}}\left(\frac{A_{h}}{A_{i}}\right)^{\frac{\rho}{1-\rho}}w_{h}^{-\frac{\rho}{1-\rho}}\right]^{-\frac{1-\rho}{\rho}}.
\end{equation}
The wedge ratios ($\omega_{h}/\omega_{u}$ and $\omega_{h}/\omega_{i}$)
can be calculated from the residuals in equations \eqref{eq: Wh/Wu}
and \eqref{eq: Wh/Ri}, respectively.

When the production function exhibits constant returns to scale, equation
\eqref{eq: morishima} can be rewritten as
\begin{equation}
\epsilon_{ab}=\frac{\partial\ln x_{a}\left(\boldsymbol{p},y,\boldsymbol{A}\right)}{\partial\ln p_{b}}-\frac{\partial\ln x_{b}\left(\boldsymbol{p},y,\boldsymbol{A}\right)}{\partial\ln p_{b}}.
\end{equation}
The Morishima elasticity of substitution can be derived as

\begin{equation}
\epsilon_{k_{i}\ell_{u}}=\epsilon_{\ell_{h}\ell_{u}}=\frac{1}{1-\sigma},
\end{equation}
\begin{equation}
\epsilon_{k_{i}\ell_{h}}=\epsilon_{\ell_{h}k_{i}}=\frac{1}{1-\rho},
\end{equation}
\begin{equation}
\epsilon_{\ell_{u}k_{i}}=\frac{\sigma-\rho}{\left(1-\sigma\right)\left(1-\rho\right)}\left[\frac{r_{i}^{-\frac{\rho}{1-\rho}}}{r_{i}^{-\frac{\rho}{1-\rho}}+w_{h}^{-\frac{\rho}{1-\rho}}\left(\left.\omega_{h}\right/\omega_{i}\right)^{\frac{\rho}{1-\rho}}\left(\left.A_{h}\right/A_{i}\right)^{\frac{\rho}{1-\rho}}}\right]+\frac{1}{1-\rho},
\end{equation}
\begin{equation}
\epsilon_{\ell_{u}\ell_{h}}=\frac{\sigma-\rho}{\left(1-\sigma\right)\left(1-\rho\right)}\left[\frac{w_{h}^{-\frac{\rho}{1-\rho}}\left(\left.\omega_{h}\right/\omega_{i}\right)^{\frac{\rho}{1-\rho}}\left(\left.A_{h}\right/A_{i}\right)^{\frac{\rho}{1-\rho}}}{r_{i}^{-\frac{\rho}{1-\rho}}+w_{h}^{-\frac{\rho}{1-\rho}}\left(\left.\omega_{h}\right/\omega_{i}\right)^{\frac{\rho}{1-\rho}}\left(\left.A_{h}\right/A_{i}\right)^{\frac{\rho}{1-\rho}}}\right]+\frac{1}{1-\rho}.
\end{equation}
The Morishima elasticity of substitution between ICT equipment and
unskilled labor is greater than that between ICT equipment and skilled
labor if and only if $\sigma>\rho$.

\subsection{Adjustment for labor composition and efficiency\label{subsec: adjustment}}

When we construct the data on wages and hours worked for skilled and
unskilled labor, we adjust for changes in labor composition and efficiency
under the assumption that workers are perfect substitutes within skill
groups in a similar way to \citet*{Autor_Katz_Kearney_RESTAT08}.
High-skilled labor constitutes skilled labor, while medium- and low-skilled
labor constitute unskilled labor. Each skill type of labor can be
divided into male and female labor, each of which can be subdivided
into young (aged between 15 and 29 years), middle (aged between 30
and 49 years), and old (aged 50 years and older) labor, in the EU
KLEMS database.\footnote{Unfortunately, labor is not categorized by age in Slovenia or by age
and gender in Portugal and Sweden. However, the main results remain
nearly unchanged irrespective of whether these three countries are
included or excluded.}

First, if no adjustment was needed, the wages for labor of skill type
$j\in\left\{ h,u\right\} $ in country $c$, sector $n$, and year
$t$ could be calculated as $\widetilde{w}_{j,cnt}=\sum_{g_{j}}\theta_{j,cnt}^{g_{j}}\widetilde{w}_{j,cnt}^{g_{j}}$,
where $\theta_{j,cnt}^{g_{j}}$ is the share of total hours worked
by group $g_{h}$ $\in$ \{male, female\} $\times$ \{young, middle,
old\} or $g_{u}$ $\in$ \{medium-skilled, low-skilled\} $\times$
\{male, female\} $\times$ \{young, middle, old\}, that is, $\theta_{j,cnt}^{g_{j}}=\widetilde{\ell}_{j,cnt}^{g_{j}}/\sum_{g_{j}}\widetilde{\ell}_{j,cnt}^{g_{j}}$.
To adjust for compositional changes, the weight is fixed at its country-specific
mean: $\overline{\theta}_{j,cn}^{g_{j}}=\sum_{t=1}^{T_{c}}\theta_{j,cnt}^{g_{j}}/T_{c}$,
where $T_{c}$ is the number of years observed for country $c$. The
composition-adjusted wages for labor of skill type $j$ in country
$c$, sector $n$, and year $t$ are calculated as $w_{j,cnt}=\sum_{g_{j}}\overline{\theta}_{j,cn}^{g_{j}}\widetilde{w}_{j,cnt}^{g_{j}}$.

Similarly, if no adjustment was needed, the hours worked by labor
of skill type $j$ in country $c$, sector $n$, and year $t$ could
be calculated as $\widetilde{\ell}_{j,cnt}=\sum_{g_{j}}\widetilde{\ell}_{j,cnt}^{g_{j}}$.
To adjust for compositional changes, the hours worked are weighted
according to time-invariant labor efficiency. The composition-adjusted
hours worked by labor of skill type $j$ in country $c$, sector $n$,
and year $t$ are calculated as $\ell_{j,cnt}=\sum_{g_{j}}(\overline{w}_{j,cn}^{g_{j}}/\overline{w}_{j,cn}^{g_{j^{\prime}}})\widetilde{\ell}_{j,cnt}^{g_{j}}$,
where the efficiency unit is measured by the country-specific mean
of wages: $\overline{w}_{j,cn}^{g_{j}}=\sum_{t=1}^{T_{c}}\widetilde{w}_{j,cnt}^{g_{j}}/T_{c}$.
The weight is normalized by the efficiency unit for the base group
$g_{j^{\prime}}$. We choose male middle-aged skilled labor as the
base group for skilled labor and male middle-aged medium-skilled labor
as the base group for unskilled labor. However, our results do not
depend on the choice of the base group.

\subsection{The rental price of capital\label{subsec: rental_price}}

Capital is classified into eight categories: (i) computing equipment,
(ii) communications equipment, (iii) software, (iv) transport equipment,
(v) other machinery and equipment, (vi) non-residential structures
and infrastructures, (vii) residential structures, and (viii) other
assets. We classify categories (i), (ii), and (iii) as ICT equipment,
(iv) and (v) as non-ICT equipment, and (vi) as capital structure.
Non-ICT capital consists of non-ICT equipment and capital structure.

As shown in equation \eqref{eq: FOC_k}, the rental price of capital
($r_{j}$) is determined by the price of investment ($q_{j}$), depreciation
rate ($\delta_{j}$), and interest rate ($\iota$) for each $j\in\{i,o\}$.
The price of investment is calculated by dividing the nominal value
by the real value of investment. The depreciation rate is calculated
as the time average of those obtained from the law of motion of capital
in a similar way to that in \citet*{Greenwood_Hercowitz_Krusell_AER97}.
Let $cpi$ denote the consumer price index, information about which
can be obtained from OECD.Stat. Following \citet{Niebel_Saam_RIW16},
the rental price of capital is calculated as
\begin{equation}
r_{j,cnt}=\delta_{j,cn}q_{j,cnt}+\iota_{ct}q_{j,cn,t-1}-\frac{1}{2}\left(\ln q_{j,cnt}-\ln q_{j,cn,t-2}\right)q_{j,cn,t-1},
\end{equation}
where the interest rate is calculated as
\begin{equation}
\iota_{ct}=0.04+\frac{1}{5}\sum_{\tau=-2}^{2}\frac{cpi_{c,t-\tau}-cpi_{c,t-\tau-1}}{cpi_{c,t-\tau-1}}.
\end{equation}
The advantage of this approach is that it does not require the assumption
of competitive markets.

\subsection{Trends in factor prices and quantities in OECD countries\label{subsec: trends_oecd}}

\begin{figure}[h]
\caption{Rental prices of ICT and non-ICT capital in the goods sector of OECD
countries\label{fig: Ri/Ro_oecd_goods}}

\begin{centering}
\subfloat[Australia]{
\centering{}\includegraphics[scale=0.35]{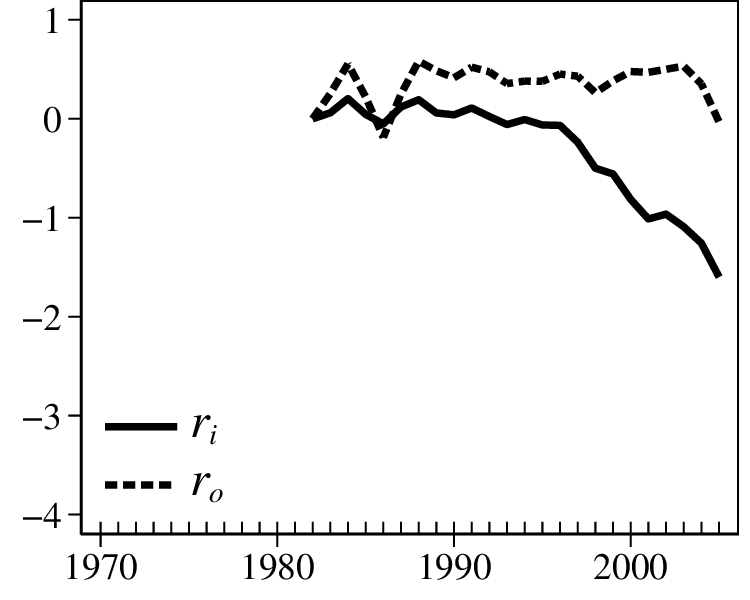}}\subfloat[Austria]{
\centering{}\includegraphics[scale=0.35]{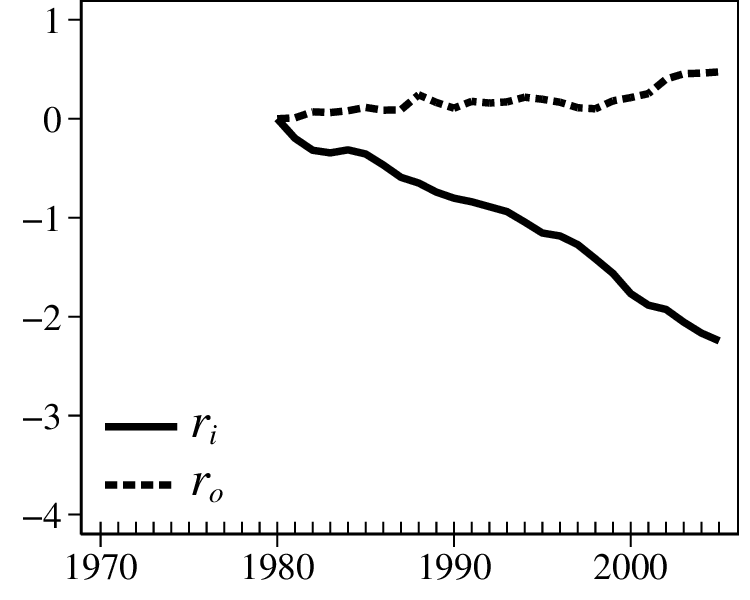}}
\par\end{centering}
\begin{centering}
\subfloat[Czech Republic]{
\centering{}\includegraphics[scale=0.35]{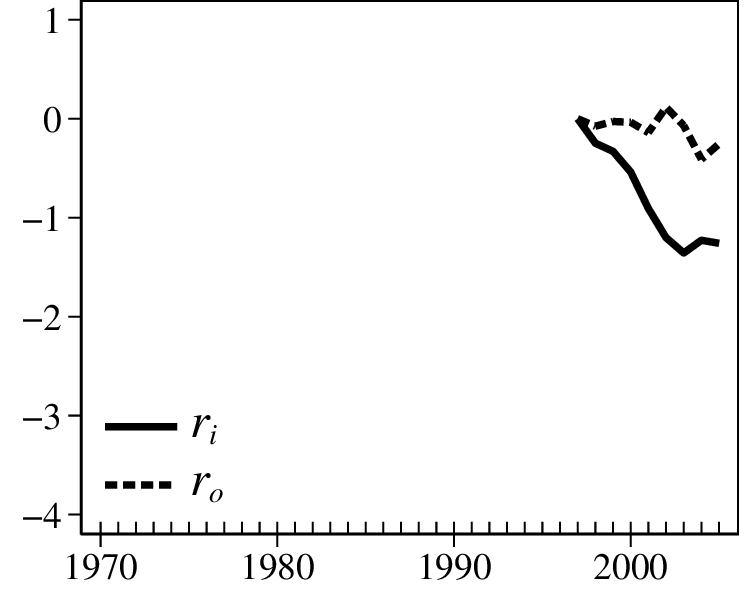}}\subfloat[Denmark]{
\centering{}\includegraphics[scale=0.35]{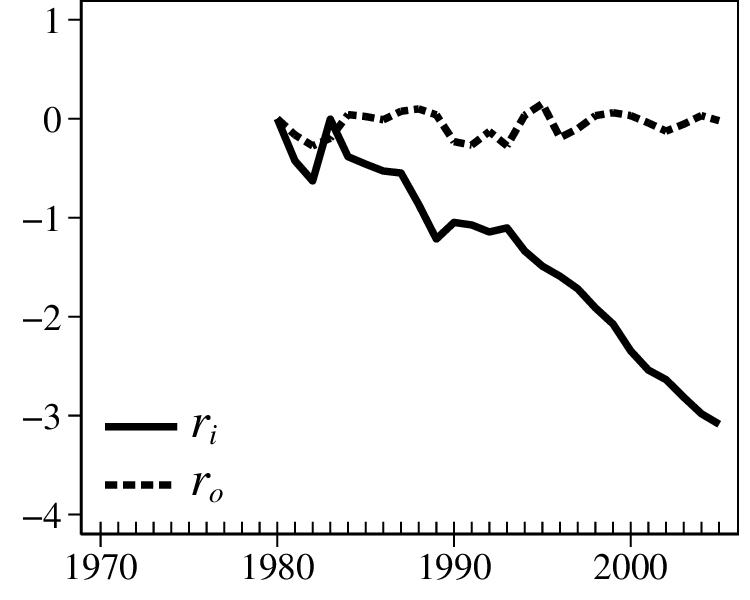}}\subfloat[Finland]{
\centering{}\includegraphics[scale=0.35]{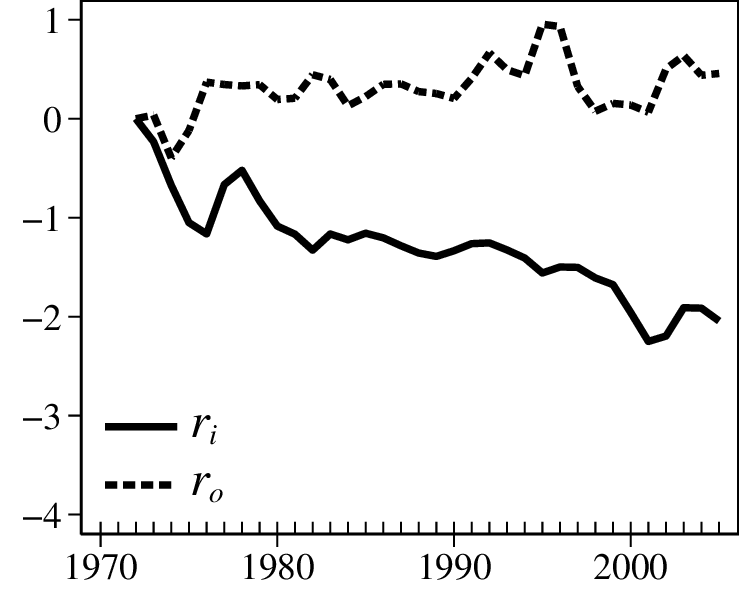}}
\par\end{centering}
\begin{centering}
\subfloat[Germany]{
\centering{}\includegraphics[scale=0.35]{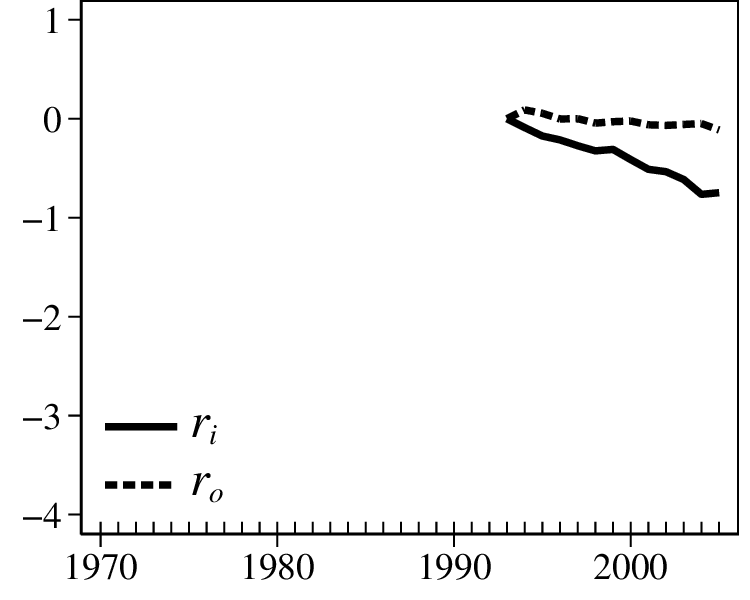}}\subfloat[Italy]{
\centering{}\includegraphics[scale=0.35]{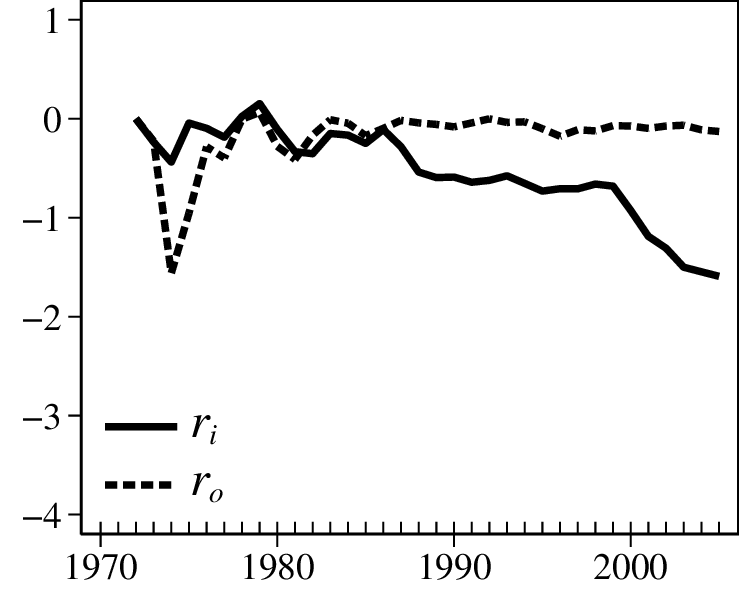}}\subfloat[Japan]{
\centering{}\includegraphics[scale=0.35]{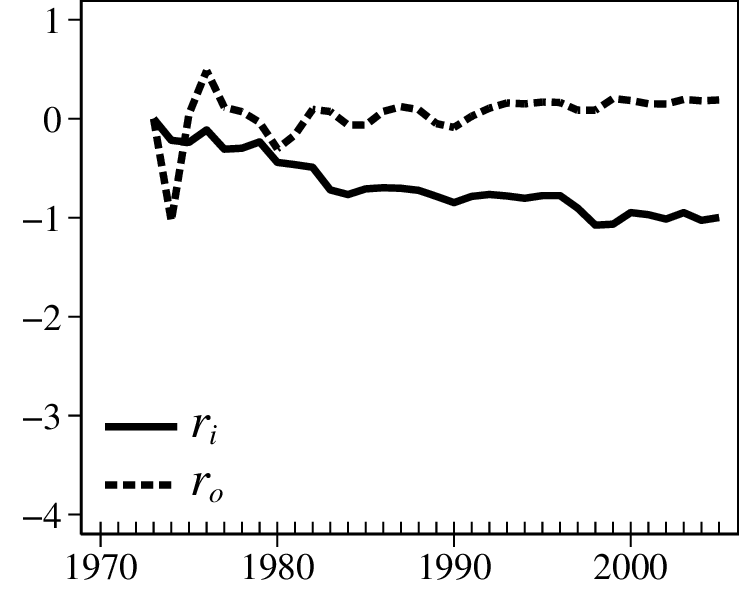}}
\par\end{centering}
\begin{centering}
\subfloat[Netherlands]{
\centering{}\includegraphics[scale=0.35]{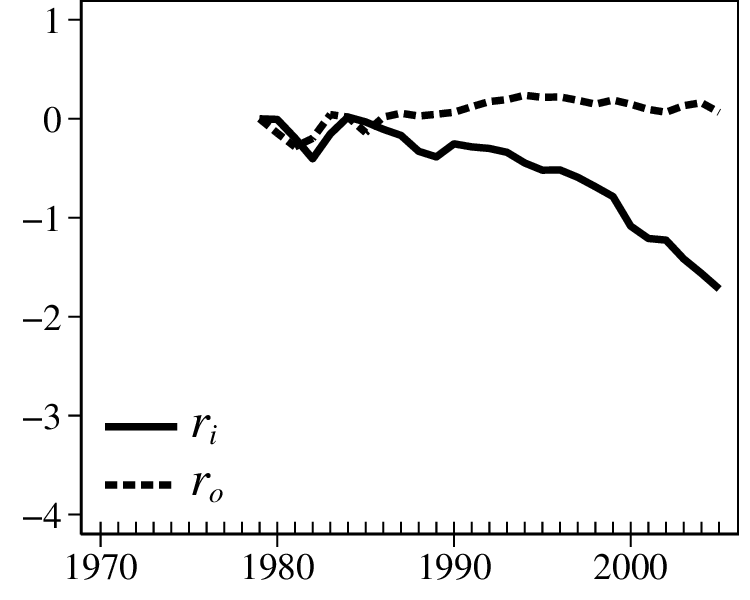}}\subfloat[Portugal]{
\centering{}\includegraphics[scale=0.35]{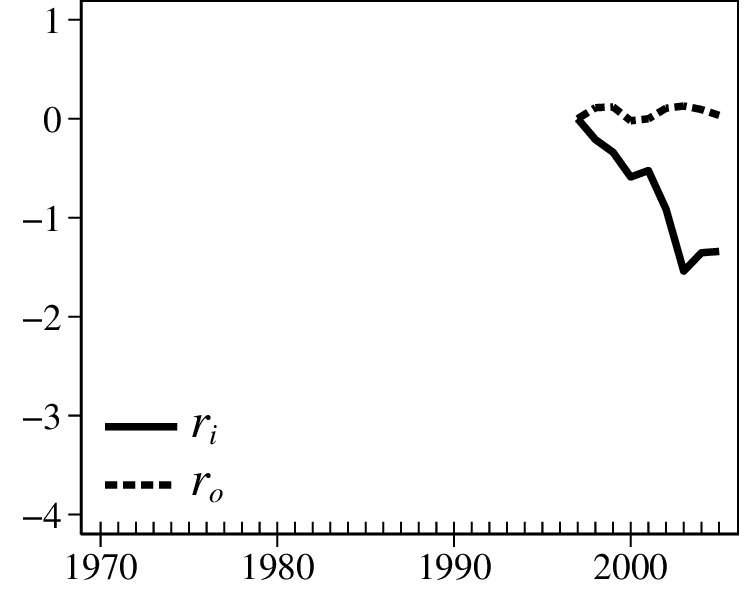}}\subfloat[Slovenia]{
\centering{}\includegraphics[scale=0.35]{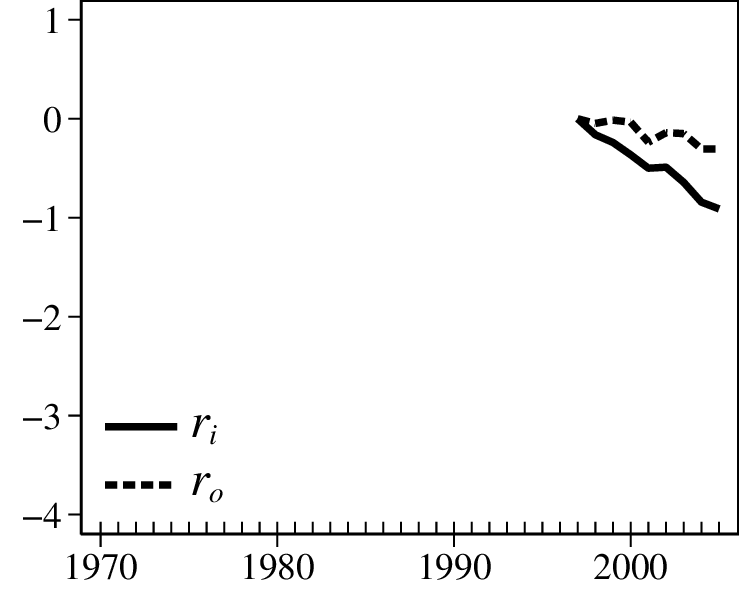}}
\par\end{centering}
\begin{centering}
\subfloat[Sweden]{
\centering{}\includegraphics[scale=0.35]{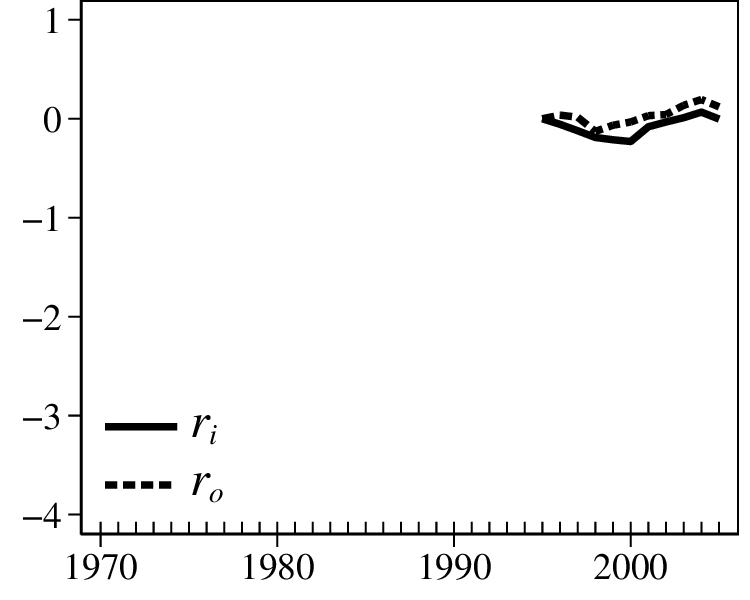}}\subfloat[United Kingdom]{
\centering{}\includegraphics[scale=0.35]{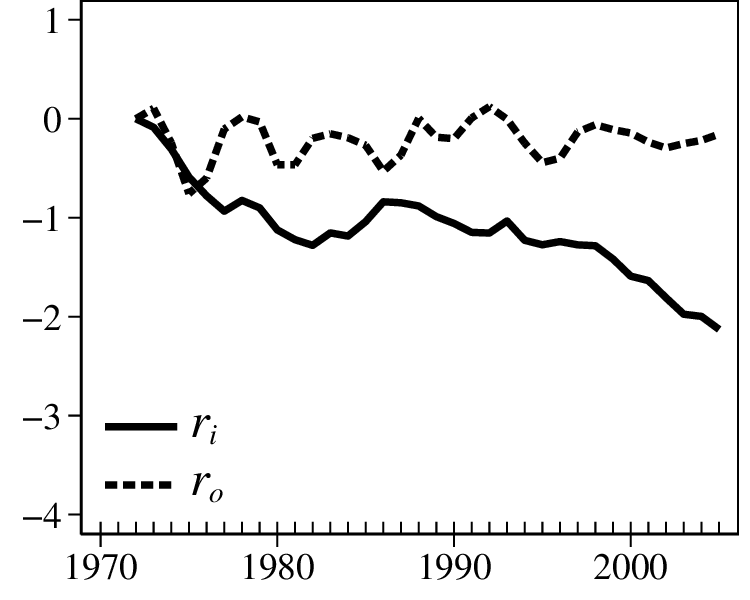}}
\par\end{centering}
\textit{\footnotesize{}Notes}{\footnotesize{}: The bold and dashed
lines indicate the rental price of ICT capital ($r_{i}$) and the
rental price of non-ICT capital ($r_{o}$), respectively. All series
are logged and normalized to zero in the initial year.}{\footnotesize\par}
\end{figure}

\begin{figure}[h]
\caption{Rental prices of ICT and non-ICT capital in the service sector of
OECD countries\label{fig: Ri/Ro_oecd_service}}

\begin{centering}
\subfloat[Australia]{
\centering{}\includegraphics[scale=0.35]{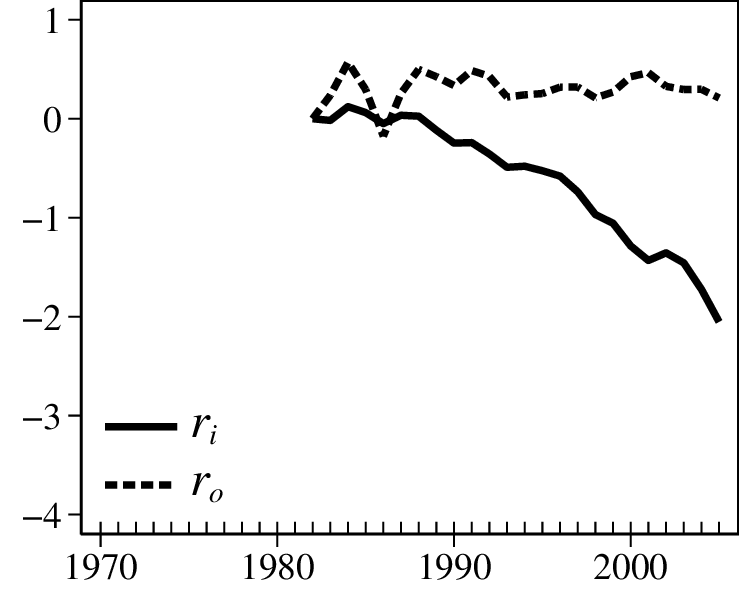}}\subfloat[Austria]{
\centering{}\includegraphics[scale=0.35]{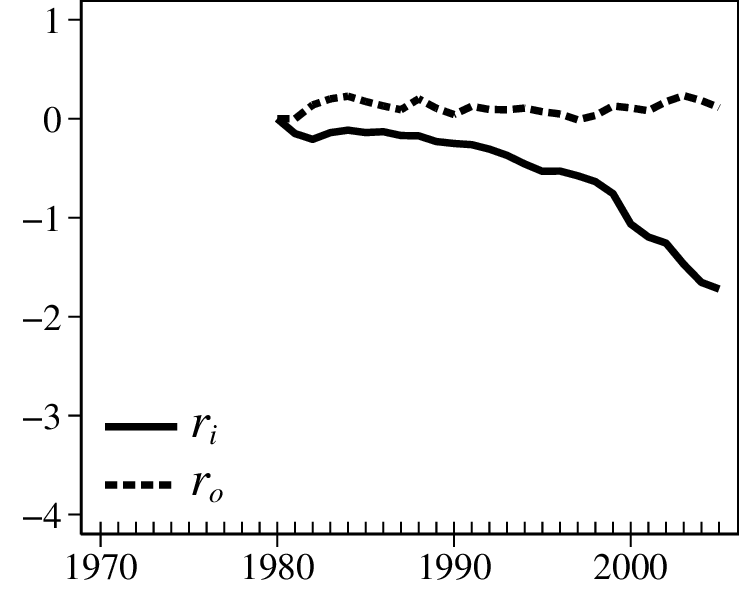}}
\par\end{centering}
\begin{centering}
\subfloat[Czech Republic]{
\centering{}\includegraphics[scale=0.35]{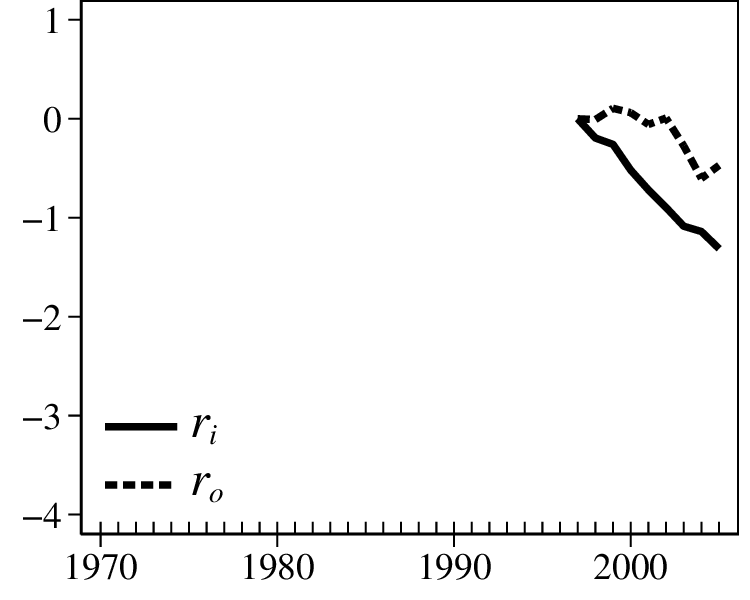}}\subfloat[Denmark]{
\centering{}\includegraphics[scale=0.35]{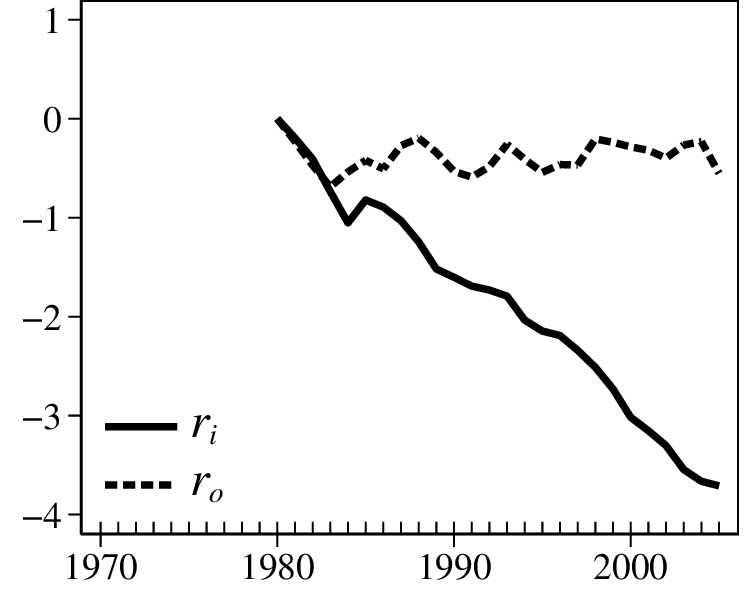}}\subfloat[Finland]{
\centering{}\includegraphics[scale=0.35]{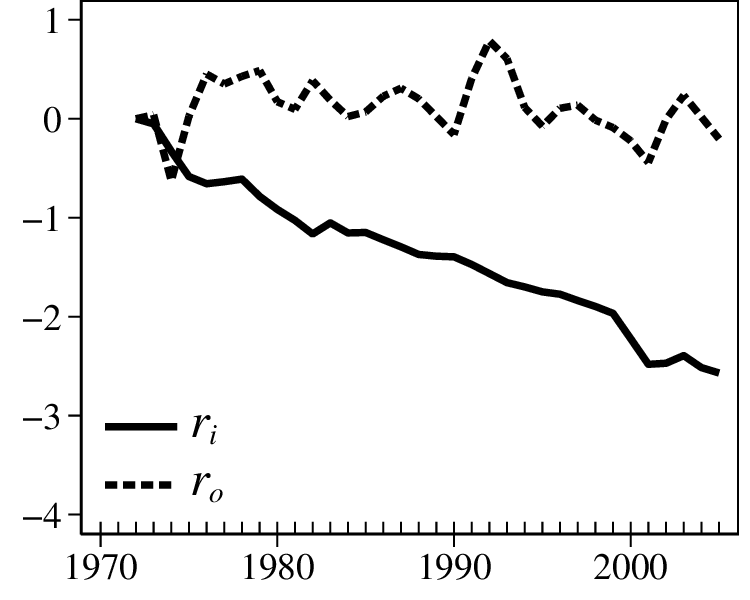}}
\par\end{centering}
\begin{centering}
\subfloat[Germany]{
\centering{}\includegraphics[scale=0.35]{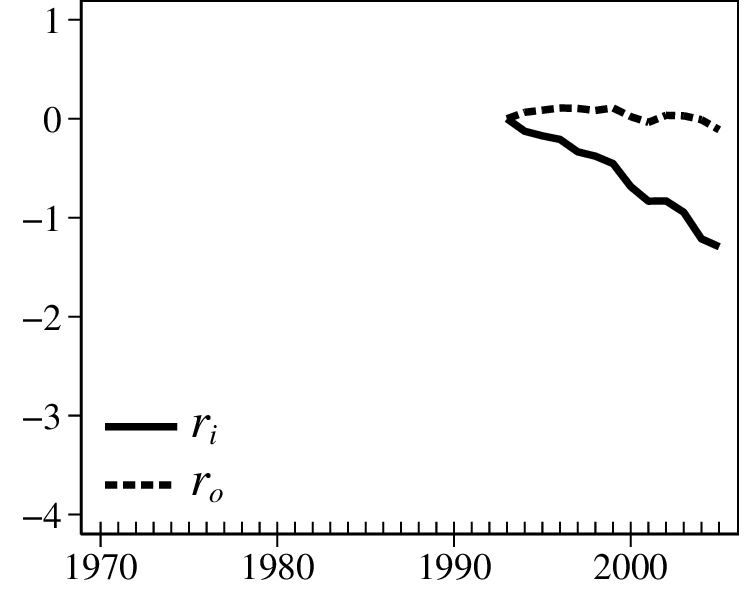}}\subfloat[Italy]{
\centering{}\includegraphics[scale=0.35]{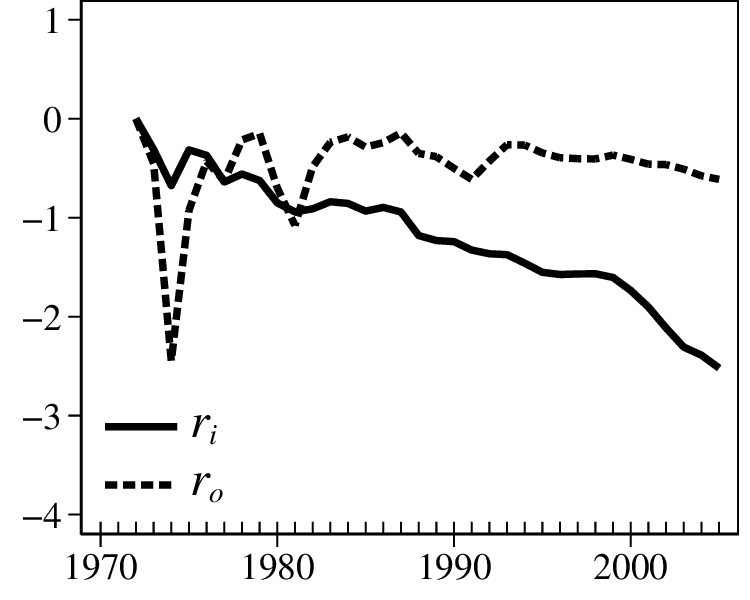}}\subfloat[Japan]{
\centering{}\includegraphics[scale=0.35]{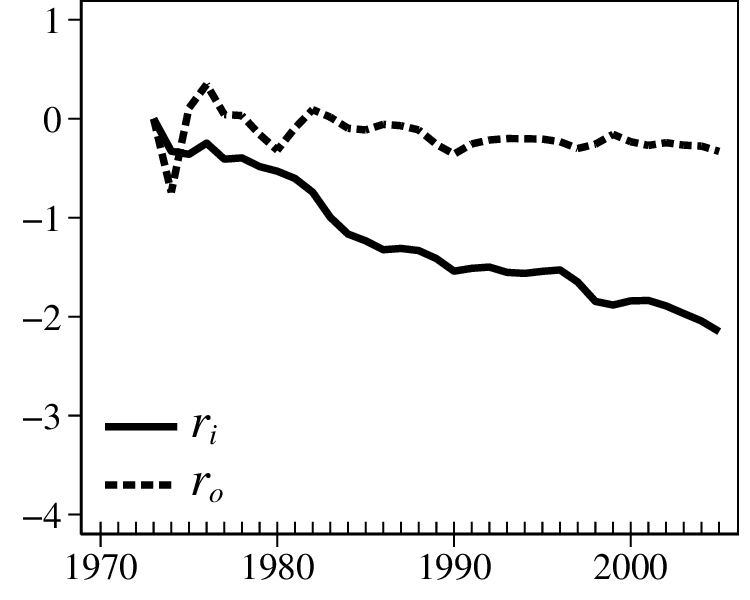}}
\par\end{centering}
\begin{centering}
\subfloat[Netherlands]{
\centering{}\includegraphics[scale=0.35]{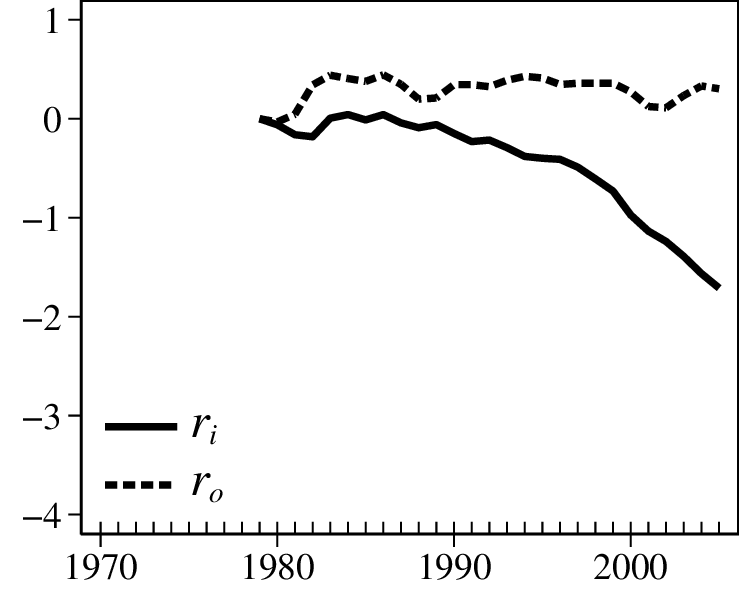}}\subfloat[Portugal]{
\centering{}\includegraphics[scale=0.35]{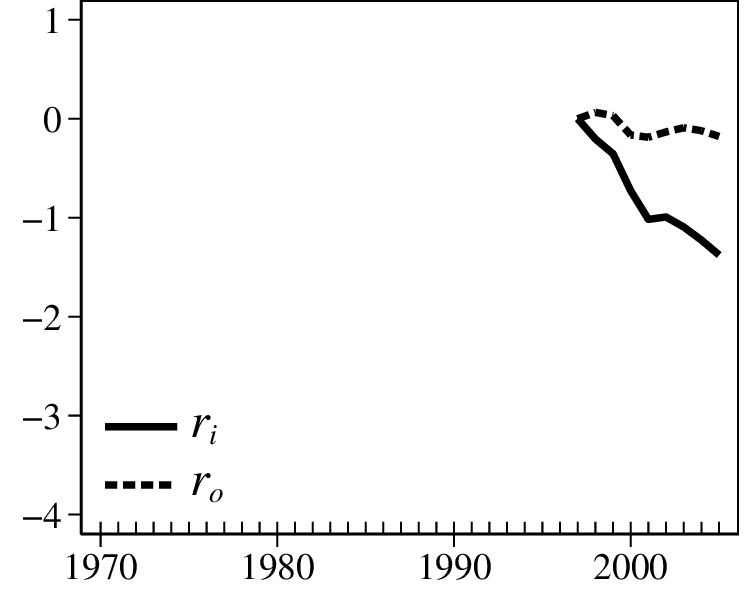}}\subfloat[Slovenia]{
\centering{}\includegraphics[scale=0.35]{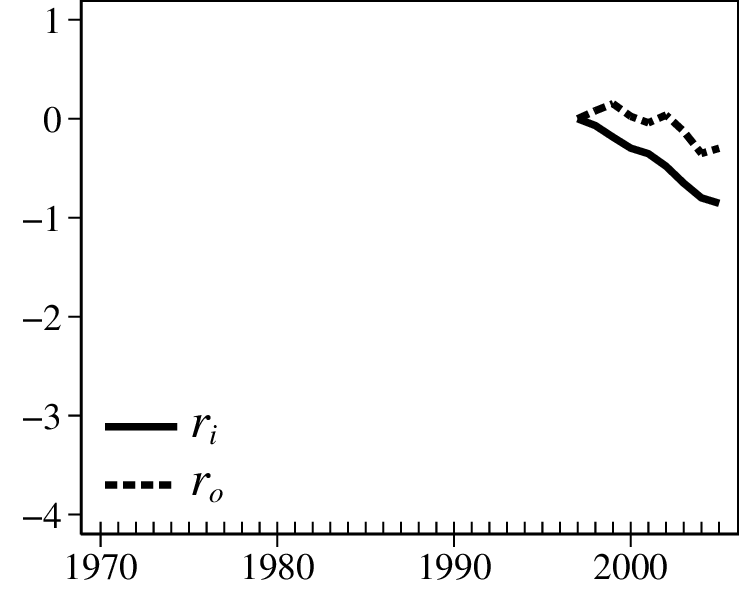}}
\par\end{centering}
\begin{centering}
\subfloat[Sweden]{
\centering{}\includegraphics[scale=0.35]{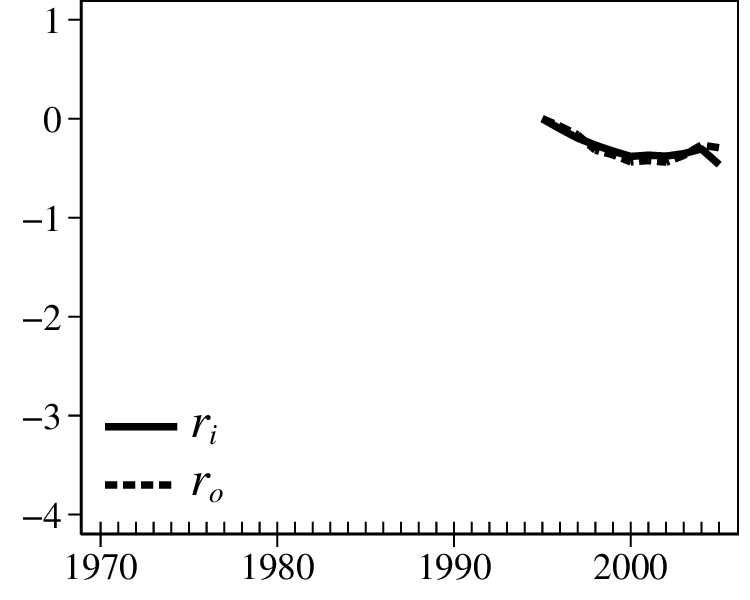}}\subfloat[United Kingdom]{
\centering{}\includegraphics[scale=0.35]{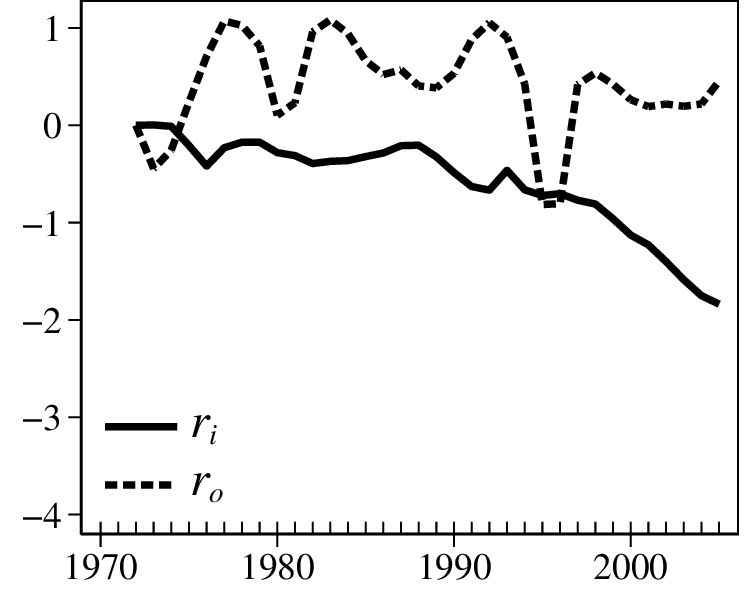}}
\par\end{centering}
\textit{\footnotesize{}Notes}{\footnotesize{}: The bold and dashed
lines indicate the rental price of ICT capital ($r_{i}$) and the
rental price of non-ICT capital ($r_{o}$), respectively. All series
are logged and normalized to zero in the initial year.}{\footnotesize\par}
\end{figure}

\begin{figure}[h]
\caption{Relative factor quantities in the goods sector of OECD countries\label{fig: LhLuKiKe_oecd_goods}}

\begin{centering}
\subfloat[Australia]{
\centering{}\includegraphics[scale=0.35]{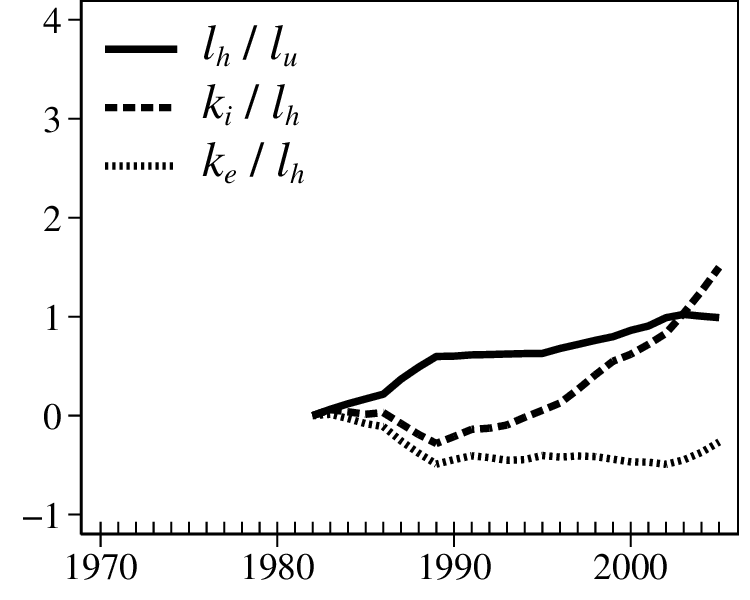}}\subfloat[Austria]{
\centering{}\includegraphics[scale=0.35]{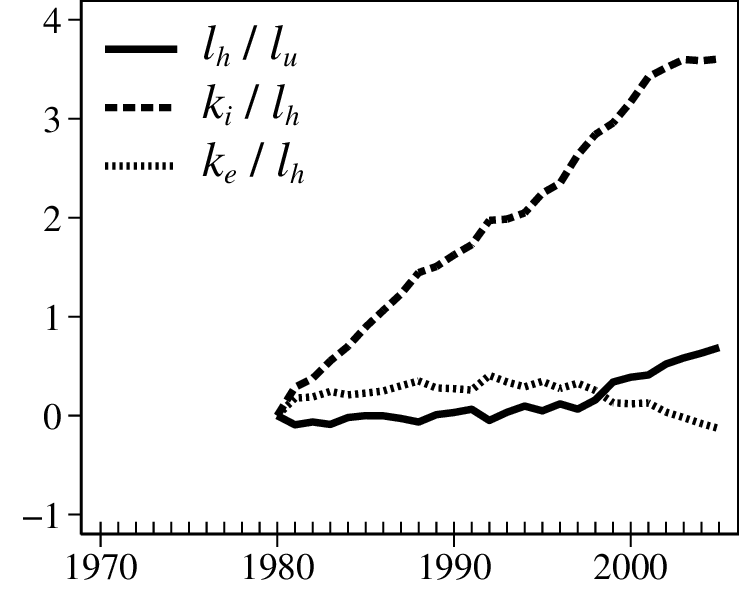}}
\par\end{centering}
\begin{centering}
\subfloat[Czech Republic]{
\centering{}\includegraphics[scale=0.35]{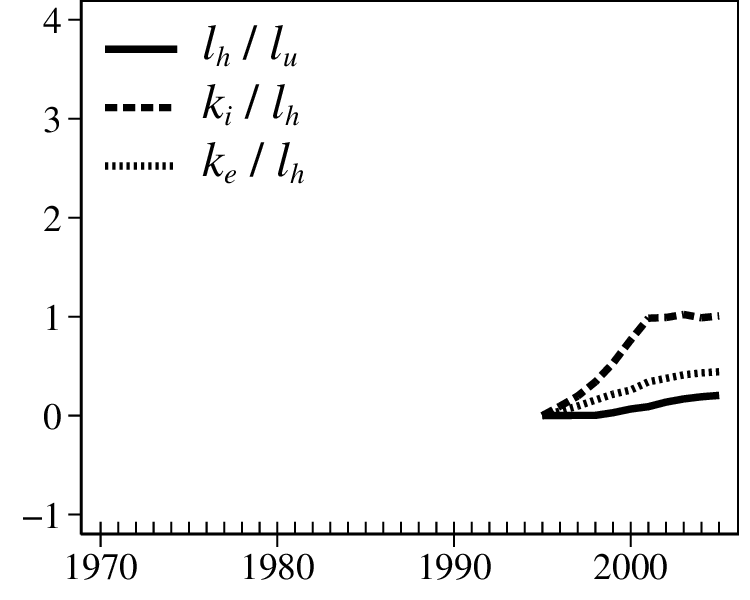}}\subfloat[Denmark]{
\centering{}\includegraphics[scale=0.35]{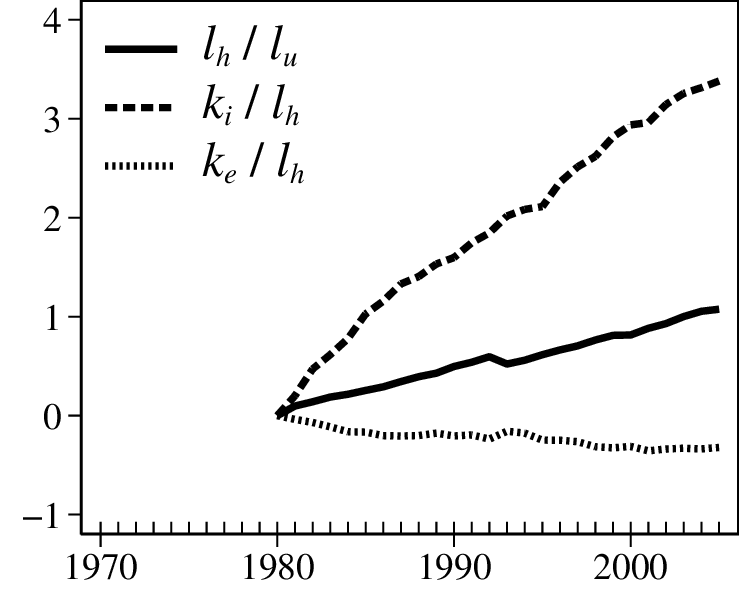}}\subfloat[Finland]{
\centering{}\includegraphics[scale=0.35]{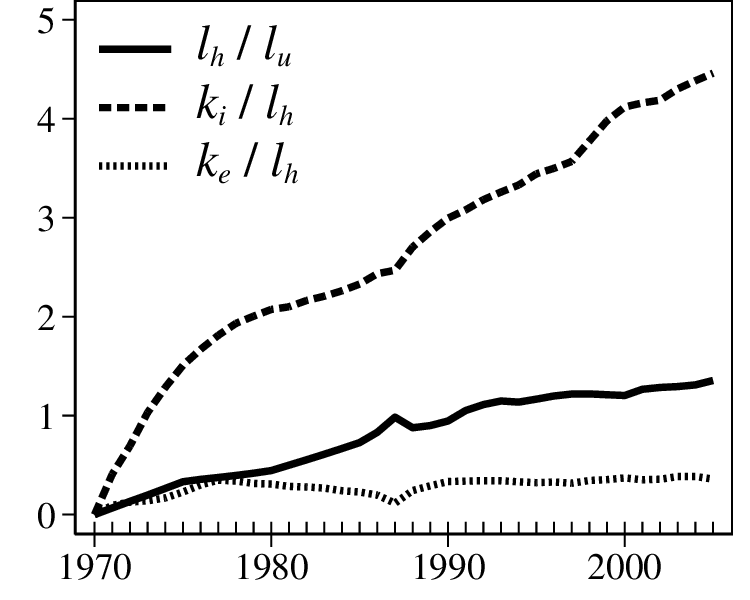}}
\par\end{centering}
\begin{centering}
\subfloat[Germany]{
\centering{}\includegraphics[scale=0.35]{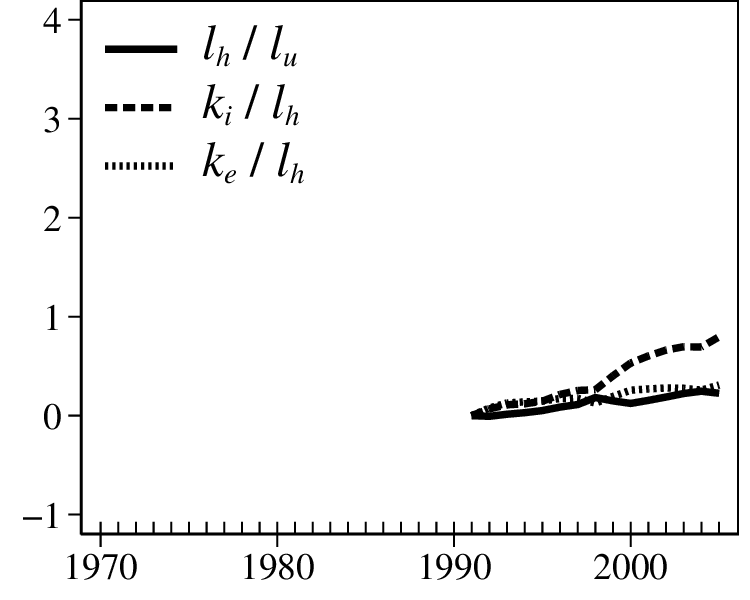}}\subfloat[Italy]{
\centering{}\includegraphics[scale=0.35]{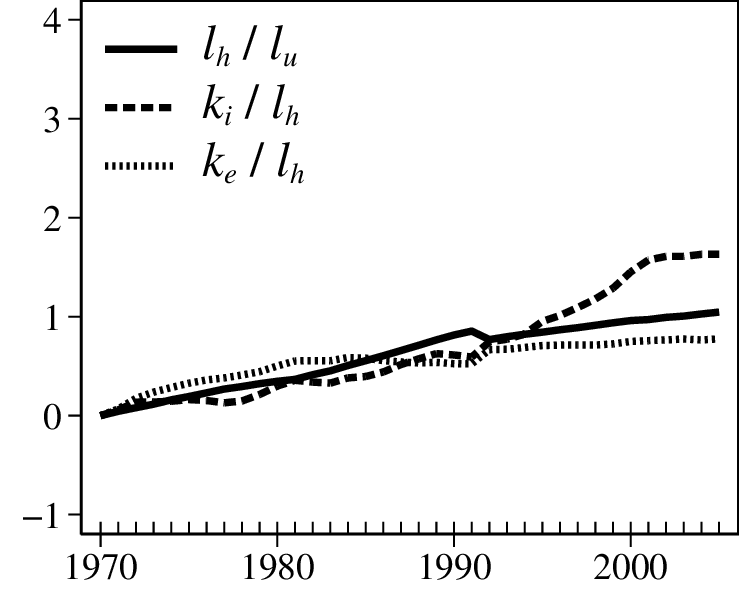}}\subfloat[Japan]{
\centering{}\includegraphics[scale=0.35]{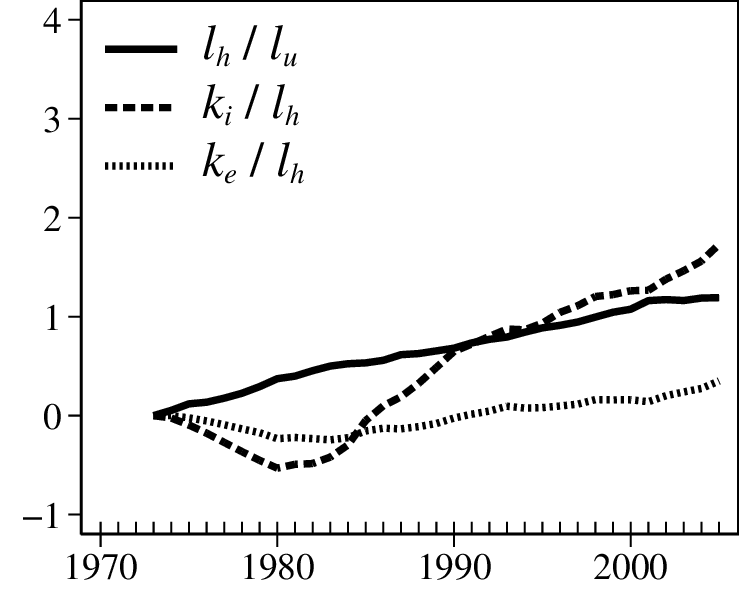}}
\par\end{centering}
\begin{centering}
\subfloat[Netherlands]{
\centering{}\includegraphics[scale=0.35]{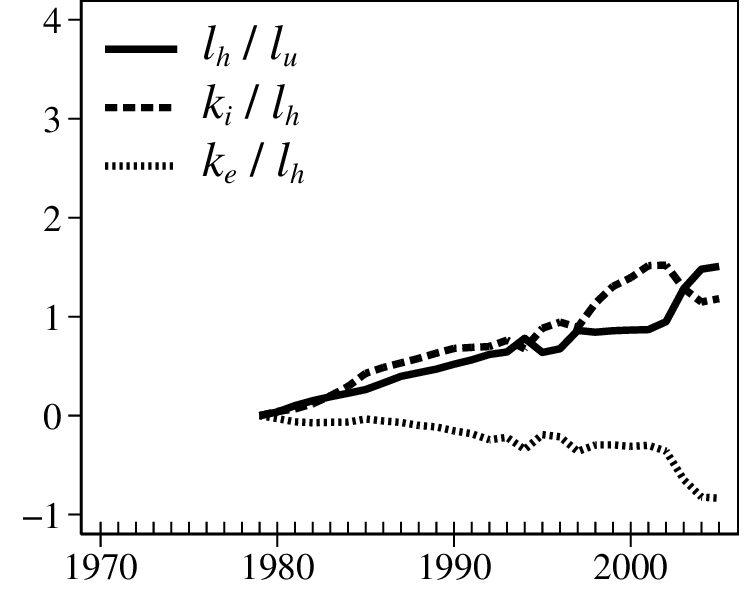}}\subfloat[Portugal]{
\centering{}\includegraphics[scale=0.35]{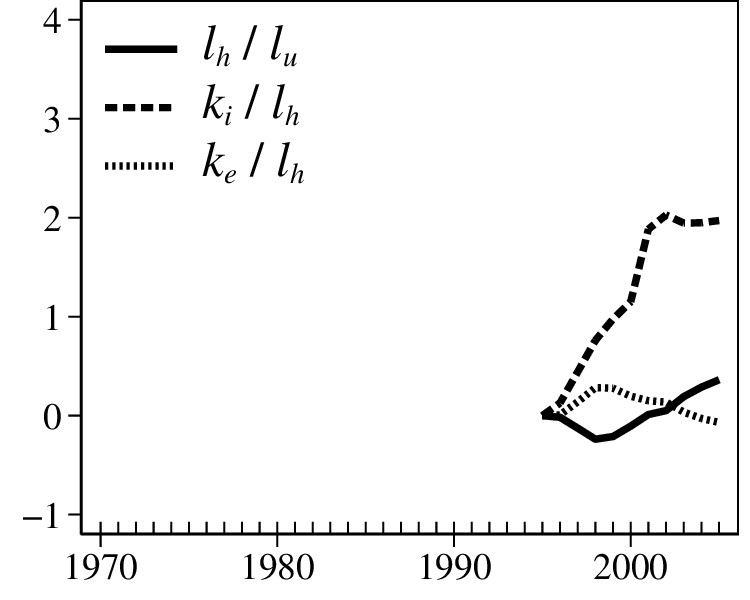}}\subfloat[Slovenia]{
\centering{}\includegraphics[scale=0.35]{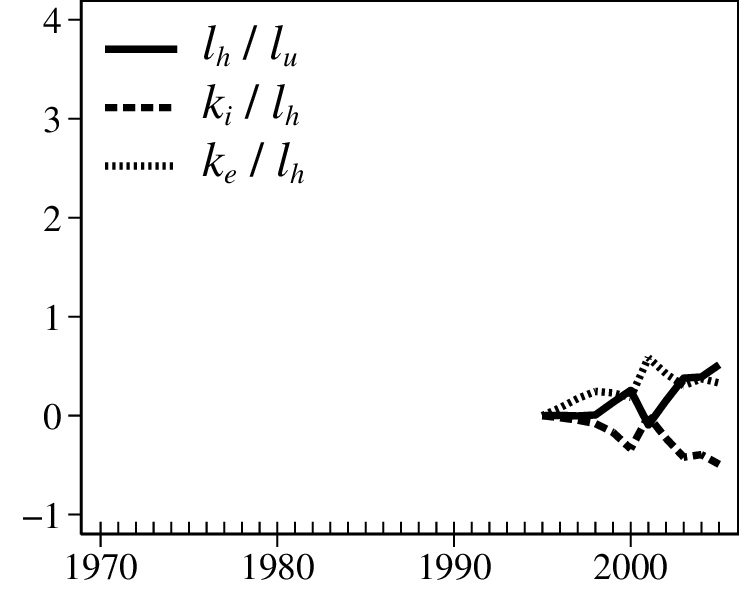}}
\par\end{centering}
\begin{centering}
\subfloat[Sweden]{
\centering{}\includegraphics[scale=0.35]{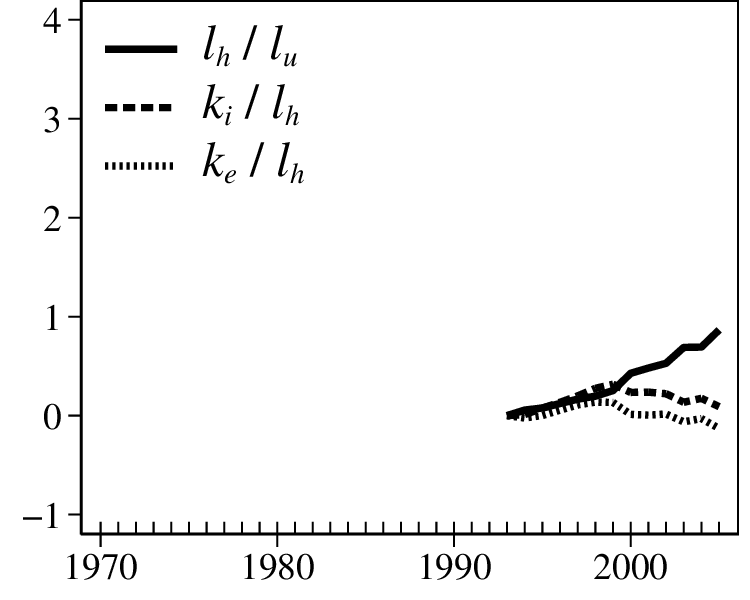}}\subfloat[United Kingdom]{
\centering{}\includegraphics[scale=0.35]{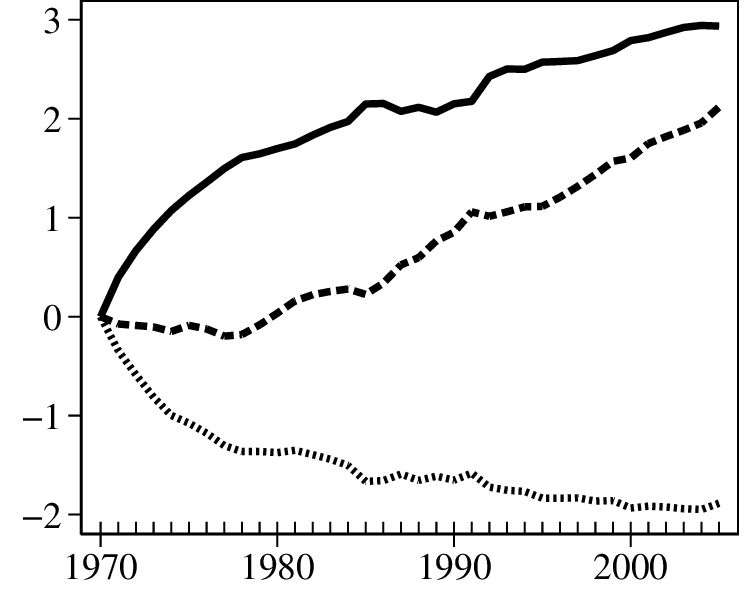}}
\par\end{centering}
\textit{\footnotesize{}Notes}{\footnotesize{}: The bold, dashed, and
dotted lines indicate the ratios of skilled to unskilled labor ($\ell_{h}/\ell_{u}$),
ICT capital to skilled labor ($k_{i}/\ell_{h}$), and capital equipment
to skilled labor ($k_{e}/\ell_{h}$), respectively. All series are
logged and normalized to zero in the initial year.}{\footnotesize\par}
\end{figure}

\begin{figure}[h]
\caption{Relative factor quantities in the service sector of OECD countries\label{fig: LhLuKiKe_oecd_service}}

\begin{centering}
\subfloat[Australia]{
\centering{}\includegraphics[scale=0.35]{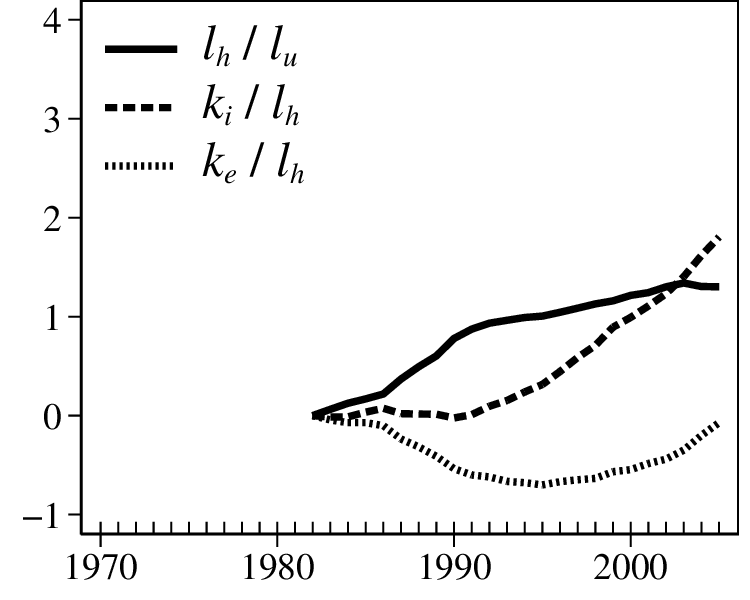}}\subfloat[Austria]{
\centering{}\includegraphics[scale=0.35]{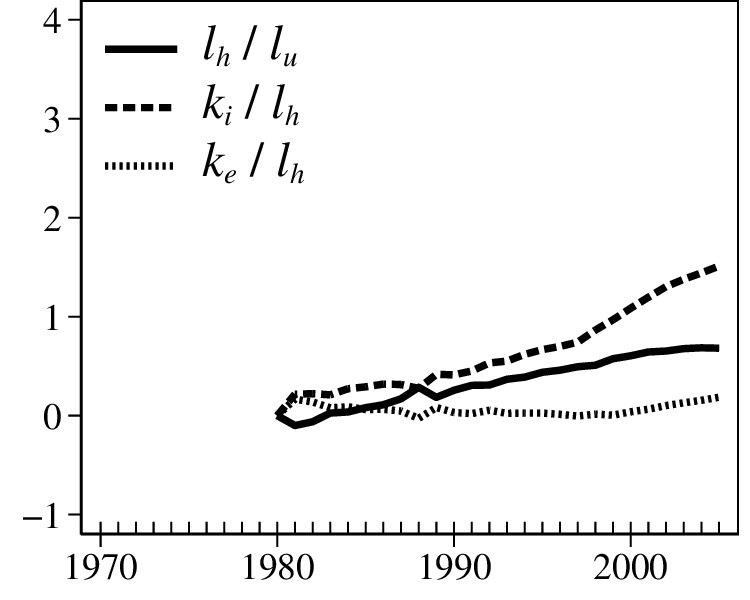}}
\par\end{centering}
\begin{centering}
\subfloat[Czech Republic]{
\centering{}\includegraphics[scale=0.35]{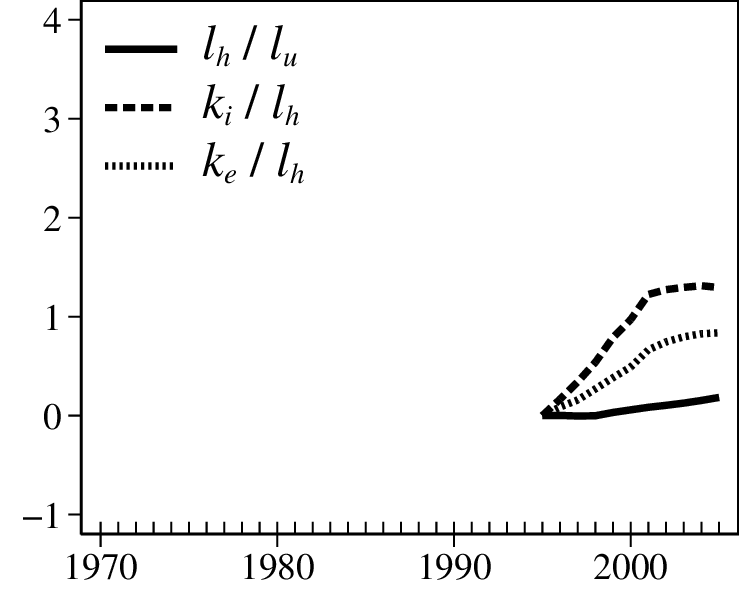}}\subfloat[Denmark]{
\centering{}\includegraphics[scale=0.35]{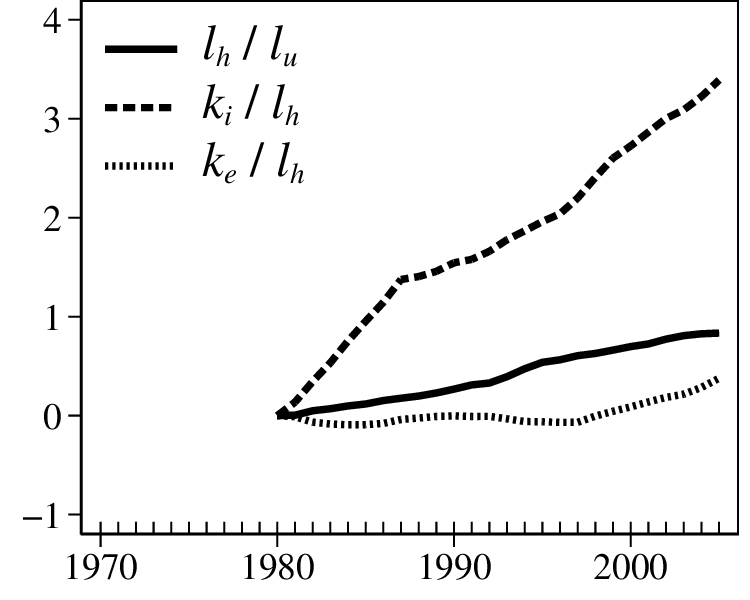}}\subfloat[Finland]{
\centering{}\includegraphics[scale=0.35]{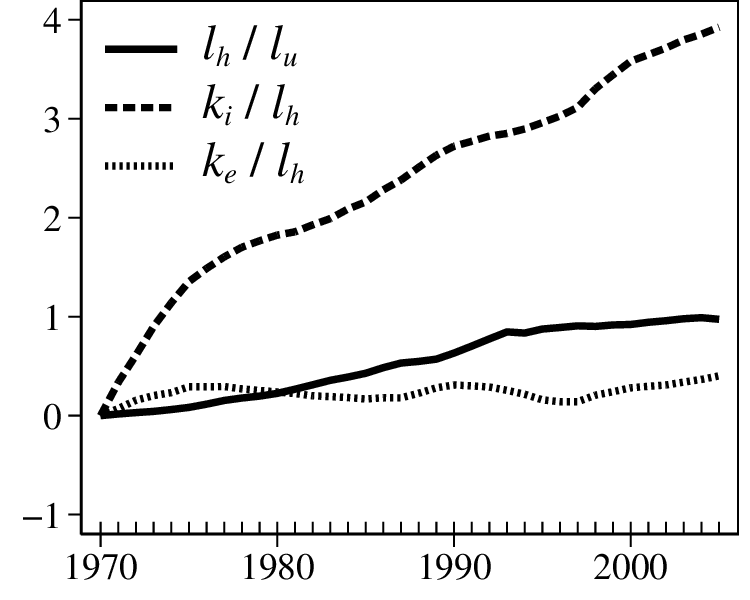}}
\par\end{centering}
\begin{centering}
\subfloat[Germany]{
\centering{}\includegraphics[scale=0.35]{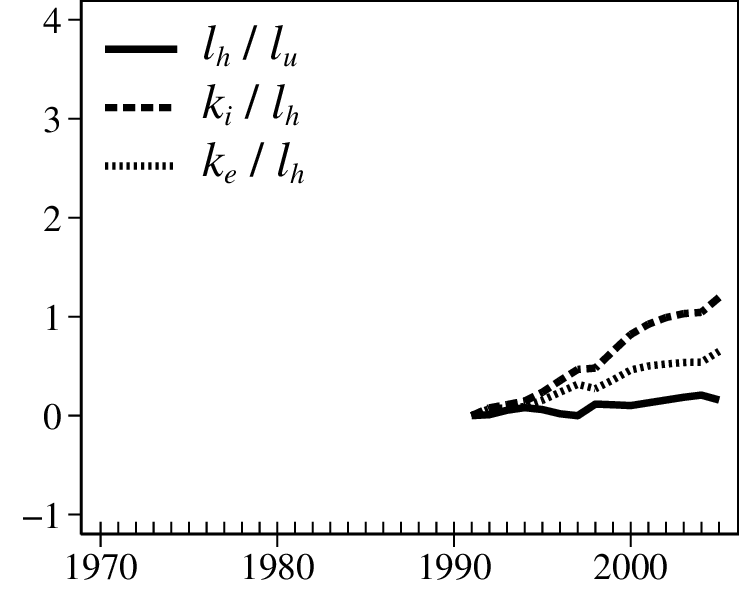}}\subfloat[Italy]{
\centering{}\includegraphics[scale=0.35]{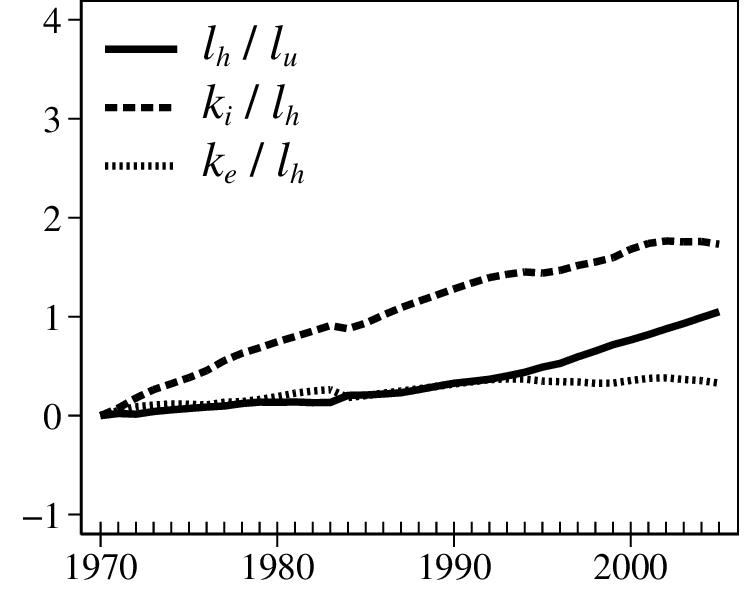}}\subfloat[Japan]{
\centering{}\includegraphics[scale=0.35]{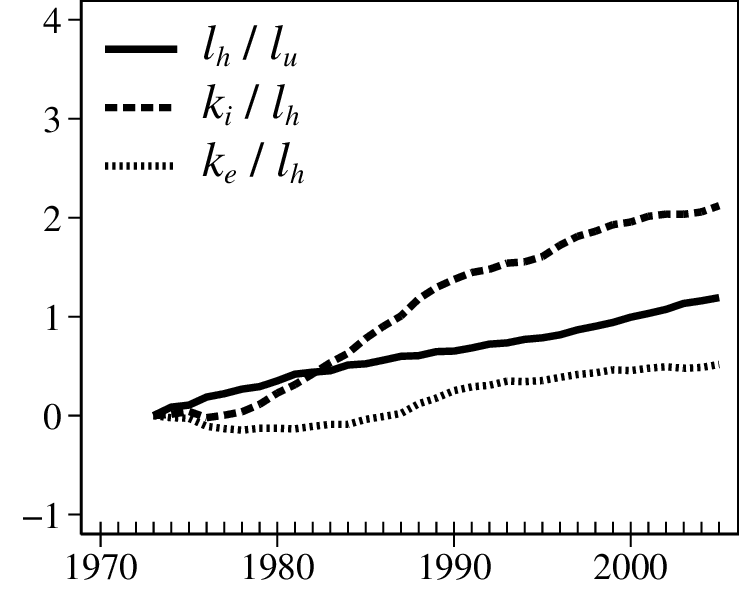}}
\par\end{centering}
\begin{centering}
\subfloat[Netherlands]{
\centering{}\includegraphics[scale=0.35]{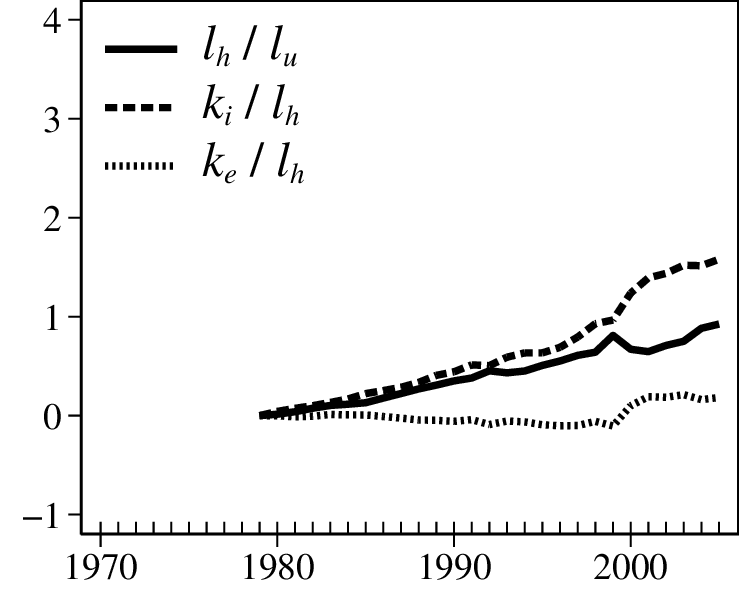}}\subfloat[Portugal]{
\centering{}\includegraphics[scale=0.35]{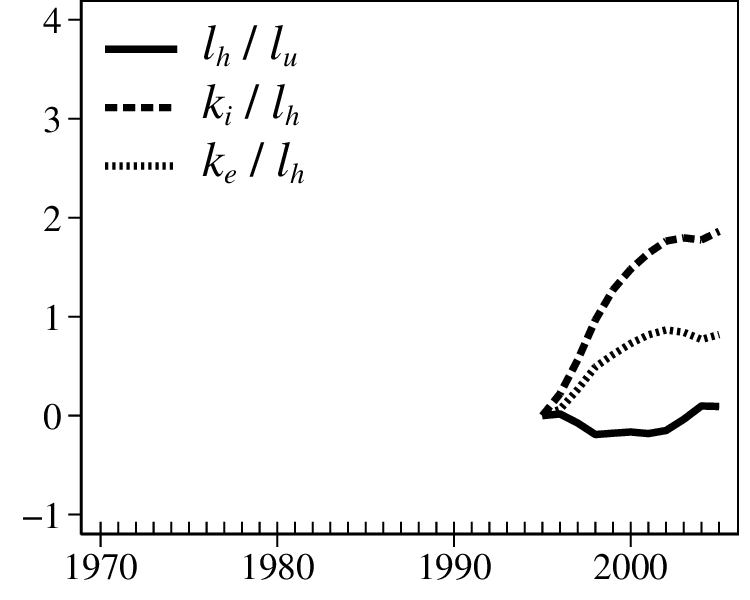}}\subfloat[Slovenia]{
\centering{}\includegraphics[scale=0.35]{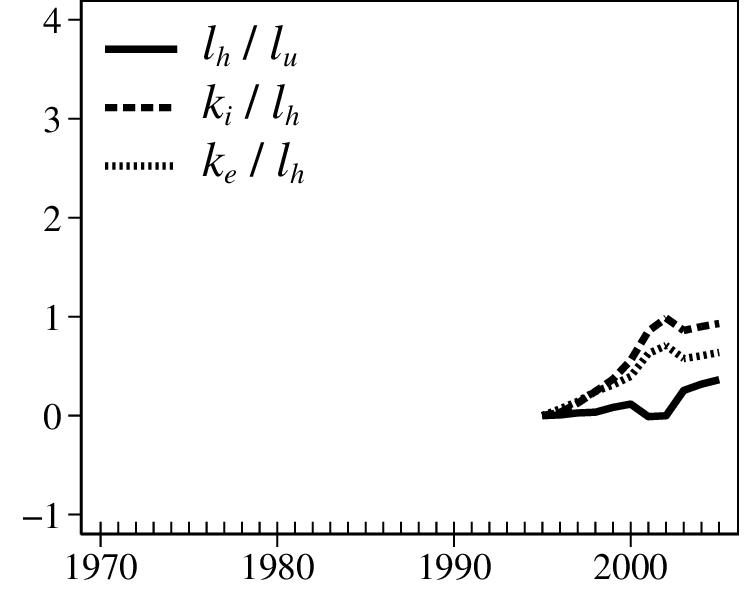}}
\par\end{centering}
\begin{centering}
\subfloat[Sweden]{
\centering{}\includegraphics[scale=0.35]{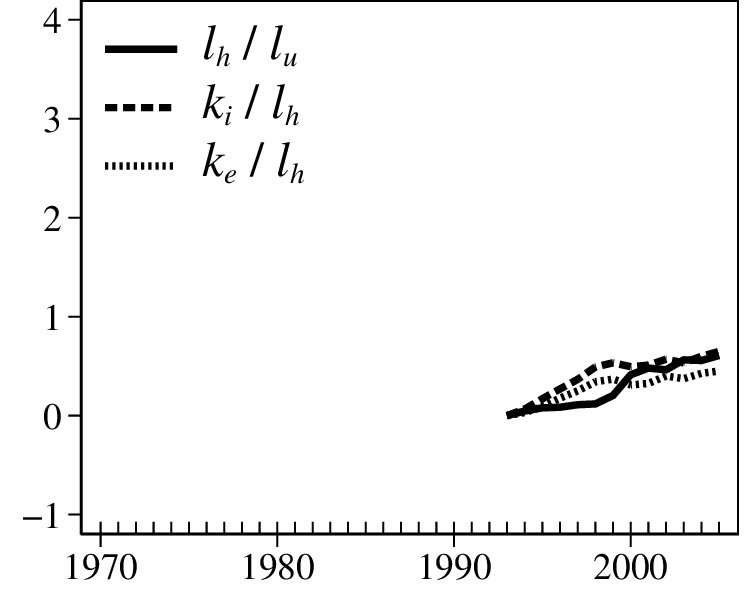}}\subfloat[United Kingdom]{
\centering{}\includegraphics[scale=0.35]{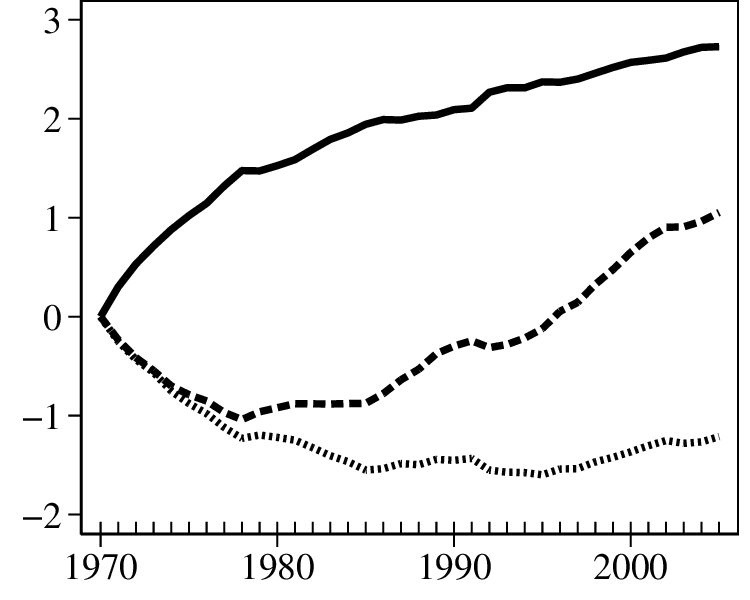}}
\par\end{centering}
\textit{\footnotesize{}Notes}{\footnotesize{}: The bold, dashed, and
dotted lines indicate the ratios of skilled to unskilled labor ($\ell_{h}/\ell_{u}$),
ICT capital to skilled labor ($k_{i}/\ell_{h}$), and capital equipment
to skilled labor ($k_{e}/\ell_{h}$), respectively. All series are
logged and normalized to zero in the initial year.}{\footnotesize\par}
\end{figure}

Figures \ref{fig: Ri/Ro_oecd_goods}\textendash \ref{fig: LhLuKiKe_oecd_service}
show the trends in the rental prices of ICT and non-ICT capital and
the relative quantities of factors in OECD countries other than the
United States.

\subsection{Shapley decomposition\label{subsec: shapley}}

We implement the Shapley decomposition to measure the contribution
of each factor to changes in the skill premium. Changes in the skill
premium can be represented as:

\begin{equation}
\Delta\ln\left(\frac{w_{h}}{w_{u}}\right)=f\left(d_{1},d_{2},\ldots,d_{\mathcal{K}}\right),
\end{equation}
where $d_{\kappa}$ for $\kappa\in\{1,2,\ldots,\mathcal{K}\}$ denotes
the determinant factor of the skill premium. The subscripts $c$,
$n$, and $t$ are suppressed for notational simplicity. Let $\Gamma(\Upsilon)$
denote the level of the skill premium if the factors, $d_{\kappa}$
for $\kappa\notin\Upsilon$, are held fixed at the initial value,
$o=(o_{1},\ldots o_{\mathcal{K}})$ denote the order in which the
factors are held fixed, and $\Upsilon(o_{\eta},o)=\{o_{\eta^{\prime}}|\eta^{\prime}>\eta\}$
denote the set of factors that remain unfixed after the $\eta$-th
factor is held fixed. The marginal contribution of the $\kappa$-th
factor can be written as
\begin{equation}
\Lambda_{d_{\kappa}}^{o}=\Gamma\left(\Upsilon\left(d_{\kappa},o\right)\cup\left\{ d_{\kappa}\right\} \right)-\Gamma\left(\Upsilon\left(d_{\kappa},o\right)\right).
\end{equation}

The Shapley decomposition is implemented by averaging the marginal
contributions of each component over all possible sequences \citep{Shorrocks_JoEI13}.
Let $\mathcal{O}$ denote the set of sequences. The Shapley decomposition
is
\begin{equation}
\Lambda_{d_{\kappa}}=\frac{1}{\mathcal{K}!}\sum_{o\in\mathcal{O}}\Lambda_{d_{\kappa}}^{o}.
\end{equation}
Changes in the skill premium can then be decomposed as
\begin{equation}
\Delta\ln\left(\frac{w_{h}}{w_{u}}\right)=\Lambda_{k_{i}}+\Lambda_{\ell_{h}}^{CSC}+\Lambda_{\ell_{h}}^{RLQ}+\Lambda_{\ell_{u}}+\Lambda_{A_{u}^{h}}+\Lambda_{A_{h}^{i}}.
\end{equation}
where $\Lambda_{k_{i}}$ and $\Lambda_{\ell_{h}}^{CSC}$ represent
the respective marginal contributions of $k_{i}$ and $\ell_{h}$
in the second term of equation \eqref{eq: Wh/Wu}, $\Lambda_{\ell_{h}}^{RLQ}$
and $\Lambda_{\ell_{u}}$ represent the respective marginal contributions
of $\ell_{h}$ and $\ell_{u}$ in the third term of equation \eqref{eq: Wh/Wu},
and $\Lambda_{A_{u}^{h}}$ and $\Lambda_{A_{h}^{i}}$ represent the
respective marginal contributions of $A_{h}/A_{u}$ in the first term
and $A_{i}/A_{h}$ in the second term of equation \eqref{eq: Wh/Wu}.
The capital\textendash skill complementarity effect can be calculated
as the sum of the first two terms, the relative labor quantity effect
can be calculated as the sum of the third and fourth terms, and the
relative factor-augmenting technology effect can be calculated as
the sum of the last two terms. The differences in changes in the skill
premium between the United States and other countries can be decomposed
as
\begin{equation}
\Delta\ln\left(\frac{w_{h,\text{US},nt}}{w_{u,\text{US},nt}}\right)-\Delta\ln\left(\frac{w_{h,cnt}}{w_{u,cnt}}\right)=\sum_{d_{\kappa}}\left(\Lambda_{d_{\kappa},\text{US},nt}-\Lambda_{d_{\kappa},cnt}\right),
\end{equation}
where $d_{\kappa}=\{k_{i},\ell_{h},\ell_{u},A_{h}/A_{u},A_{i}/A_{h}\}$,
and $\Lambda_{\ell_{h}}=\Lambda_{\ell_{h}}^{CSC}+\Lambda_{\ell_{h}}^{RLQ}$.
\end{document}